\documentclass[journal]{IEEEtran} % compsoc
\usepackage{tikz}
\usepackage{amsmath}

\usepackage{filecontents}

\usepackage{tikz}
\usepackage{amsmath}
\usepackage{filecontents}
\usepackage{multirow}
\usepackage{subfigure}
\usepackage{booktabs}
\usepackage{color}
\usepackage{algorithm}  
\usepackage{algorithmicx}  
\usepackage[noend]{algpseudocode}
\usepackage{amssymb}
\usepackage{caption}
\usepackage{subcaption}
\usepackage{graphicx}
\usepackage[normalem]{ulem}
\usepackage{threeparttable}
\usepackage{pifont}
\usepackage{hyperref}
\usepackage{soul}
\usepackage{enumitem}
\makeatletter

\newcommand{\Rmnum}[1]{\expandafter\@slowromancap\romannumeral #1@}
\makeatother
\newcommand{\CN}{\ensuremath{\mathcal{N}}}

\newcommand{\BR}{\ensuremath{\mathbb{R}}}

\definecolor{darkblue}{RGB}{70,130,180}

\definecolor{darkblue2}{RGB}{55, 126, 184}

\useunder{\uline}{\ul}{}

\usepackage[english]{babel}

\begin{document}

\title{FLTracer: Accurate Poisoning Attack Provenance in Federated Learning} %Autonomous, Plan-and-Execute

\author{Xinyu Zhang\textsuperscript{$*$}, Qingyu Liu\textsuperscript{$*$}, Zhongjie Ba,~\IEEEmembership{Member,~IEEE}, Yuan Hong,~\IEEEmembership{Senior Member,~IEEE}, Tianhang Zheng, Feng Lin,~\IEEEmembership{Senior Member,~IEEE}, Li Lu,~\IEEEmembership{Member,~IEEE}, and Kui Ren,~\IEEEmembership{Fellow,~IEEE}
        % <-this % stops a space
% \thanks{Xinyu Zhang, Qingyu Liu, Zhongjie Ba, Feng Lin, Li Lu, and Kui Ren are with the School of Cyber Science and Technology, Zhejiang University, Hangzhou, Zhejiang, China, 310027 and ZJU-Hangzhou Global Scientific and Technological Innovation Center, No.733 Jianshe San Road, Xiaoshan District, Hangzhou, Zhejiang, China, 311200. E-mail: {xinyzhang53, qingyuliu, zhongjieba, flin, li.lu, kuiren}@zju.edu.cn.}
% \thanks{Yuan Hong is with the University of Connecticut. E-mail: yuan.hong@uconn.edu.
% }
% \thanks{Tianhang Zheng is with the University of Toronto. E-mail: th.zheng@mail.utoronto.ca.
% }
% \thanks{Corresponding author: Zhongjie Ba.}
\thanks{\textsuperscript{$*$}Equal Contribution. }  %\href{https://github.com/Eyr3/FLTracer}{} % Corresponding author.
% }
% \author{Xinyu Zhang, Qingyu Liu, Zhongjie Ba, Yuan Hong, Tianhang Zheng, Feng Lin, Li Lu, and Kui Ren
        % <-this % stops a space
}

\markboth{}% IEEE TRANSACTIONS ON INFORMATION FORENSICS AND SECURITY
{Shell \MakeLowercase{\textit{et al.}}: A Sample Article Using IEEEtran.cls for IEEE Journals}


\maketitle

%requirement- https://signalprocessingsociety.org/publications-resources/information-authors

\begin{abstract}
%-------------------------------------------------------------------------------

% [Not more than 200 words, and preferably closer to 150.]
% has emerged as
Federated Learning (FL) is a promising distributed learning approach that enables multiple clients to collaboratively train a shared global model. However, recent studies show that FL is vulnerable to various poisoning attacks, which can degrade the performance of global models or introduce backdoors into them. In this paper, we first conduct a comprehensive study on prior FL attacks and detection methods. The results show that all existing detection methods are only effective against limited and specific attacks. Most detection methods suffer from high false positives, which lead to significant performance degradation, especially in not independent and identically distributed (non-IID) settings. 
To address these issues, we propose FLTracer, the first FL attack provenance framework to accurately detect various attacks and trace the attack time, objective, type, and poisoned location of updates. Different from existing methodologies that rely solely on cross-client anomaly detection, we propose a Kalman filter-based cross-round detection to identify adversaries by seeking the behavior changes before and after the attack. Thus, this makes it resilient to data heterogeneity and is effective even in non-IID settings. To further improve the accuracy of our detection method, we employ four novel features and capture their anomalies with the joint decisions. Extensive evaluations show that FLTracer achieves an average true positive rate of over $96.88\%$ at an average false positive rate of less than $2.67\%$, significantly outperforming SOTA detection methods.
\footnote{Code is available at \url{https://github.com/Eyr3/FLTracer}.}

% For different attacks,  To further reduce false positives caused by data heterogeneity, FLTracer exploits task and domain consistency detection to monitor variations of adversaries’ feature patterns across clients and rounds. 

\end{abstract}

\vspace{-3mm}
\section{Introduction}

% \vspace{-1.5mm}
\IEEEPARstart{F}{ederated} learning (FL)~\cite{mcmahan2017communication}, a machine learning technique that allows multiple parties to collaborate on data analysis without revealing their private data, has been deployed at scale in the last decade~\cite{hard2018federated, Designin30:online}. 
%, e.g.,  Gboard and personalized Siri~\cite{}. 
Under FL, a central server (e.g., cloud server) distributes a global model to multiple clients (e.g., smartphones and IoT devices). These clients individually train local models based on the global model with their own data. Then, the central server collects local model updates, aggregates them into a new global model, and starts the next round. However, recent studies show that malicious clients can apply a variety of poisoning attacks to the global model by disrupting their local models: \emph{Untargeted attacks} that attempt to slow down the learning process, decrease the overall accuracy of the task, or even make the global model unusable~\cite{2019Local, shejwalkar2021manipulating, shejwalkar2021back}. \emph{Backdoor attacks} (aka. targeted attacks) that aim to inject the backdoor into a model so that it incorrectly predicts Trojan data samples (i.e., samples containing attacker-chosen triggers) as the attacker-chosen class, without affecting the predictions for clean data samples~\cite{bagdasaryan2020backdoor,bagdasaryan2020blind,zhang2022neurotoxin}. %unmodified

To better understand the security threat posed to FL by poisoning attacks, we conduct a comprehensive empirical study of existing attacks. Our study assesses the effectiveness, stability, and robustness of 14 untargeted and 16 backdoor attacks using 9 metrics (see $\S$~\ref{sec:performance_attack}).
% and resistance to state-of-the-art (SOTA) detection methods (including Byzantine-robust aggregation~\cite{blanchard2017machine}, updates anomaly detection~\cite{shejwalkar2021manipulating, nguyen2022flame}, and model sanitization~\cite{liu2018fine}). 
Our findings indicate that the majority of untargeted attacks lead to a significant decrease in the accuracy or stability of the global model. Furthermore, most backdoor attacks can rapidly inject backdoors into the global model within approximately 30 training rounds, which persists for a duration exceeding 200 rounds.

\begin{figure*}[!ht]
  % \vspace{-9mm}
  \centering
  \includegraphics[width=\linewidth]{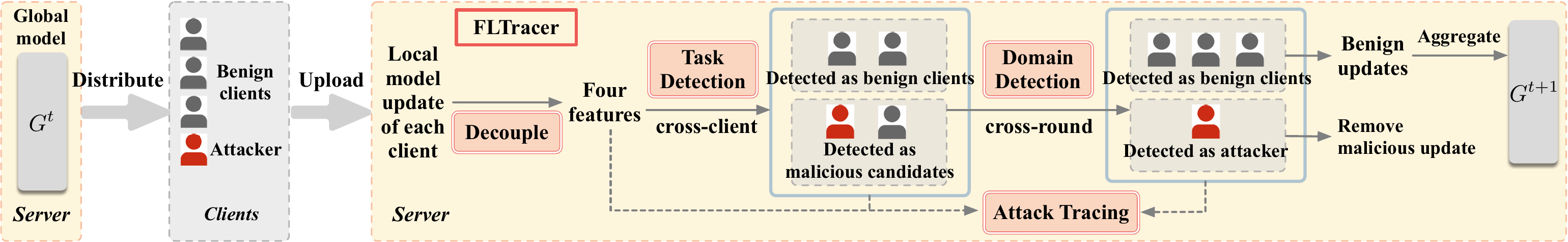}
  \vspace{-5mm}
  \caption{Overview of Attack Provenance (FLTracer), which accurately detects adversaries and traces the attack.} % The server aggregates sanitized updates and generate a new global model.
  \vspace{-6mm}
  \label{fig:intro}
\end{figure*}

However, existing detection methods suffer from limitations in accurately detecting attacks, especially in non-IID settings, for two primary reasons. 
First, adversaries can employ various stealthy attacks, such as untargeted adaptive attacks~\cite{qi2022towards} that modify small portions of update parameters, and backdoor attacks~\cite{bagdasaryan2020blind, xie2019dba} that employ covert strategies involving data and code modifications. These stealthy attacks present difficulties for current detection methods~\cite{blanchard2017machine, shejwalkar2021manipulating} that rely solely on identifying anomalous updates, resulting in a low true positive rate (TPR). The introduction of malicious updates can compromise the accuracy of the global model or inject backdoors. Second, in non-IID settings, the presence of highly heterogeneous feature and label distributions causes significant variation in benign client updates~\cite{zhu2021federated}. Consequently, existing cross-client detection methods~\cite{nguyen2022flame} are prone to misclassifying and removing these highly heterogeneous updates from training, resulting in a high false positive rate (FPR). The lack of unique data from misclassified benign clients further diminishes the accuracy of the global model.

In this paper, we introduce \emph{attack provenance} in FL, which accurately detects adversaries and traces the attack time, objective, type, and poisoned location of model updates. To our best knowledge, we propose the first FL attack provenance framework, FLTracer, as illustrated in Fig.~\ref{fig:intro}. Regarding detection, FLTracer aims to minimize the introduction of poisoned data from a wide range of attacks (achieving high TPR) while maximizing benign data utilization (for low FPR), particularly in non-IID settings. Specifically, FLTracer extracts features from the model updates of each client and applies cross-client and cross-round detection to identify them. Based on the detection results, FLTracer aggregates the benign updates and provides attack tracing for malicious updates. 

FLTracer addresses the aforementioned two limitations. 
Firstly, we introduce a novel update decoupling technique to generate four new sensitive features that represent the impact of various attacks from both value and function perspectives. Detecting these fine-grained features with \emph{joint decisions} is more accurate than detecting updates directly, especially for stealthy attacks. Secondly, to further reduce the FPR of backdoor attacks in non-IID settings, we propose task and domain detection to capture \emph{intrinsic differences} between adversary and benign clients. The detections are based on the following observations: (1) the adversary's training task is highly inconsistent with benign clients due to backdoor attacks involving learning both original and backdoor tasks, and (2) the adversary's feature domain (such as training data and code) changes drastically before and after the attack, in contrast to benign clients. 
Therefore, we propose task similarity ($\mathrm{TSim}$) to measure the training task differences across clients in the same round, as well as domain distance ($\mathrm{DDist}$) and a new \emph{Kalman filter-based} domain detection to monitor and capture domain changes of the same client across rounds.

Our detection is resilient to data heterogeneity and effective in non-IID settings, as demonstrated by the following analysis. In real-world FL scenarios, such as autonomous driving and multiple edge devices, the data distribution collected by vehicles, security cameras, and other IoT devices tends to remain relatively constant or change slowly w.r.t. time, weather conditions, light, etc~\cite{malki2022machine}. Different from prior works~\cite{zhang2022fldetector, cao2023fedrecover}, which consider benign clients' data distribution to remain constant, our framework also accommodates asymptotic distributions. For benign clients with fixed data distribution, we use the newly proposed Kalman filter to estimate the domain state of each client. Adversaries can be easily identified by their abrupt changes in domain state when launching attacks. For benign clients with gradually changing data, the parameters of the Kalman filter are dynamically updated to adapt to the evolving stream state of each client. Similarly, adversaries can be distinguished from benign clients by their abrupt state changes rather than gradual transitions. In cases where certain benign clients' data undergoes abrupt changes, we sacrifice those clients to ensure that no poisoned data is introduced.
% (see $\S$~\ref{sec:kf_detect})

Furthermore, our framework can trace attack attributions based on our detailed analysis of multiple features. Once an adversary is detected, the attack tracing can be provided: \romannumeral 1) \emph{the timing of the attack}, \romannumeral 2) \emph{the objective and type of the attack}, and \romannumeral 3) \emph{the specific poisoned location of model updates}.

Therefore, our contributions are summarized as follows:

\begin{itemize}[leftmargin=*]
\setlength\itemsep{0.4em}
%To the best of our knowledge, this paper is the first to  large-scale experimental
\item We conduct a comprehensive study of current FL poisoning attacks, including 14 untargeted and 16 backdoor attacks, along with 4 detection methods, in IID and non-IID settings. We evaluate these attacks using 9 metrics and demonstrate that existing attacks pose severe threats to FL. %Moreover, we discover a new adaptive strategy that further enhances targeted attacks' stealthiness. %during the experiment,  and makes them harder to detect

% We perform the first systematic experimental study of existing poisoning attacks and detection methods under FL, including untargeted and targeted attacks. Our assessment of untargeted attacks includes classical and adaptive attacks against Byzantine-robust AGRs. To understand the risks of targeted attacks, we evaluate not only the attack strategies proposed under FL, but also those proposed under centralized learning. In addition, we propose a new convolutional replacement strategy that can be combined with existing targeted attacks to make them more stealthy. We provide nine evaluation metrics to describe the attack feasibility in terms of effectiveness, robustness, and resistance to existing detection methods. 

\vspace{-1.5mm}
\item To the best of our knowledge, we propose the first FL attack provenance framework called FLTracer. It not only accurately detects adversaries with minimum global model performance drop, but also provides attack tracing. FLTracer utilizes 4 new sensitive features to analyze the different crucial impacts of attacks on model updates, enabling the accurate detection of a wide range of attacks. To avoid the effects of data heterogeneity, FLTracer further leverages $\mathrm{TSim}$ and $\mathrm{DDist}$ to capture intrinsic anomalies of adversaries across clients and rounds, respectively.

% We provide the first attack  in FL  that can generate a result and an in-depth explanation of the detection results.% from different perspectives, including feature visual interpretation and feature importance ranking.
\vspace{-1.5mm}
\item We conduct extensive experiments to evaluate FLTracer on 6 types of datasets and 7 types of model architectures. The results show high accuracy in IID and non-IID settings. In the non-IID settings, FLTracer achieves an average TPR of over $96.88\%$ and an average FPR of less than $2.67\%$, significantly outperforming state-of-the-art (SOTA) detection methods. It is worth noting that our detection is instantaneous, with detection validity limited to the current round.
% in $11$ untargeted and five targeted attacks 
% In particular, FLTracer successfully prevents $40$ backdoor injections out of $40$ backdoor attacks without sacrificing the accuracy of the global model. 

\end{itemize}
% Our detection system , achieving  and 2\% in most untargeted and targeted attacks, and achieving accuracy comparable to the benchmark removing malicious clients in most attacks. 
%(MNIST, EMNIST, CIFAR10, and GTSRB) (SimpleNet, AlexNet, ResNet18, VGG16, and ResNet34)

% In addition, most of evaluation metrics for attacks are limited. For example, accuracy reduction and attack success rate are usually used to evaluate untargeted and targeted attacks, respectively. It is insufficient for describing the impact of attacks on the global model. 

\vspace{-2mm}
\section{Background}
\label{sec:background}

% In this section, we will briefly go through basic notations of federated learning. The important notations are summarized in Table~\ref{tab:notation}
% \begin{table}[h]
% \caption{Notation.}
% \begin{tabular}{l|l}
% Term & Description                \\ 
% \hline
% \hline
% % $D_i$       & client $i$-th local dataset \\
% $G^t$       & global model parameters at round $t$ \\
% %$L_i^t$     & local model of client $i$ at round $t$ \\
% $E$         & local epochs \\
% $lr$        & local learning rate \\
% $\eta$      & global learning rate \\
% $\nabla l$  & gradient of the classification loss $l$ \\
% $\theta, \theta^*$    & model update, disturbed model update \\
% $m$         & number of categories of the model \\
% % $r$    & representation layer             \\
% % $c$   & classification layer \\
% % $w$    & weight in update           \\
% % $b$    & bias in update             \\
% $\mu$       & mean of an distribution \\
% $\sigma$    & standard deviation of an distribution \\
% \end{tabular}
% \label{tab:notation}
% \end{table}

% \vspace{-2mm}
\subsection{Federated Learning}
% \vspace{-1.5mm}
% Federated learning (FL) is a machine learning technique that enables multiple parties to collaborate on data analysis and prediction without revealing their private data~\cite{mcmahan2017communication}. 
FL generates a shared model by iteratively training the global model on the central server, with the cooperation of $N$ distributed clients. In each round, the central server aggregates a global model from several local models trained individually by the clients using their local data. 
Specifically, at round $t$, the central server first broadcasts the current global model parameters $G^t$ to a randomly selected subset of $n$ clients. Then, each selected client initializes its local model parameters $L_i^{t}\!=\!G^t$ and updates $L_i^{t}$ with its local dataset. In this process, the selected client updates the local model individually by the stochastic gradient descent optimization for $E$ local epochs with a local learning rate. After obtaining $L_i^{t+1}$, each client calculates and sends the local model update $\theta_i^{t+1} \!=\! L_i^{t+1} - G^t$ to the server. Finally, the central server collects all updates and aggregates the new global model as $G^{t+1} \!=\! G^t+\eta \sum_{i=1}^{n} \theta_i^{t+1}$, where $\eta=\frac{1}{n}$ is the global learning rate. The new global model will be iteratively trained until achieving the predefined high accuracy on the test dataset or number of rounds. 

%Federated learning is a machine learning technique that enables multiple clients to collaboratively train a joint model, each with a private local dataset. The distributed datasets of $N$ clients could be independently and identically distributed (\textit{IID}) or not independently and identically distributed (\textit{non-IID}). Using the client's local dataset $D_i$, a local model is trained by each client. Eventually, the global model obtained by local model aggregation performs well on test data from both the central server and the distributed clients. 
%

\vspace{-3mm}
\subsection{Poisoning Attacks on Federated Learning}
\label{sec:poisonattack}
% Poisoning attacks attempt to disrupt, slow down, or mislead the training of the model. They can be divided into untargeted and targeted attacks depending on the adversary's objectives.
%~\cite{kairouz2021advances}

% \vspace{-1.5mm}
\noindent\textbf{Untargeted attacks:} intend to slow down the training and eventually decrease the accuracy of the global model, including three main types. 
\textit{Add noise attack}~\cite{blanchard2017machine, wu2020federated} affects the training of the global model by altering the local update. It transforms the update $\theta_i$ to $\theta_i^*=\theta_i+\delta$, where $\delta$ is a random noise. 
\textit{Sign-flipping attack}~\cite{li2019rsa} is another local update modification attack. It flips the signs of parameters in an update, i.e., $\theta_i^*=-\theta_i$. It has a greater impact on the training than add noise attack. 
%while keeping the absolute values of the parameters unchanged
\textit{Dirty label attack}~\cite{biggio2012poisoning} in centralized learning is a targeted attack, where the adversary misleads the training samples from one source class to another target class. However, under FL, it behaves like an untargeted attack, %reducing the accuracy of the global model 
since the aggregation of the central server relieves this misdirection. 
%Increasing the number of source labels and target labels or even randomly labeling training samples can also lead a significant decline in the accuracy and recall of the global model.

% There are independent attacks and collusive attacks based on whether the malicious clients collude. In independent attack (e.g., \textit{dirty label attack}, \textit{add noise attack}, and \textit{sign-flipping attack}), malicious clients can poison a model by itself. In collusive attacks(e.g., \textit{Shejwalkar's attack}\cite{shejwalkar2021manipulating}, \textit{Fang's attack}\cite{2019Local}, \textit{Baruch's attack}\cite{baruch2019little}), several malicious clients collaborates to make a more stealthy attack.

%For instance, in an image classification task, the adversary embeds white triangles in a number of "car" images during the training phase, and labels these images as "bird"s. At the end of training, the learned model incorrectly assumes that the white triangles are part of the "bird" feature and predicts the target label "bird". Meanwhile, the model performs normally on the unmodified "car" images. 
\vspace{0.5mm}
\noindent\textbf{Backdoor attacks (aka. targeted attacks):} aim to inject a backdoor into a model, such that the model predicts Trojan data as the attacker-chosen class, while making unaffected predictions on clean data. The process of the backdoor attack consists of a Trojan data synthesis phase and a backdoor injection phase.
% (i.e., data contain triggers) , also known as backdoor attacks, 
For the synthesis of Trojan data, the adversary embeds the trigger into the clean data and mislabels these data as the target category. There are two main types of backdoor triggers. %, as shown in Fig.~\ref{fig:triggers} in $\S$~\ref{sec:expsetup}. 
A \textit{patch trigger} is a small pattern consisting of a few pixels, such as a $3\times 3$ checkerboard~\cite{gu2017badnets} or a reverse-engineered trigger~\cite{liu2018trojaning}. 
A \textit{perturbation trigger} is a perturbation of the same size as the training image, such as a random noise perturbation~\cite{chen2017targeted}, a sinusoidal signal perturbation~\cite{barni2019new}, or a reflected image~\cite{liu2020reflection}. 
% \textit{Distributed triggers}~\cite{xie2019dba} are designed for FL that decomposes a pattern trigger into several independent patterns and embeds the different patterns into several client datasets. 

For the backdoor injection phase, the adversary uses synthetic Trojan data to train backdoor models, including three injection strategies.
\textit{BadNets attack}~\cite{gu2017badnets} employs a classic injection strategy by training a local model with Trojan data containing backdoors and then poisoning the global model. %It requires multiple attack rounds to successfully inject backdoors because the average aggregation weakens the attack. 
% \textit{Model replacement attack} (MRA)~\cite{bagdasaryan2020backdoor} addresses the weaknesses of aggregation by scaling up the malicious updates. Modifying the scaling factor allows malicious updates to survive from the average aggregation. This injection strategy achieves a $100\%$ backdoor accuracy of the global model even in a single round. 
\textit{Model replacement attack} (MRA)~\cite{bagdasaryan2020backdoor} is a one-shot backdoor attack that amplifies the malicious updates by using a scaling factor before uploading to the server.
\textit{Distributed backdoor attack} (DBA)~\cite{xie2019dba} is an FL-specific injection strategy that splits the backdoor trigger into multiple independent triggers and embeds the different triggers in multiple client datasets.

\vspace{0.5mm}

\noindent\textbf{Adaptive attacks:} are proposed to counter existing detection and defense methods. Experimental evidence has demonstrated that existing methods exhibit inadequate detection capabilities against adaptive attacks (see $\S$~\ref{sec:performance_attack}). \textbf{Adaptive untargeted attacks} include Manipulating the Byzantine (\textit{MB attack})~\cite{shejwalkar2021manipulating}, \textit{Fang attack}~\cite{2019Local}, and \textit{LIE attack}~\cite{baruch2019little}, to counter Byzantine-robust aggregations. They require the cooperation of multiple malicious clients for effective execution and formulate an optimization problem to construct a malicious update that not only mimics the benign updates, but also significantly diminishes the model's performance. For \textbf{Adaptive backdoor attacks}, \textit{Blind backdoor attack}~\cite{bagdasaryan2020blind} utilizes multi-task learning and the multiple gradient descent algorithm to balance both the original and backdoor tasks, achieving high performance on both tasks while maintaining stealthiness. We propose the \textit{Feature Replacement Attack} (FRA) to manipulate only the weights of feature extraction layers, which are observed to be the critical updates for backdoor attacks. This approach aligns with the findings of the \textit{Subnet replacement attack}~\cite{qi2022towards}, which effectively injects covert backdoors into models by modifying a limited set of model parameters.

\vspace{-3mm}
\subsection{Kalman Filter} \label{pre:kalman}

% \vspace{-1.5mm}
% The Kalman filter\cite{kalman1960new} is a recursive solution to the discrete-data linear problem. By continuously updating the estimate of the system state, the Kalman filter can provide a more accurate prediction of the actual state of the system. 
The Kalman filter~\cite{kalman1960new} is a recursive solution that utilizes the state-space model of a linear system to optimally estimate the system's state by integrating past observations with a dynamic model of the system's behavior. It is suitable for applications involving time-varying dynamical systems. 
% Its applications include robotics, navigation, signal processing, and econometrics. 
At its core, the Kalman filter uses a state-space model that represents the underlying system to be estimated. The \textit{state vector}, denoted as $x^t$, at \textit{time step} $t$ can be estimated and predicted from the state vector $x^{t-1}$ at the previous moment, and the control input $u^t$, expressed as $x^t = A x^{t-1} + B u^t + w$. Here, $w$ is the \textit{process noise}. The \textit{state transition pattern} $A$ and the \textit{control input matrix} $B$ are system parameters that describe the system's behavior, physical laws, and the relationships between the state variables $x$ and the \textit{control inputs} $u$. It is worth noting that in time-varying dynamical systems, $A$ and $B$ may change over time. In such cases, the parameters can be 
% preferentially 
estimated or updated based on the available historical measurement data and the estimation process. Then, the Kalman filter can be used to predict the next state vector. 
\vspace{-1mm}
\section{Problem Formulation} 
\label{sec:problem}

% \vspace{-2mm}
\subsection{System Model}
\begin{figure}[h]
\vspace{-3mm}
  \centering
  \includegraphics[width=0.85\linewidth]{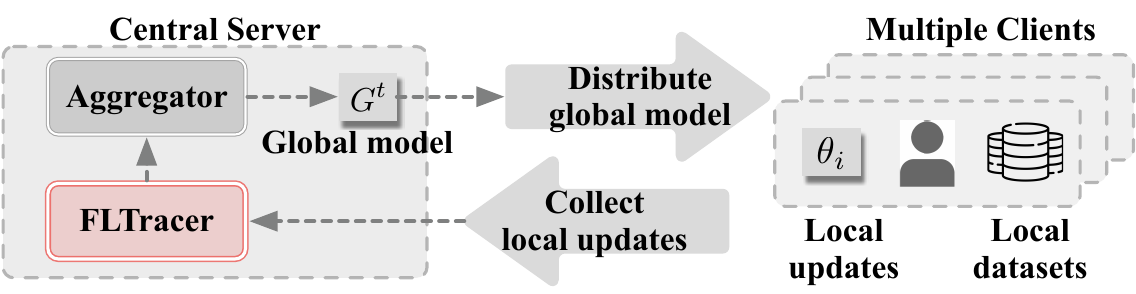}
  \vspace{-2mm}
  \caption{The system model} % The left and right sides indicate two entities. The three colored arrows indicate three channels.
  \label{img:system_model}
  \vspace{-4mm}
\end{figure}

% attackers only in threat model
%Federated Learning enables the central server to collaboratively train a shared model with multiple clients, each with a private local dataset. To produce the shared model, the central server communicates with participating clients and updates the global model parameters iteratively.
% \vspace{-1.5mm}
The FL system commonly involves two entities: an aggregator deployed on the central server and multiple clients. The aggregator distributes the global model and aggregates updates. The clients provide local updates trained on their datasets. To prevent updates of malicious clients from joining the global model, we develop FLTracer, deployed on the central server, as depicted in Fig.~\ref{img:system_model}. FLTracer has no restriction on datasets, i.e., the training data can be {IID} or {non-IID}, regardless of whether it is among benign or malicious clients. We assume that the distribution of a client's data across training rounds remains unchanged or gradually changes over time, which relaxes the assumption of unchanged data distribution in~\cite{cao2023fedrecover, zhang2022fldetector}. 
Specifically, we discuss a practical scenario where clients' real-world driving dataset changed over time in $\S$~\ref{sec:exp_diff_system}.

%\uline{In some scenarios, even though the real-time data collected by the client may change over time, its distribution does not change drastically, in which case FLTracer is effective.}}

% The proposed system includes three communication channels: 
% \textit{The distributing channel} transmits the global model from the aggregator to clients. \textit{The collecting channel} collects updates from selected clients and transmits them to FLTracer. After receiving updates, FLTracer detects and removes malicious updates. \textit{The uploading channel} delivers sanitized updates from FLTracer to the aggregator and is a zero-traffic channel deployed on the central server.
% The aggregator distributes the model structure and parameters to each client during initialization. The aggregator selects a set of clients during the training and sends the new global model parameters to them. 

\vspace{-2mm}
\subsection{Threat Model} 
\label{sec:threatmodel}
% \vspace{-1.5mm}
Following the threat model in the previous works~\cite{blanchard2017machine, bagdasaryan2020backdoor, nguyen2022flame, cao2020fltrust, shejwalkar2021manipulating}, we consider a strong adversary who can compromise up to $50\%$ of the clients. However, the adversary cannot compromise the central server. The compromised clients will attempt to poison the global model, i.e., degrade its performance and inject backdoors. Table~\ref{tab:attacktax1} summarizes the adversary's objectives and attack types. The details of the attacks are described in $\S$~\ref{sec:poisonattack}. Specifically, we consider five types of untargeted attacks for the adversary who wants to degrade the performance of the global model, and five types of backdoor attacks for backdoor injection. Both untargeted and backdoor attacks include adaptive attacks (w.r.t. existing defenses). 
For the central server, we assume it is honest and fully trusted. Thus, the inference attacks and corresponding countermeasures are not considered.
% These injection strategies can be combined with three types of triggers: patch, perturbation, and distributed triggers.
%(such as decreasing the overall accuracy of the task, slowing down the learning process, or even making the global model unusable)

Regarding the adversary's capabilities, we consider a strong adversary who has full control over the compromised clients and knows the detailed design of our detection mechanism and the aggregation algorithm. Specifically, the adversary has five capabilities.
\ding{172}~{Update modification (modif.)}: It can modify updates sent to the central server. \ding{173}~{Dataset modif.}: It controls the local dataset of the compromised clients. Both the data and associated labels can be modified. \ding{174}~{Code modif.}: The adversary is able to modify the local learning process by modifying the loss function. \ding{175}~{Hyperparameter modif.}: It can modify the hyperparameters to optimize the attacks, such as the number of epochs and the local learning rate. \ding{176}~{Timing control}: The adversary can decide the optimal time to launch attacks. %We summarize the adversary's capabilities required for each attack in Table~\ref{tab:attacktax1} in column 3. 
Table~\ref{tab:attacktax1} summarizes the capabilities of each attack.

\begin{table}[t]
\setlength\tabcolsep{1.5pt}
\centering
\scriptsize
\caption{Specific attacks we considered with their corresponding objective, type, and capability.} %“-” denotes the BAE-Insert attack cannot be performed on LSTM.
\vspace{-2mm}
% \resizebox{\linewidth}{!}
% {
\begin{threeparttable}
\begin{tabular}{|c|l||c|l|}
\hline
Objective & Attack type - capability & Objective & Attack type - capability \\
\hline
\multirow{5}{*}{\begin{tabular}[c]{@{}c@{}}Degrading \\ performance\\      (Untargeted \\ attacks)\end{tabular}} & Add noise - \ding{172}\ding{175}\ding{176} &  \multirow{5}{*}{\begin{tabular}[c]{@{}c@{}}Inject the \\ backdoor\\      (Backdoor \\ attacks)\end{tabular}} & BadNets attack (BN) - \ding{173}\ding{175}\ding{176}  \\
 & Sign-flipping - \ding{172}\ding{175}\ding{176} &  & Model replacement(MRA) -\ding{172}\ding{173}\ding{175}\ding{176}  \\
 & Dirty   label - \ding{173}\ding{175}\ding{176}  &  & Distributed backdoor(DBA) - \ding{173}\ding{175}\ding{176} \\
 
 & MB attack$^*$ - \ding{172}\ding{174}\ding{175}\ding{176} &  & Blind backdoor(Blind)$^*$- \ding{173}\ding{174}\ding{175}\ding{176} \\
 & Fang attack$^*$ - \ding{172}\ding{174}\ding{175}\ding{176} &  & Feature replacement(FRA)$^*$-\ding{172}\ding{173}\ding{175}\ding{176} \\
 \hline
\end{tabular}
\begin{tablenotes} 
\item[] ''$*$'' denotes adaptive untargeted attacks or adaptive backdoor attacks.  
\item[] $\sharp$ with the capabilities of \ding{175} and \ding{176}, all attacks will be more effective. 
\end{tablenotes}
\end{threeparttable}
% }
\label{tab:attacktax1}
\vspace{-5mm}
\end{table}

% For the central server, we assume it is honest and fully trusted. Thus, inference attacks and corresponding countermeasures are not considered.
\vspace{-3mm}
\subsection{Design Goal}
\label{sec:design_goal}
% \vspace{-1.5mm}
We aim to design a framework that can accurately detect a variety of attacks and provide attack tracing in both IID and non-IID settings, achieving the following goals: 
% \begin{itemize}
%     \item     
% \end{itemize}

\vspace{0.5mm}
\noindent\textbf{Accuracy:} The framework should guarantee the accuracy of the global model against SOTA poisoning attacks (including adaptive) in both IID and non-IID settings. Specifically, the detection should achieve a high TPR and a low FPR, so that the vast majority of benign clients can join the training. 

\vspace{0.5mm}
\noindent\textbf{Backdoor-proof:} The framework should prevent backdoor injection into the global model even if the adversary performs a strong and stealthy backdoor attack. 
% guarantee the security of the global model, i.e.,

% \YH{to me, seems ``effectiveness'' and ``security'' are somewhat vague, can be more concrete, e.g., accuracy and backdoor-proof}

\vspace{0.5mm}
\noindent\textbf{Provenance:} The framework should provide the attack tracing, including the attack time, objective, type, and specific poisoned location of the model update. 

\vspace{-1mm}
\section{Key Observations for Attack Provenance} \label{sec:keyobservation}

\begin{figure}[!h]
  \centering
  % \vspace{-3mm}
  \includegraphics[width=\linewidth]{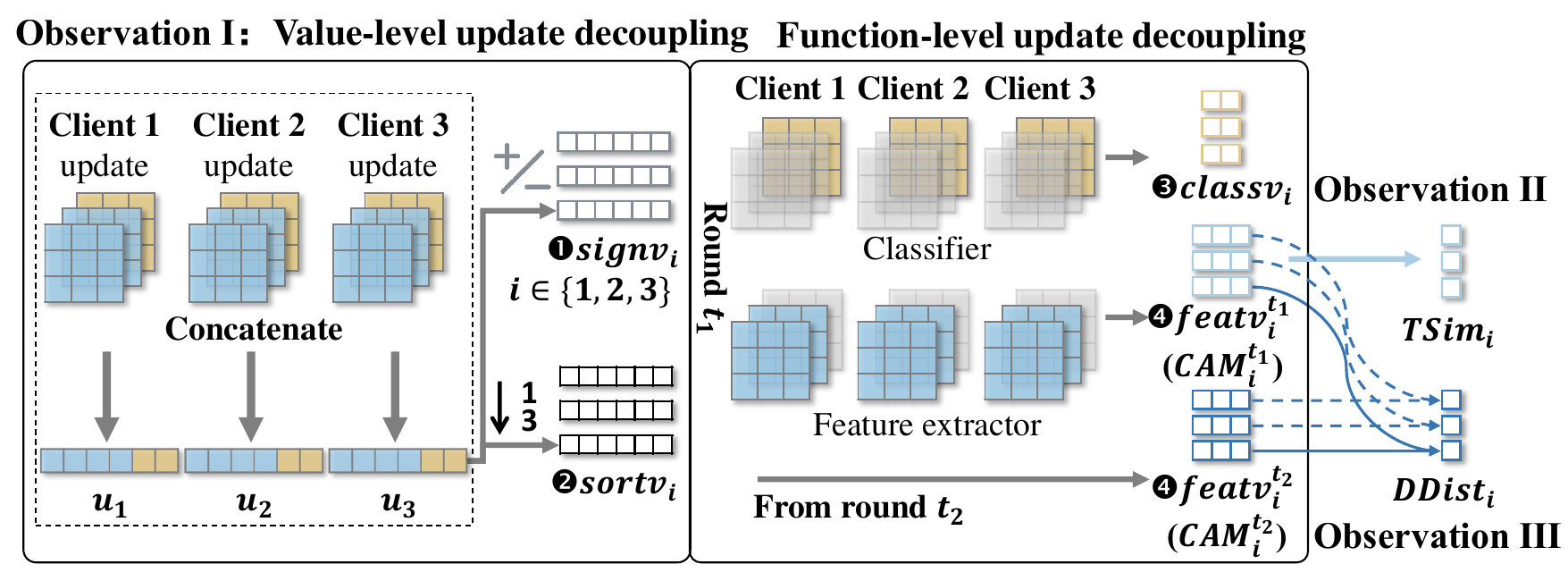}
  \vspace{-5mm}
  \caption{Overview of the key observations for FLTracer}
  \label{fig:observation}
  \vspace{-4mm}
\end{figure}

% \vspace{-1.5mm}
In this section, we present three key observations of FLTracer, as depicted in Fig.~\ref{fig:observation}. To achieve attack provenance, we propose four new features (see $\S$~\ref{sec:observation1}) to analyze update anomalies at different critical locations of the update. To further improve the detection accuracy of backdoor attacks, we introduce $\mathrm{TSim}$ (see $\S$~\ref{sec:observation2}) and $\mathrm{DDist}$ (see $\S$~\ref{sec:observation3}) to capture intrinsic differences between malicious and benign clients from task and domain perspectives, respectively. 
% task anomalies between clients within the same round and domain anomalies between rounds within the same client, respectively. 

\vspace{-3mm}

\vspace{-3mm}
\subsection{Poisoned Portions Analysis} \label{sec:observation1}
% \vspace{0.5mm}
% \noindent
\textbf{Observation~\Rmnum{1}:} % (decoupling) 
Previous methods~\cite{blanchard2017machine, shejwalkar2021manipulating} typically distinguish between malicious and benign updates by detecting anomalies in the concatenation of the update values of the client $i$ ($u_i$). However, our results in $\S$~\ref{sec:performance} show that these methods are only effective for certain specific attacks for two primary reasons: First, the magnitude of the update values is larger in shallow layers compared to deep layers, leading to the limited reflection of deep layer anomalies in $u_i$. 
Second, adaptive attacks such as FRA selectively modify critical updates, causing the unmodified normal values in $u_i$ to mask the anomalies in the modified portions. Therefore, we decouple updates into four sensitive features to detect anomalies, as decoupled training outperforms joint training~\cite{kang2019decoupling}. 
% drawing inspiration from the superior performance of decoupled training over joint training~\cite{wang2021decoupling, kang2019decoupling},

\textbf{Update decoupling at the value and function level can analyze the specific poisoned portions of model updates by various attacks.} As described in $\S$~\ref{sec:threatmodel}, adversaries have the ability to execute a variety of attacks, either by directly manipulating updates or by using stealthy strategies such as modifying the dataset or code during training.

% For attacks that directly modify updates, we decouple the concatenation of entire updates ($u_i$) into symbols and numeric values. \ding{182}~{Sign vector (\texttt{signv})} is the sign matrix of $u_i$, reflecting the adversary's modification of the update symbols. \ding{183}~{Sort vector (\texttt{sortv})} is a vector of the same size as $u_i$, and each element in $\texttt{sortv}_i$ represents the order of the corresponding element ($u_i$) among $u_{\{i\in [n]\}}$. For example, for the $1$st element, we sort the $1$st elements of $n$ clients in ascending order. If the result of the sorting is from client $s_1$ to client $s_n$, then the $1$st element for each client is $\texttt{sortv}_i^1=s_i$, where $i$ denotes the client $i$. The \texttt{sortv} reflects the change in the order of all parameters of a client. This is a more robust property, where each parameter contributes equally to the anomaly of an update, thus balancing anomalies in shallow and deep layers. 

% we decouple the symbol and numeric value of each element in the concatenation of entire updates. 
% is the sign matrix of $u_i$, 
% is the order matrix of $u_i$ among $u_{\{i\in [n]\}}$. 
For attacks that directly modify updates, we decouple the concatenation of updates ($u_i$) into symbolic and numeric values.
\ding{182}~{Sign vector (\texttt{signv})} is the concatenation of the signs of each element in $u_i$, reflecting the adversary's alterations to the update symbols.
\ding{183}~{Sort vector (\texttt{sortv})} is the same size as $u_i$, and each element in $\texttt{sortv}_i$ indicates the order of the corresponding element ($u_i$) within $u_{\{i\in[n]\}}$. 
Specifically, for the $1$st element, we arrange the $1$st element of $n$ clients in ascending order. If the sorting result ranges from client $s_1$ to client $s_n$, then the $1$st element for each client $i$ is $\texttt{sortv}_i^1\!=\!s_i$. The \texttt{sortv} captures changes in the order of all elements for a client. This property ensures a higher level of robustness, as each element contributes equally to the anomaly of an update, thereby balancing anomalies in both shallow and deep layers. 

For attacks that modify the dataset and code, the changed training process results in an alteration in the model's functionality. Thus, we decouple the updates regarding the classifier and the feature extractor. 
% For attacks that modify the dataset and code, we observe that anomalies can be detected within different functional blocks of the model. 
\ding{184}~{Classifier vector (\texttt{classv})} is the concatenation of weights and biases of the classification layer, capturing the adversary's modification of the decision boundary. 
% \ding{185}~{Feature extractor vector (\texttt{featv})} extracts the fine-grained anomaly of the feature extractor, reflecting the adversary's modifications to the feature space. For feature extractors in simple neural networks, such as fully-connected neural networks, \texttt{sortv} is sufficient for detection. For other $k$-layer neural networks (such as Transformer), each encoder layer is considered a module, and all parameters in the encoder layer are concatenated. The anomaly score ($\rho_i^\kappa\in\BR$) of each encoder layer is computed using Principal Component Analysis (PCA). All anomaly scores of the encoder layer are concatenated as $\texttt{featv}_i=\{\rho_i^1,\cdots, \rho_i^k\}\in\BR^k$. 
\ding{185}~{Feature extractor vector (\texttt{featv})} extracts the fine-grained anomaly of the feature extractor, indicating the adversary's modifications to the feature space. For $k$-layer neural networks, such as Transformer~\cite{dosovitskiyimage}, Principal Component Analysis (PCA)~\cite{wold1987principal} is employed to compute anomaly scores ($\rho_i^\kappa\!\in\!\BR$) for each encoder layer (or block) within the feature extractor. All anomaly scores of the encoder layer are concatenated as $\texttt{featv}_i=\{\rho_i^1,\cdots, \rho_i^k\}\!\in\!\BR^k$. 

Especially for the $k$-layer CNN, the detection of backdoors is challenging (see $\S$~\ref{sec:performance_attack}). Considering that the convolution kernel serves as a processing unit~\cite{zeiler2014visualizing}, we introduce a novel feature extractor, called the Convolution Anomaly Matrix (\texttt{CAM}). This matrix contains the anomaly vectors ($A^\kappa_i$) from all convolutional layers. The $\texttt{CAM}_i$ for client $i$ is computed as 

\vspace{-6mm}

\begin{align}
\small
\label{eq:cam}  \texttt{CAM}_i&=\{A^1_i, A^2_i, \cdots, A^k_i\},\\
\label{eq:a}    A^\kappa_i&= w^\kappa_i \circledast w^\kappa_{\{i\in [n]\}}.
\end{align}
\vspace{-5mm}

\noindent Here, $w^\kappa_i\!\in\! \BR^{C^\kappa_{i} \!\times\! C^\kappa_{2} \!\times\! L^\kappa_1 \!\times\! L^\kappa_2}$ denotes the convolution kernels in layer $\kappa$, where $C^\kappa_{1}, C^\kappa_{2}, L^\kappa_1\!\times\! L^\kappa_2$ represents the numbers of output channels, input channels, and kernel size. The operation $\circledast$ calculates the anomaly vector $A^\kappa_i\!\in\! \BR^{C^\kappa_{1} \times C^\kappa_{2}}$ for client $i$ among all participants (see Appendix~\ref{sec:details_CAA}). \emph{In this paper, we describe the detection approach for the most complex features (i.e., $\texttt{CAM}_i$ and $A_i$). For the relatively simpler features $\texttt{featv}_i$ and $\rho_i$, our detection can use them instead of $\texttt{CAM}_i$ and $A_i$.} %Alg.~\ref{alg:cam}

\begin{figure}[]
% \vspace{-3mm}
  \centering
  \includegraphics[width=\linewidth]{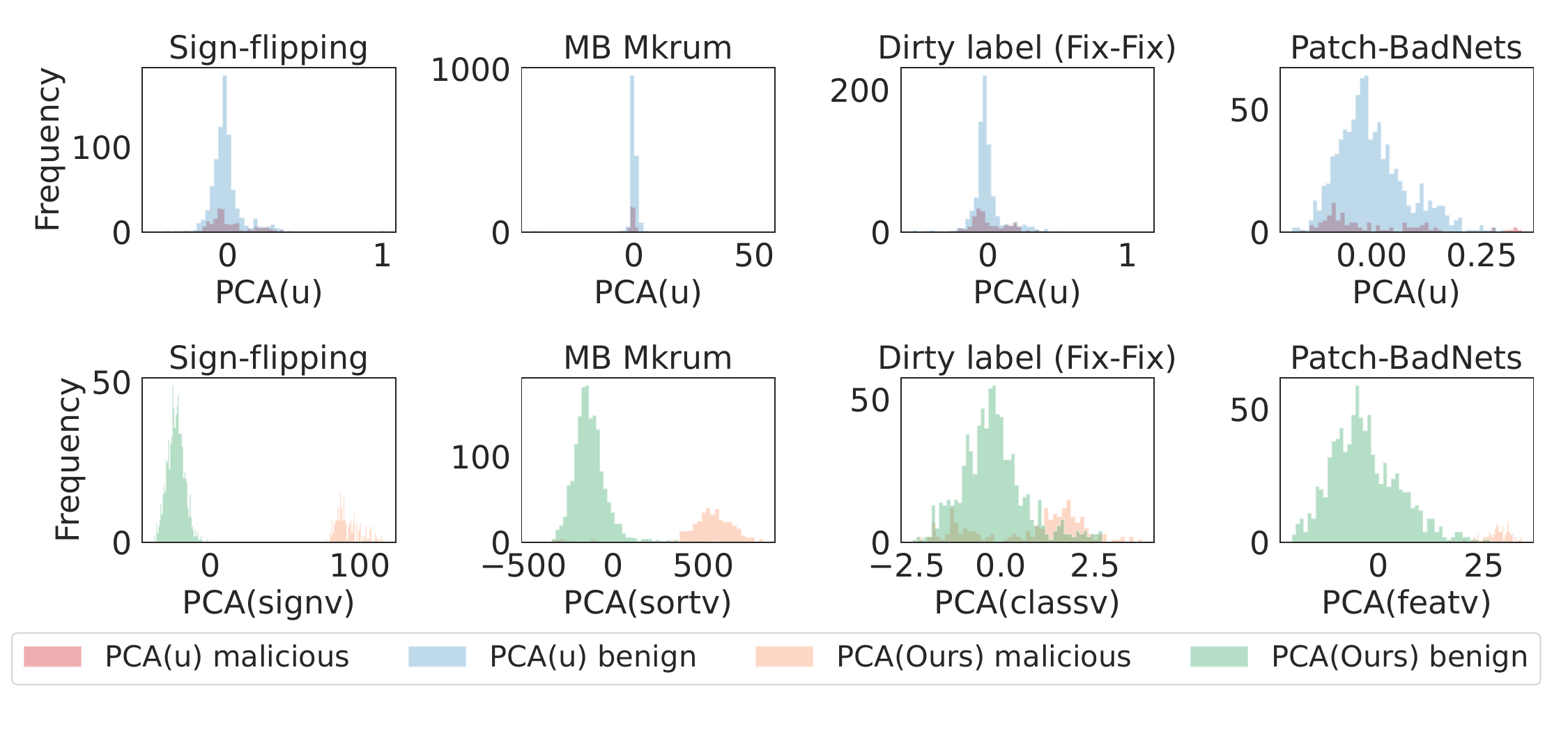}
  \vspace{-6mm}
  \caption{The distribution of dimensionality reduction results for the four features of malicious and benign clients. The top and bottom rows show the results for $u$ and our features.}
  \vspace{-6mm}
  \label{fig:diff_MT}
\end{figure}

To illustrate the advantages of our features over the feature in the previous work~\cite{shejwalkar2021manipulating} (i.e., $u$), we perform four attacks on CIFAR10 in the non-IID settings. Fig.~\ref{fig:diff_MT} shows the distribution of the PCA results for these four features. Our features discriminate the malicious and benign clients better than $u$.

\vspace{-3mm}
\subsection{Task Consistency Analysis} \label{sec:observation2}
% \vspace{0.5mm}
% \noindent 
\textbf{Observation~\Rmnum{2}:} % (task detection) 
In FL, the central server collaborates with $N$ clients to perform the shared task, resulting in a consistent pattern within the convolution kernels of their local model. However, in a backdoor attack, this pattern deviates from the normal fixed pattern because the models are trained for both backdoor and regular tasks. To distinguish between backdoor and normal models, we leverage the similarity of convolution kernels across clients. For two convolution feature vectors $A_1$ and $A_2$, we define the {\textit{Task Similarity}} ($\mathrm{TSim}\in \BR$) as \par

\vspace{-3mm}

\begin{small}
\begin{equation}
\label{eq:tsim}  \mathrm{TSim}_{A_1,A_2}=\frac{(A_1-\overline{A_1})\cdot (A_2-\overline{A_2})}{\Vert A_1-\overline{A_1} \Vert \  \Vert A_2-\overline{A_2} \Vert},
\end{equation}
% \vspace{-1mm}
\end{small}%
where $\overline{A_i}\in \BR$ is the average of $A_i$. Here we use an adjusted cosine similarity metric, which mitigates the effect of different client data domains~\cite{post}. In essence, TSim focuses solely on assessing the correlation between convolution kernels within a convolutional layer, regardless of the overall value of $A_i$. 

\begin{figure}[]
  \centering
  % \vspace{-4mm}
  \includegraphics[width=\linewidth]{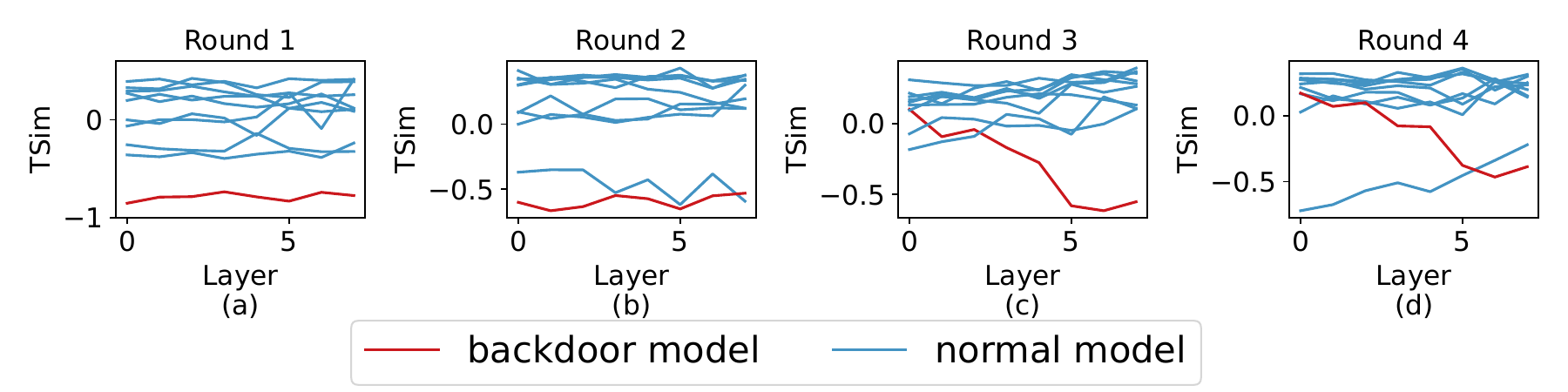}
  \vspace{-5mm}
  \caption{TSim of different layers in the backdoor and normal models each round. Lower TSim indicates more anomalous.}
  \label{fig:deep_conv}
  \vspace{-5mm}
\end{figure}

\textbf{A successful backdoor attack involves compromising the deep layers' weights in the feature extractor.} To analyze this, we specifically examine the variations between the layers in backdoor models and normal models. We extract the \texttt{CAM}s for each participant and plot the TSim in each layer (see Fig.~\ref{fig:deep_conv}). Notably, the deep layers of the backdoor models exhibit more anomalies, suggesting that the backdoor features tend to be embedded in the deeper layers. This phenomenon can be attributed to the fact that the model's shallow layers mainly capture texture information, whereas the deeper layers primarily contain structural details and possess a larger receptive field~\cite{zeiler2014visualizing}. 
It is worth noting that some normal models with highly heterogeneous data exhibit similar anomalies to the adversary (see Fig.~\ref{fig:deep_conv} (b) and (d)). 

Therefore, in $\S$~\ref{sec:detect}, we present a Task Detection approach to assess the consistency of tasks among participants within a round. Models that incorporate additional tasks, such as backdoor models or models trained with highly diverse data, are identified as potential malicious candidates. Furthermore, we introduce Observation~\Rmnum{3} to differentiate between backdoor models and models trained using highly heterogeneous data.

%A larger TSim indicates that the training tasks of the convolutional layers of model $1$ and model $2$ are more similar. 

% This is because training these samples is also considered to inject an edge-case backdoor into the model. %in $\S$~\ref{sec:cas}

% According to the feature replacement attack, the backdoor can be injected only when the feature extractor's weights are modified. When we freeze the feature extractor's weights, the backdoor attack almost fails no matter how many rounds the bias and classifier are trained, as shown in Fig.~\ref{fig:fixed_conv}. 

% Relying solely on Task detection is insufficient, as it leads to the inadvertent deletion of numerous edge clients. Therefore, it is crucial to employ advanced observations further to differentiate and exclude them accurately.

\vspace{-3mm}
\subsection{Domain Consistency Analysis} \label{sec:observation3}
% \vspace{0.5mm}
% \noindent 
\textbf{Observation~\Rmnum{3}:} % (domain detection)
In a consistent data distribution, each client maintains fixed domain features. The fluctuation in the domain features of each processing unit (i.e., convolution kernel) between two consecutive rounds represents the domain variation of a client. 
% The convolutional layer acts as a feature extractor for the NN, and each processing unit (i.e., convolution kernel) represents a part of the domain variation. Thus, we accumulate the differences of all convolution kernels between two rounds to characterize the domain variation of a client.
We define the \textit{Domain Distance} ($\mathrm{DDist} \in \BR$) between two convolution feature vectors of client $i$ in different rounds $t_1$ and $t_2$ as: % (i.e., $A_i^{t_1}$ and $A_i^{t_2}$) 
\begin{equation}
\small
\setlength{\abovedisplayskip}{0.05cm}
\begin{aligned}
\label{eq:ddist}   \mathrm{DDist}_{A_{i}^{t_1}, A_{i}^{t_2}} =\frac{1}{len(A_i^{t_1})} \cdot {\sum_{j=1}^{len(A_i^{t_1})} (A_{i,j}^{t_1}-A_{i,j}^{t_2})},
\end{aligned}
\setlength{\belowdisplayskip}{0.05cm}
\end{equation}
where $j$ denotes each element in $A_i^{t_1}$ and $A_i^{t_2}$.
The average anomaly value (${1}/len(A_i^{t_1}$)) is used to balance anomalies for feature maps of different sizes.

% The domain features of the malicious client change dramatically before and after the attack due to the dataset and code modification, especially in backdoor attacks. 

\begin{figure}[!ht]
\vspace{-4mm}
  \centering
   \includegraphics[width=\linewidth]{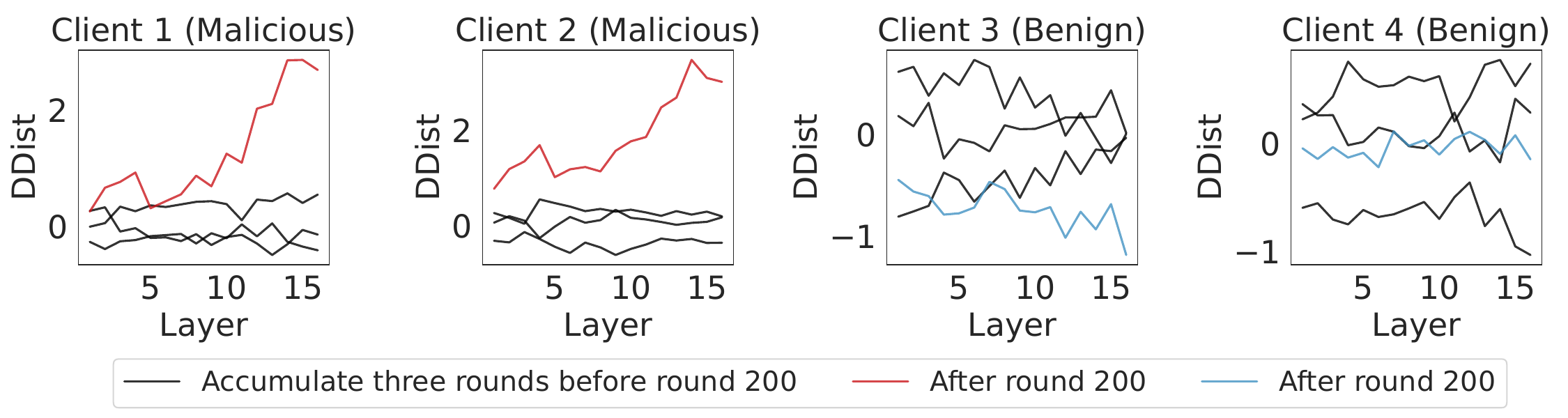}
   \vspace{-6mm}
  \caption{DDist of different layers for malicious and benign clients before and after the attack (launched at round 200).} %The left and right represent the DDist of malicious and benign clients.
  \vspace{-4mm}
  \label{fig:timesim}
\end{figure}

% \noindent 
\textbf{In non-IID settings, we accurately identify malicious clients through domain distance analysis. This approach rests on the observation that benign clients' domains remain consistent across rounds, whereas malicious ones exhibit inconsistency.} 
%The update features of a benign client are consistent across rounds, while the update features of malicious clients are inconsistent.} 
A common strategy for backdoor adversaries is to train a model normally and then launch a backdoor attack after the model converges~\cite{bagdasaryan2020backdoor}. 
The introduction of a backdoor trigger in the training data results in a sudden and drastic shift in the data domain. 
% resulting in a drastic variation in the domain features. 
Conversely, the data domain of benign clients stays constant or undergoes gradual fluctuations over time. This divergence in behavior allows for anomaly detection by scrutinizing the domain consistency of a given client across rounds.
% The perturbation on the dataset can cause a sudden change of domain features across rounds.
To show the differences between malicious and benign clients, we plot their $\mathrm{DDist}$ across rounds (see Fig.~\ref{fig:timesim}). The $\mathrm{DDist}$ of malicious clients remains consistent prior to the attack, but exhibits drastic variation after the attack, clearly diverging from that of the benign clients.
% {Fig. shows the DDist of malicious and benign clients before and after the attack. The DDist of benign clients remains consistent in different rounds. In contrast, the DDist of malicious clients remains consistent before the attack but is greatly different after the attack, distinguishable from benign clients.}

% Clients who maliciously modify the dataset and code during training are considered malicious. 

% We also discuss the case when the targeted attack is launched at the first round of training, as illustrated in $\S$~\ref{}.

% For each client, Domain detection maintains a reference feature $\overline{V}_i$ and calculates the DDist of $\overline{V}_i$ with the current $V_i$. A large DDist indicates that the feature domain of the client changes dramatically as the number of training rounds increases.

% \vspace{-2mm}

% as a constraint condition of Task detection (Observation ~\Rmnum{2}) 
Therefore, we propose a Kalman filter-based Domain Detection approach for detecting domain shift in clients across rounds in $\S$~\ref{sec:kf_detect}. Clients are identified as malicious if they exhibit anomalies in both $\mathrm{TSim}$ and $\mathrm{DDist}$. Additionally, the consistency of $\mathrm{DDist}$ can help distinguish potential malicious candidates identified by Task detection (due to their highly heterogeneous data) from actual malicious clients.

\begin{figure*}[!ht] 
  \centering
  % \vspace{-mm}
  \includegraphics[width=0.98\linewidth]{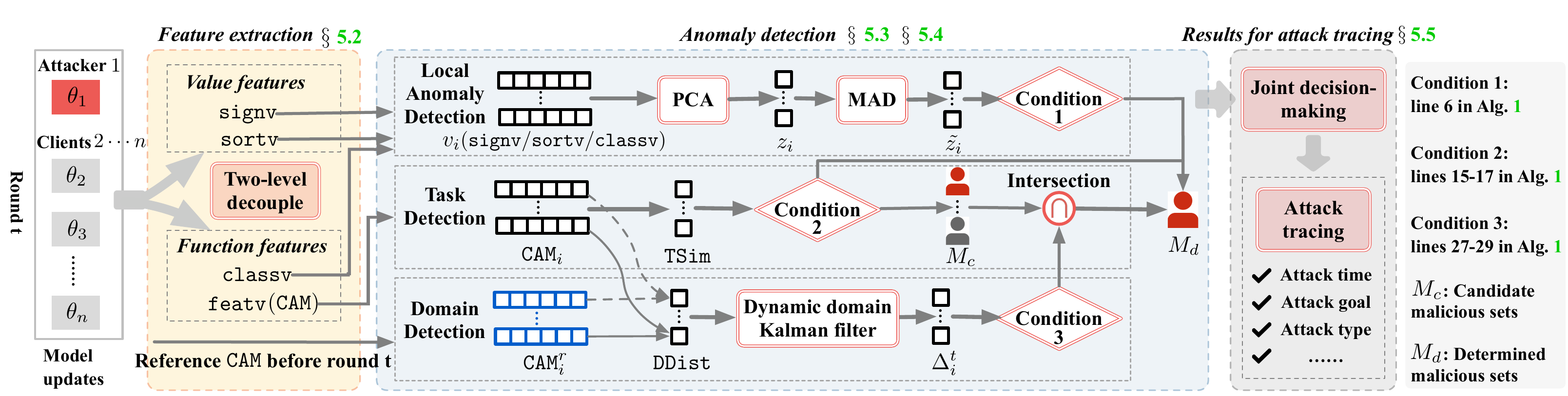}
  \vspace{-1mm}
  \caption{FLTracer (Attack Provenance) Framework. Upon receiving the client updates $\theta_{\{i\in[n]\}}$, FLTracer uses a two-level decoupling technique to generate four features representing different attacks (see $\S$\,\ref{sec:decouple}). Then, it uses Local Anomaly Detection to identify anomalies in \texttt{signv}, \texttt{sortv}, and \texttt{classv}. For the complex feature, \texttt{featv} (\texttt{CAM}), FLTracer monitors task and domain anomalies concurrently by Task and Domain Detection (see $\S$\,\ref{sec:detect} and $\S$\,\ref{sec:kf_detect}). In the results phase, FLTracer employs joint decision-making to accurately identify adversaries and provides attack tracing to support the detection process (see $\S$\,\ref{sec:results}). } %The validity of our detection is limited to the current round. 
  \label{fig:overview}
  \vspace{-4mm}
\end{figure*}

\vspace{-3mm}
\subsection{Attack Tracing Analysis} 
% \vspace{0.5mm}
% \noindent \textbf{Attack Tracing:}
We present an analysis of attack tracing using newly introduced features and observations. Firstly, FLTracer is capable of identifying the adversary and pinpointing the time of attack when any feature detects an anomaly. Secondly, FLTracer enables the analysis of attack objectives and types in $\S$~\ref{sec:threatmodel} by detecting anomalies within the features (see Table~\ref{tab:explainattack}). The Add noise, MB attack, and Fang attack all alter the update. To capture their anomalies concerning the order of each element in the update, we employ the \texttt{sortv}. For the Sign-flipping attack, which preserves the absolute values of the update while flipping only the symbols, \texttt{signv} proves effective in its detection. 
For the Dirty label attack, which manipulates dataset labels without affecting the training data itself, we employ \texttt{classv}. This attack primarily modifies decision boundaries (i.e., classifier), with minimal impact on the feature space (i.e., feature extractor).
% \st{modifies the labels of the dataset while preserving the data during training. This attack mainly alters the classifier but leaves the feature extractor almost unchanged.}
For most backdoor attacks (e.g., BadNets, DBA, Blind, and FRA attacks) that inject backdoors into the original feature space, we use \texttt{featv} (\texttt{CAM}) to analyze the feature extractor of the backdoor model. Specifically, for the MRA attack, although it injects a backdoor by modifying the training data, it also directly scales up the updates of the backdoor model. Thus, detection using \texttt{sortv} is sufficient. 
% \st{For the BadNets and Blind backdoor attacks, which can inject backdoors by covertly modifying the dataset and code, we use \texttt{featv} (\texttt{CAM}) to analyze the feature extractor of the backdoor model. The low TSim of certain layers suggests a large task gap between these layers and other participants, and backdoor features are likely embedded in these layers. In addition, by continuously monitoring the $\mathrm{DDist}$ of \texttt{CAM}, we can determine the number of rounds in which the adversary modifies the dataset or code.}

\begin{table}[!ht]
\centering
\footnotesize
\vspace{-2mm}
\setlength\tabcolsep{2pt}
\caption{Attack Tracing}
\vspace{-2mm}
\resizebox{\linewidth}{!}{
% \begin{threeparttable}
\begin{tabular}{lll}
\toprule
Attack Type   & Crucial Impact  & Feature  \\ \midrule

\begin{tabular}[c]{@{}l@{}}Add noise, MB attack,\\ Fang attack \end{tabular} & \begin{tabular}[c]{@{}l@{}}Alter the values of the update, thus change\\ their order among the participating clients\end{tabular} & \begin{tabular}[c]{@{}l@{}}\texttt{sortv} \end{tabular} \\ \cmidrule{2-3}

Sign-flipping  & \begin{tabular}[c]{@{}l@{}} Only flip the symbols of the update \end{tabular}  & \texttt{signv}  \\ \cmidrule{2-3}

Dirty label & Mainly modify the classifier & \texttt{classv}  \\ \cmidrule{2-3}

\begin{tabular}[c]{@{}l@{}}BadNets, DBA, \\ Blind, FRA \end{tabular} & \begin{tabular}[c]{@{}l@{}} Mainly modify the feature extractor,\\ leads to the anomaly of TSim and DDist\end{tabular} & \begin{tabular}[c]{@{}l@{}}\texttt{featv} \\ (\texttt{CAM}) \end{tabular}  \\ \cmidrule{2-3}

MRA  & Scale up the values of the update & \texttt{sortv}
\\ \bottomrule
\end{tabular}
% \vspace{-1mm}
% \begin{tablenote}
%     \item \ding{182}: Degrade the performance; \ding{183}: Inject the backdoor
% \end{tablenote}
% \end{threeparttable}
}
\label{tab:explainattack}
\vspace{-4mm}
\end{table}

\vspace{-1mm}
\section{System Design} \label{sec:systemdesign}

% \vspace{-1.5mm}
% \subsection{Overview of FLTracer} 
% \vspace{-1.5mm}
In this section, we present the framework of FLTracer, as depicted in Fig.~\ref{fig:overview}.
% Fig.~\ref{fig:overview} depicts the three phases of FLTracer: feature extraction, anomaly detection, and results for attack tracing. 
% Upon receiving the client updates $\theta_{\{i\in[n]\}}$, FLTracer uses a two-level decoupling technique to generate four features representing different attacks (see $\S$\,\ref{sec:decouple}). Then, it uses Local anomaly detection to identify anomalies in \texttt{signv}, \texttt{sortv}, and \texttt{classv}. For the complex feature, \texttt{featv} (\texttt{CAM}), FLTracer monitors task and domain anomalies concurrently by Task and Domain detection (see $\S$\,\ref{sec:detect} and $\S$\,\ref{sec:kf_detect}). In the results phase, FLTracer employs joint decision-making to accurately identify adversaries and provides attack tracing to support the detection process (see $\S$\,\ref{sec:results}). 
We define key notations in Table~\ref{tab:notation}.

\vspace{-3mm}
\subsection{Value and Function Level of Decoupling}
\label{sec:decouple}
% \vspace{-1.5mm}
Based on Observation~\Rmnum{1}, four newly proposed features represent the specific poisoned portions of different attacks on updates from the perspectives of value characteristics and model functions. For further information on feature extraction (labeled \ding{182} to \ding{185}), please refer to $\S$~\ref{sec:observation1}.

% \vspace{-5mm}
\begin{table}[]
\setlength\tabcolsep{2pt}
\centering
\scriptsize
\caption{Key notations}
\vspace{-2mm}
\resizebox{\linewidth}{!}{
% \begin{threeparttable}  
\begin{tabular}{@{}ll@{}}
\toprule
Term      & Description               \\
\midrule
$N, m$ & Total number of clients, the total number of malicious clients \\
$n$ & Participating clients (participants) each round \\
%$m$ & Total number of malicious clients \\
$t$, $i$ & Round index, Client index \\
% & Client index \\
$G^t$ & Global model at round $t$ \\
$\theta_i^t$ & Local model update of client $i$ at round $t$ \\
$u_i$ & The concatenation of update values of client $i$ \\
$k$ & Total number of the layers \\
$\kappa$ & Layer index \\
$w^\kappa_i$ &  Values of convolution kernels of client $i$ in layer $\kappa$ \\
$A_i^{\kappa} $ & Anomaly vector of the client $i$ in layer $\kappa$ \\ %\in \BR^{C_1^\kappa\times C_2^\kappa}
$\alpha_i^\kappa \in \BR $ & Task similarity of client $i$ in layer $\kappa$ \\
% $T_i=\{\alpha_i^\kappa\}$ & Task similarity set of client $i$ \\
$\beta_i^{\kappa, t} \in \BR $ & Domain distance of client $i$ in layer $\kappa$ at round $t$ \\ 
% $D_i=\{\alpha_i^\kappa\}$ & Domain distance set of client $i$ \\
$\texttt{CAM}_p \!=\!\{A_p^{\kappa}\}$ & Fixed feature pattern calculated by participants each round \\ 
$\texttt{CAM}_i^r \!=\!\{A_i^{\kappa, r}\}$ & Reference feature pattern of client $i$ \\ 
$\delta_i^\kappa$ & Difference between predicted and actual domain distance \\
\bottomrule
\end{tabular}
% \begin{tablenotes}
%     \item{(E)-Effectiveness; (S)-Stability; (R)-Robustness}
% \end{tablenotes}
% \end{threeparttable}
}
\label{tab:notation}
\vspace{-4mm}
\end{table}

% Firstly, for attacks that change the update sign, we extract \texttt{signv}, consisting of the sign of each element in the updates. Secondly, for attacks that alter the update value, we extract \texttt{sortv}, consisting of the rank of each element among all participants. \texttt{sortv} is a robust feature that balances the anomalies in shallow and deep layers. Then we extract the weights and bias of the classifier as \texttt{classv}, characterizing the modification of the decision boundary. Lastly,  we construct \texttt{featv} to represent the anomalies of the feature extractor. 
% Especially for CNN, we design a set of anomaly scores called \texttt{CAM} for the convolution kernels in all convolution layers, as expressed in Eq.(\ref{eq:cam}). Specifically, we use PCA and Mahalanobis Distance~\cite{de2000mahalanobis} to obtain the anomalies of a convolution kernel, and the combination of all anomalies in a convolution layer is represented as the anomaly vector $A_i^\kappa$ (i.e., the operation $\circledast$ in Eq.(\ref{eq:a})). More details about \texttt{CAM} can be found in Appendix~\ref{sec:details_CAA}. 
% To save memory and computational cost, we only need to compute about 1.5\% of all convolution kernel anomaly scores to distinguish malicious clients accurately.%We refer the readers to Alg.~\ref{alg:cam} for more details about CAA. 

\vspace{-2mm}
\subsection{Anomaly Detection} \label{sec:detect}
% \st{For simple neural networks, such as fully-connected neural networks, the first three features (i.e., \texttt{signv}, \texttt{sortv}, and \texttt{classv}) are sufficient to capture most of the attacker's modifications to the model.} 
% \vspace{-1.5mm}
We propose Local anomaly detection to capture the anomalies of \texttt{signv}, \texttt{sortv}, and \texttt{classv}. For backdoor attacks, we analyze the feature extractor from the consistency of the task and domain perspectives. We propose Task and Domain detection to detect anomalies in \texttt{CAM}. Our method is applicable to other neural networks without convolution layers by replacing \texttt{CAM} and $A$ with \texttt{featv} and $\rho$ in the detection.
% analyze the feature [qy: space consistency

\vspace{1mm}
\noindent\textbf{Local Anomaly Detection:}
%  we first decouple the updates into three features (i.e., \texttt{signv}, \texttt{sortv}, and \texttt{classv}) based on value and function. Then we introduce a general algorithm for detecting these features, called the Local anomaly detection method.
% \textbf{Identify adversaries via Local anomaly detection.} % generic cross-client detection method in EM1-EM3, which is designed for untargeted attacks and simple targeted attacks, such as model replacement backdoor attacks. 
% As depicted in Fig.~\ref{fig:overview}, we extract $\texttt{signv}_i$ (or $\texttt{sortv}_i$ or $\texttt{classv}_i$) for each client $i$, then detect their anomalies.  %~\cite{hampel1974influence} 
As described in the \textsc{Local-} \textsc{AnomalyDetect} function of Alg.~\ref{alg:detect_alg}, PCA is utilized to decrease the feature dimensionality (denoted as $v_i$) and acquire $z_i$. Then the median absolute deviation (MAD) algorithm~\cite{rousseeuw1993alternatives} is employed to compute the anomaly scores of the $z_i$, which is robust in the presence of malicious clients. The anomaly scores ($\tilde{z}_i$) are represented by the following Eq.(\ref{eq:mad}). Assuming a normal distribution, a value of $b=1.4826$ is applied, and any client whose $\tilde{z}_i>\lambda$ (where $\lambda=2.5$) is considered to have a $99.38\%$ probability of being a malicious client~\cite{rousseeuw1993alternatives}.

\vspace{-2mm}

\begin{equation}
\setlength{\abovedisplayskip}{0.1cm}
\small
\label{eq:mad} \tilde{z}_i=f_{\mathrm{MAD}}(z_i)=\frac{\lvert z_i- \mathrm{med}_q z_q \rvert}{\xi}, \xi=b\ \mathrm{med}_i\{\lvert z_i- \mathrm{med}_q z_q \rvert\}
\end{equation}

\vspace{-1mm}

\noindent\textbf{Task Detection to Select Malicious Candidates:} This method identifies discrepancies in task similarity among $n$ participants in a round, as described in the \textsc{TaskDetect} function of Alg.~\ref{alg:detect_alg}. Specifically, we require $\texttt{CAM}_{\{i \in [n]\}}$ to create a fixed feature pattern $\texttt{CAM}_{p}\!=\!\{A^1_{p}, A^2_{p}, \cdots, A^k_{p}\}$ (which represents the median of $n$ \texttt{CAM}s) for this round.  
Then we compute the set of task similarities ($T_i$) between $\texttt{CAM}_i$ and pattern $\texttt{CAM}_{p}$ per layer, as expressed below: 

\vspace{-3mm}

\begin{small}
\begin{align}
\label{eq:ts}  T_i&=\{\alpha^1_i, \alpha^2_i, \cdots, \alpha^k_i\}, %\  \alpha_i^\kappa=\mathrm{Similarity}(A^\kappa_{p}, A_i^\kappa) 
\\ 
\label{eq:adjcos}  \alpha_i^\kappa&=\mathrm{TSim}_{A_p^\kappa,A_i^\kappa}=\frac{(A^\kappa_{p}-\overline{A^\kappa_{p}})\cdot (A^\kappa_{i}-\overline{A^\kappa_{i}})}{\Vert A^\kappa_{p}-\overline{A^\kappa_{p}} \Vert \ \Vert A^\kappa_{i}-\overline{A^\kappa_{i}} \Vert}
\end{align} 
\end{small}%

% \vspace{-2mm}

\noindent where $\alpha_i^\kappa \!\in\![-1,1]$ is adopted from Eq.(\ref{eq:tsim}). A client is considered a malicious candidate if their $T_i$ value is abnormal.
From $\S$~\ref{sec:observation2}, we summarize three conditions that indicate abnormal $T_i$. First, $\min(T_i) \leq \tau, \tau\!\sim\! -1$ (see Fig.~\ref{fig:deep_conv}(a)), which suggests that certain layers in the $\texttt{CAM}_i$ of the malicious client differ significantly from benign ones. 
Second, $\max(T_i) \leq 0$ (see Fig.~\ref{fig:deep_conv}(b)), which indicates a negative correlation in the similarity between $\texttt{CAM}_i$ and the pattern $\texttt{CAM}_{p}$ for each layer.
Third, the slope of the line connecting from $\alpha_i^1$ to $\alpha_i^k$ (in $T_i$) is less than $0$ (see Fig.~\ref{fig:deep_conv}(c) and (d)). This implies that the similarity ($\alpha$) of the malicious client decreases as the layer goes deeper, in opposition to the trend of the benign one.

\vspace{1mm}
\noindent\textbf{Domain Detection to Determine Malicious Clients:} 
This method identifies malicious clients among malicious candidates by monitoring a client's change in domain distance across rounds based on $\S$~\ref{sec:observation3}. % Observation~\Rmnum{3} in  
We accumulate and construct a reference feature pattern $\texttt{CAM}^r_i\!=\!\{A_i^{1,r}, A_i^{2,r}, \cdots, A_i^{k,r}\}$ for each client. $\texttt{CAM}^r_i$ is the average of multiple \texttt{CAM}s for client $i$ in a few rounds. Then, we calculate the set of domain distances ($D_i^t$) between $\texttt{CAM}_i$ and $\texttt{CAM}^r_i$ by layer at round $t$, as: %\par
\vspace{-3mm}

\begin{small}
\begin{align}
\label{eq:dd}   D_i^t&=\{\beta^{1,t}_i, \beta^{2,t}_i, \cdots, \beta^{k,t}_i\}, %\  \beta_i^\kappa=\mathrm{DDist}(\overline{A^\kappa_i}, A_i^\kappa)
\\
\label{eq:footsim}    \beta_i^{\kappa,t}&=\mathrm{DDist}_{A^{\kappa,t}_i, A_i^{\kappa,r}} =\frac{1}{len(A^{\kappa,r}_i)}\cdot \sum_{j=1}^{len(A^{\kappa,r}_i)}(A^{\kappa,t}_{i,j} - A_{i,j}^{\kappa,r}) 
\end{align}
\end{small}%
% \vspace{-1mm}
where $\beta_i^{\kappa,t}$ is adopted from Eq.(\ref{eq:ddist}). Specifically, we present a Kalman filter-based method to monitor changes in domain distance between rounds to determine malicious clients.

\begin{algorithm}[t]
\small
  \caption{Anomaly Detection Functions}  
  \begin{algorithmic}[1]
  % \Require
  %   $\theta_{\{i\in [n]\}}$: updates of $n$ clients; $k$: total number of conv. layers
  %   \Ensure 
  %   $M_d, M_c$: Set of determined and candidate malicious clients
    
    \Function{LocalAnomalyDetect}{$v_{\{i\in [n]\}}$}
        \State Initialize $M_d = \emptyset$    
        % \State $v_{\{i\in [n]\}} \gets \mathrm{\textsc{Decouple}}(u_{\{i\in [n]\}})$
        \State $z_{\{i\in [n]\}} \gets \mathrm{PCA}(v_{\{i\in [n]\}}, \mathrm{componenets}=1)$ %, components=1
        \State $\tilde{z}_{i} \gets f_{\mathrm{MAD}}(z_{i})$ using Eq.(\ref{eq:mad}), $i\in[n]$ %\textsc{MadScore}(z_{\{i\in [n]\}})
        \For{$i\in [n]$}
            \If{$\tilde{z}_i > \lambda$} $M_d\gets M_d \cup \{i\}$
            \EndIf
        \EndFor
        \State \Return $M_d$ \Comment{{\small Determined malicious sets}}
    \EndFunction
    %\State

\vspace{0.5mm}
    
    \Function{TaskDetect}{$\texttt{CAM}_{\{i\in [n]\}}, k$}
    \State Initialize $M_d, M_c \gets \emptyset$ 
        % \State $\texttt{CAM}_{\{i\in [n]\}} \gets$ Compute using $\theta_{\{i\in [n]\}}$, $k$ and Alg.~\ref{alg:cam}
        \For{layer $\kappa \in [k]$}
            % \State $A^\kappa_{\{i\in [n]\}} \gets$ Extract from $\texttt{CAM}_{\{i\in [n]\}}$
            \State $A^\kappa_{p} \gets \mathrm{med}_i A^\kappa_i$  
            \State $\alpha^\kappa_i \gets \mathrm{TSim}(A^\kappa_{p}, A^\kappa_i)$ using Eq.(\ref{eq:adjcos}), $i \in [n]$
        \EndFor
        \State $T_i \gets$ Generated using $\alpha^\kappa_i$ and Eq.(\ref{eq:ts}), $i\in[n]$  %$$\{\left. ds^\kappa_i \right| \kappa \in [k]\}$ \Comment{{\small Model deviation}}
        \For{$i \in [n]$}
            \If{$\mathrm{min}(T_i) \leq \tau$}  $M_d \gets M_d \cup \{i\}$
            \ElsIf{$\mathrm{max}(T_i)\leq 0$} $M_c \gets M_c \cup \{i\}$
            \ElsIf{$\mathrm{slope}(T_i) \leq 0$}  $M_c \gets M_c \cup \{i\}$
            \EndIf
        \EndFor
        \State \Return $M_d, M_c$  \Comment{{\small Determined and candidate malicious sets}}
    \EndFunction
    %\State

\vspace{0.5mm}
    
    \Function{DomainDetect}{$\texttt{CAM}_{\{i\in [n]\}},\texttt{CAM}^r_{\{i\in [n]\}},k,M_c, M_d$}
        \For{layer $\kappa \in [k]$}
            \State $\beta^\kappa_i \gets \mathrm{DDist}(A^{\kappa,r}_{i}, A^{\kappa,t}_i)$ using Eq.(\ref{eq:footsim}), $i \in [n]$
        \EndFor
        \State $D_i^t \gets$ Generated using $\beta^\kappa_i$ and Eq.(\ref{eq:dd}), $i\in[n]$  %$$\{\left. ds^\kappa_i \right| \kappa \in [k]\}$ \Comment{{\small Model deviation}}
        \State $\hat{D_i^t} \gets$ Estimated using Eq.(\ref{eq:predict_state}), $i\in[n]$
        \State $\Delta_i^t=\{\delta_i^{\kappa, t}\} \!\gets\!$ Calculated using $D_i^t, \hat{D_i^t}$ and Eq.(\ref{eq:delta_domain}), $i\!\in\![n]$
        % \State $\gamma_i^t = \mathrm{slope}(\Delta_i^t)$, $i\in[n]$ %and $\Gamma^t=\gamma_{\{i\in[n]\}}$
        \For{$i \in [n]$}
            \If{$\mathrm{slope}(\Delta_i^t) > 0$} 
            \If{$f_{\mathrm{MAD}}(\mathrm{slope}(\Delta_i^t))>\lambda$} $M_c \gets M_c \cap \{i\}$
            \ElsIf{$f_{\mathrm{MAD}}(\delta_i^{\kappa,t}) > \lambda$} $M_c \gets M_c \cap \{i\}$
            \ElsIf{$\mathrm{mean}(\Delta_i^t) > 0$}  $M_c \gets M_c \cap \{i\}$
            \EndIf
            \EndIf
        \EndFor
        \State \Return $M_d\gets M_d \cup M_c$ \Comment{{\small Determined malicious sets}}
    \EndFunction
  \end{algorithmic}
  \label{alg:detect_alg}
\end{algorithm}

% \vspace{-2mm}

%~ in Appendix\ref{sec:CAM_ccd}

% To visualize the difference of cross-client, Fig.~\ref{fig:CAA_croc} illustrates the \texttt{CAM} and $T$ for ten clients in a round, including nine benign and one malicious client. The $T_i$ of the malicious client has an extremely large $\alpha$ in layers 14-16. In addition, the $\alpha$ of the malicious client becomes larger as the layer goes deeper, a trend that is opposite to that of the benign client. 

% % the deviation score ($ds$)
% % The model deviation ($T_i$) of client $i$ is the combination of $ds$ of all convolution layers, which can be expressed as 

% % differs in two ways: certain layers  and the trend of $ds$ across layers is opposite to the benign. These observations make sense and are explained as follows. First, a layer with extremely large $ds$ indicates that $ds$ is out of the normal range. This client can be considered as a definite malicious client. Second, the $ds$ value  This is because deeper layers are more important than the shadow layers in backdoor tasks. Therefore, adversaries tend to modify the deeper convolution layers to achieve effective backdoor injection. These clients can be selected as malicious candidates. 

\vspace{-2mm}

\subsection{Kalman Filter-based Domain Detection} \label{sec:kf_detect}
% \vspace{-1.5mm}
We present a Kalman filter-based method to model the domain state of each client and detect malicious clients by comparing the predicted domain and actual domain states. If the difference exceeds certain conditions (explained later), we classify the client as malicious. Refer to the Kalman filter in $\S$~\ref{pre:kalman}, which models a system at time step $t$ with the equation $x^t=A x^{t-1}+B u^t+w$. Here, the core is to define each component of the system, including the state vector $x$, state transition pattern $A$, control input matrix $B$, control input $u$, and noise $w$. In addition, to ensure consistency with FL training, it is necessary to satisfy the following equations for the local model update $\theta_i^t$ and global model update $G^t$:
% We define the transition pattern of each client's domain state by fitting the domain state between different rounds of normal training. This mainly reflects the distribution fluctuation of the client's dataset. According to $\S$~\ref{sec:observation3}, the data domain of the malicious client undergoes a sudden and drastic shift due to the poisoned training data, whereas the benign client maintains its transition pattern. We utilize the Kalman filter and the cumulative historical reference set to model and predict domain state, enabling anomaly detection via discrepancies between predictions and the actual values. Based on the training process of FL, we have:
\vspace{-2mm}
\begin{align}
\small
    \label{eq:state_theta} \theta_i^t =&\ \mathbf{F^1_i} \cdot \theta_i^{t-1} + \mathbf{F^2_i} \cdot G^{t-1} + \mathbf{H}_i \circ {\Omega}_i + \mathbf{W}_i \\
    \label{eq:state_g} G^t =&\hspace{7.4em} G^{t-1} + \hspace{2.2em} {\Omega}_i + \mathbf{P}_i \cdot\Theta_i^t 
\vspace{-3mm}
\end{align}
where ${\Omega}_i$ represents the impact of client $i$'s datasets on updates and $\Theta_i^t=\sum_{l=1}^n \theta_l^t - \theta_i^t$ denotes the sum of updates from other clients. Client-specific parameters include $\mathbf{F_i^1}, \mathbf{F_i^2}, \mathbf{H}_i, \mathbf{W}_i$, and $\mathbf{P}_i$. Eq.(\ref{eq:state_theta}) indicates that each client $i$'s update is related to their update at the previous round ($\theta_i^{t-1}$), their dataset (${\Omega}_i$), and the starting training model ($G^{t-1}$). Eq.(\ref{eq:state_g}) indicates that the global model is related to the previous round's global model ($G^{t-1}$), the training data of client $i$ (${\Omega}_i$), and the updates of other participants, except client $i$ ($\Theta_i^t$).

\vspace{1mm}
\noindent\textbf{Dynamic Domain State for FL:}  \label{sec:kf_model}
To model the domain state, we first calculate the domain distance set of $\theta_i^t$, $G^t$, and $\Theta_i^t$ and obtain $D_i^t$, $D(G)_i^t$, and $D(\Theta)_i^t$ using Eq.(\ref{eq:dd}). We define the state vector as $X_i^t = [D_i^t, D(G)_i^t]\in\BR^{2k\times 1}$ and the control input as $Z_i^t=D(\Theta)_i^t\in\BR^{k\times 1}$. According to Eq.(\ref{eq:state_theta}) and Eq.(\ref{eq:state_g}), the dynamic domain state of the FL for each client $i$ is modeled as follows, where $\circ$ denotes the Hadamard product:
\begin{equation}
\footnotesize
\setlength{\arraycolsep}{2.5pt}
\label{eq:domain_state}
X_i^t % \begin{bmatrix}\theta_i^t \\ G^t \end{bmatrix} 
\!=\! \begin{bmatrix} \mathbf{F^1_i} \!&\! \mathbf{F^2_i} \\ \mathbf{0}_{k\times k} \!&\! \mathbf{1}_{k\times k} \end{bmatrix} \!\cdot\! X_i^{t-1} %\begin{bmatrix}\theta_i^{t-1} \\ G^{t-1} \end{bmatrix} 
\!+\!\begin{bmatrix} \mathbf{H}_i \\ \mathbf{1}_{k\times 1} \end{bmatrix} \circ  {\Omega}_i 
+\begin{bmatrix} \mathbf{0}_{k\times 1} \\ \mathbf{P}_i \end{bmatrix} \!\cdot\! \begin{bmatrix} \mathbf{0}_{k\times 1} \\ Z^t_{i} \end{bmatrix}
\!+\!\begin{bmatrix} \mathbf{W}_i \\ \mathbf{0}_{k\times 1} \end{bmatrix} 
\end{equation}
Here, $\mathbf{F^1_i}, \mathbf{F^2_i}\in\BR^{k\times k}$ are the state transition patterns for dynamics of the system, ${\Omega}_i\in \BR^{k\times 1}$ is the dataset anomaly matrix for identifying each client's dataset, $\mathbf{H}_i\in \BR^{k\times 1}$ and $\mathbf{P}_i\in \BR^{k\times k}$ are the control input matrices to model the effect of the control input $Z_i^t$ on the state, and $\mathbf{W}_i$ is the process noise reflecting the uncertainty in the system dynamics. 

To fit these parameters for each client, we accumulate three rounds of updates and global models for all clients and compute the corresponding $\{X_i^t\}$ and $\{Z_i^t\}$. We then fit the parameters $\mathbf{F^1_i}, \mathbf{F^2_i}, {\Omega}_i, \mathbf{H}_i, \mathbf{P}_i$, and $\mathbf{W}_i$ for each client $i$ using a one-layer fully connected neural network. 

In the case where the client's data distribution changes slowly over time, we need to update the dataset parameter ${\Omega}_i$ each round and leave the other parameters unchanged.

\vspace{1mm}

\noindent\textbf{Domain State Prediction and Detection:}
After obtaining the above parameters, given Eq.(\ref{eq:state_theta}), we can predict the domain state of client $i$ at round $t$ with the input $D_i^{t-1}$ and $G^{t-1}$:

\vspace{-2mm}

\begin{equation}
\small
\label{eq:predict_state}
    \hat{D_i^t} = \mathbf{F^1_i} \cdot D_i^{t-1} + \mathbf{F^2_i} \cdot G^{t-1} + \mathbf{H}_i \circ {\Omega}_i + \mathbf{W}_i .
\end{equation}

\vspace{-1mm}

The difference between the predicted domain state ($\hat{D_i^t}$) and the actual domain state (${D_i^t}$) represents the change in the domain of client $i$, which can be expressed as

\vspace{-2mm}

\begin{equation}
\small
\begin{aligned}
\label{eq:delta_domain}
    \Delta_i^t =D_i^t- \hat{D_i^t}
    % =\{\beta_i^{1,t}, \cdots, \beta_i^{k,t}\} -\{\hat{\beta}_i^{1,t}, \cdots, \hat{\beta}_i^{k,t}\} \\
    =\{\delta_i^{1,t}, \cdots, \delta_i^{k,t}\}
\end{aligned}
\end{equation}

\vspace{-1mm}

\noindent where $\delta_i^{\kappa,t}=\beta_i^{\kappa,t}-\hat{\beta}_i^{\kappa,t}$ represents the domain state anomaly of client $i$ in the layer $\kappa$. We summarize four conditions based on $\S$~\ref{sec:observation3} to select abnormal $\Delta_i^t$, as described in the \textsc{Domain-} \textsc{Detect} function of Alg.~\ref{alg:detect_alg}. First, the slope of the line connecting from  $\delta_i^{1,t}$ to $\delta_i^{k,t}$ is greater than $0$ (see Fig.~\ref{fig:timesim}), which means that the malicious client's domain changes become larger as the layer gets deeper. This condition is consistent with our Observation~\Rmnum{2} and similar to the condition in line 17 of Alg.~\ref{alg:detect_alg}. Second, the $\mathrm{MAD}$ value (using Eq.(\ref{eq:mad})) for the slope is abnormal, and $\lambda$ is equal to that in line 6 of Alg.~\ref{alg:detect_alg}. Third, the $\mathrm{MAD}$ value for the domain change of any layer is abnormal. Fourth, the average of $\delta_i^{\kappa,t}$ is greater than $0$, meaning that the actual domain state for the malicious client is always greater than the predicted state. This is because the malicious client's training data changes drastically compared to before.

\vspace{-2mm}
\subsection{Results for Attack Tracing} \label{sec:results}
 % \vspace{-1.5mm}

After receiving the detection results of four features, the joint decision module identifies clients as malicious if any feature detects anomalies, and determines the remaining clients as benign. Then, \emph{FLTracer determines the launch time of the attack and utilizes the detection details of each feature to trace the attack's objective, type, and poisoned portions of updates.} 

The type and objective of attacks are discerned as follows: anomalies in \texttt{signv}, \texttt{sortv}, \texttt{classv}, and \texttt{featv} (\texttt{CAM}) correspond to sign-flipping, adaptive untargeted, dirty label, and backdoor attacks, respectively. The anomalies in the first three vectors indicate a reduction in the performance of the global model (untargeted), while in \texttt{featv}, they indicate the injection of a backdoor into the model (backdoor). 
We evaluate the percentage of directly traced attack objectives and types, with FLTracer achieving a success rate of $93.75\%$ and $80\%$ for tracing untargeted and backdoor attacks, as well as $84.38\%$, $78.00\%$, $95.00\%$, and $80.00\%$ for tracing sign-flipping, adaptive untargeted, dirty label, and backdoor attacks, respectively. 
Next, we analyze the poisoned portions of updates in the attacks. Any anomalies in \texttt{signv}, \texttt{sortv}, \texttt{classv}, and \texttt{featv} indicate that the opponent altered the symbols, sorting, classifier, and feature extractor, respectively. Specifically, in \texttt{featv}, certain layers exhibit a large Tsim, which implies the possible embedding of a backdoor within those layers. 
In situations where there are multiple adversaries in FL, utilizing attack tracing can identify colluding clients launching similar attacks.

\section{Evaluation} \label{sec:evaluation}

% \vspace{-2mm}

\subsection{Experimental Setup} \label{sec:evalation_setup}

\noindent\textbf{Datasets and Data distribution:} 
We evaluate FLTracer on multiple datasets for different learning tasks, including three image classification datasets (MNIST~\cite{lecun1998gradient}, EMNIST~\cite{cohen2017emnist}, and CIFAR10~\cite{krizhevsky2009learning}), a traffic sign recognition dataset (GTSRB~\cite{Houben-IJCNN-2013}), a human activity recognition dataset (HAR~\cite{anguita2013public}), and a real-world driving dataset (BDD100K~\cite{bdd100k}). 
% MNIST~\cite{lecun1998gradient} is a 10-class class-balanced digit image classification dataset. 
% EMNIST~\cite{cohen2017emnist} is a 47-class class-imbalanced digit image classification dataset. 
% CIFAR10~\cite{krizhevsky2009learning} is a 10-class class-balanced color image classification dataset. 
% German Traffic Sign Recognition Benchmark (GTSRB)~\cite{Houben-IJCNN-2013} is a 43-class class-imbalanced traffic sign dataset with varying light conditions and rich backgrounds.
% Human Activity Recognition (
HAR is a 6-class class-imbalanced signal dataset representing human activity collected from smartphones of 30 real-world users. 
BDD100K is a large-scale driving video dataset of 100K real-world videos collected from diverse locations. We use its 100K image dataset, each of which is sampled at the 10th second from the video. We perform the 10-class road object classification task on these class-imbalanced data, where the distribution of the data changes slowly w.r.t weather conditions, scene types, and different times of the day.

\begin{figure}[t]
  \centering
  \subfigure{
  \includegraphics[width=0.8\linewidth]{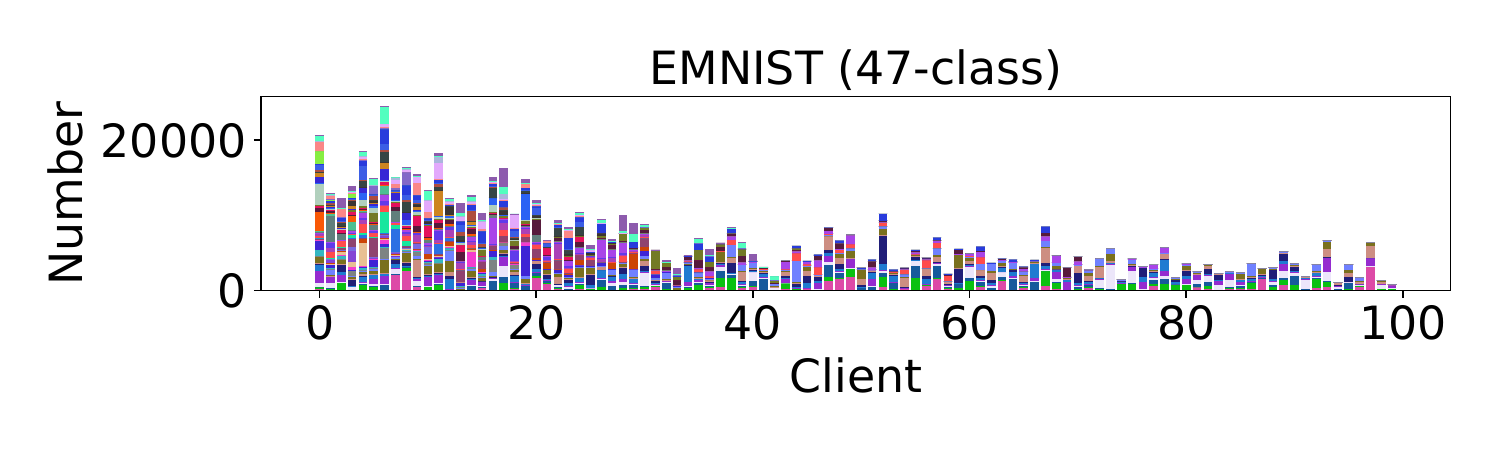}
  }
  \vskip -3.5mm
  \subfigure{
  \includegraphics[width=0.8\linewidth]{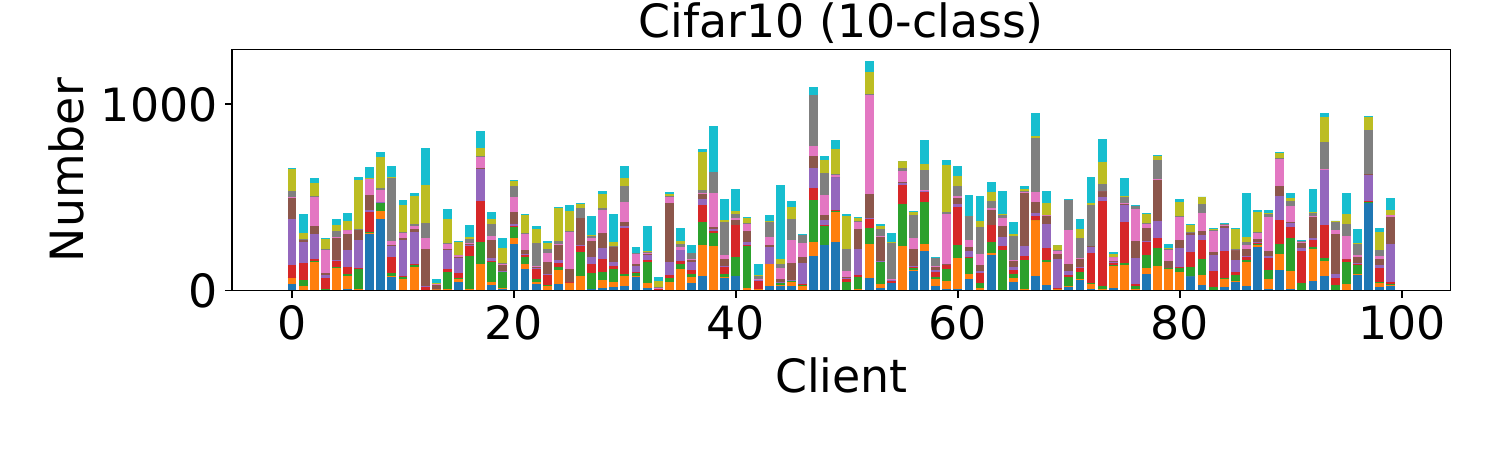} 
  }
  \vskip -4.5mm
  \subfigure{
  \includegraphics[width=0.8\linewidth]{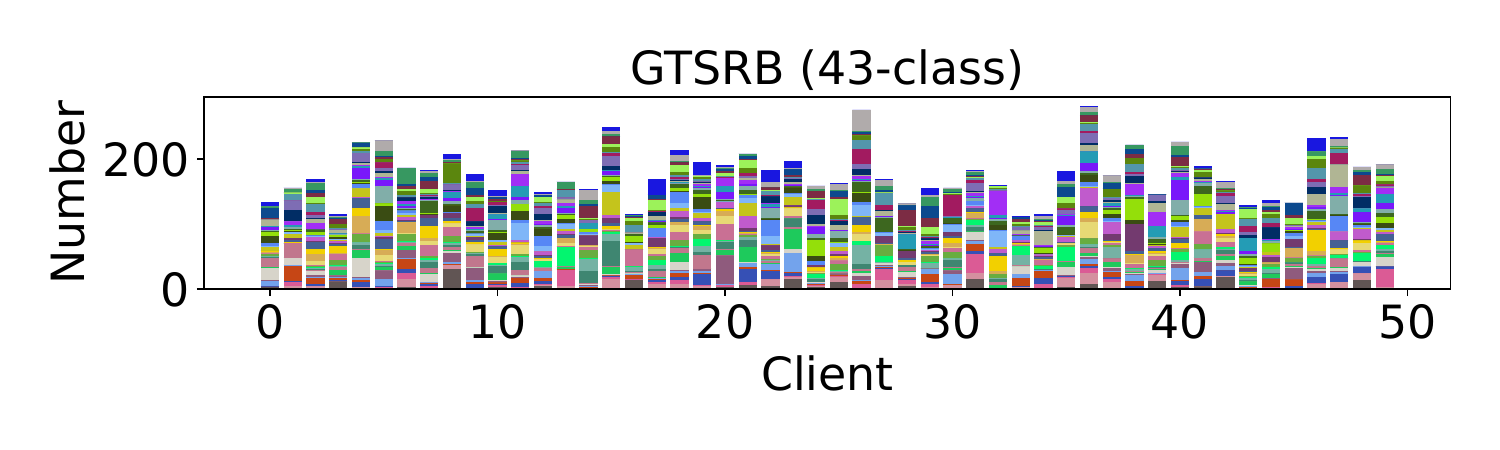} 
  }
  \vskip -4mm
  \subfigure{
  \includegraphics[width=0.8\linewidth]{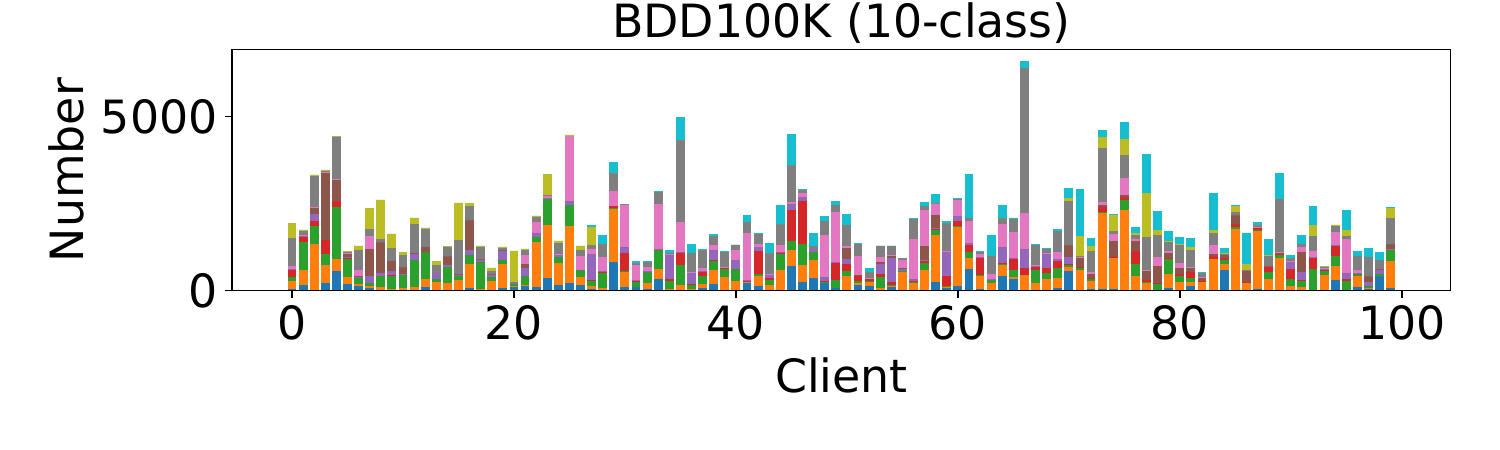} 
  }
\vspace{-6mm}
  \caption{Data distribution of $N$ clients derived from the Dirichlet distribution with $d=0.5$.}
\vspace{-6mm}
  \label{fig:data_distribution}
\end{figure}

We generate both IID and non-IID data distributions for each dataset, as listed in Table~\ref{tab:dataset1} (column 6).
To simulate real non-IID settings, we use a Dirichlet distribution~\cite{minka2000estimating} with the degree of non-IID ($d$) set to $0.5$, following the setup in~\cite{shejwalkar2021back, bagdasaryan2020backdoor, cao2020fltrust, xie2019dba}. We utilize the code in \cite{bagdasaryan2020backdoor, bagdasaryan2020blind} to distribute both the labels and data of the dataset, as depicted in Fig.~\ref{fig:data_distribution}. Note that a smaller value of $d$ indicates more heterogeneous data, and we compare the impact of non-IID degrees with one class per client, $d\!=\!0.5$, $d\!=\!5$ and $d\!\!\to\infty$. 
Since HAR is derived from real-world users, we do not need to distribute the data, and consider each user a client. The BDD100K dataset is partitioned into three subsets (i.e., daytime, dusk, and nighttime) and images in each part are further distributed using a $d\!=\!0.5$. We emulate client training based on real-world changes in data distribution, starting with daytime data, then adding dusk data, and finally adding nighttime data.

% Following the setup in~\cite{shejwalkar2021back, bagdasaryan2020backdoor, xie2019dba}, we use a Dirichlet distribution~\cite{minka2000estimating} with $d=0.5$ to simulate the data distribution in non-IID settings, as shown in Fig.~\ref{fig:data_distribution}. 

\vspace{0.5mm}

\noindent\textbf{Model architectures:}
We conduct experiments on 7 model architectures, as listed in Table~\ref{tab:dataset1} (column 7), including fully connected-based, convolution-based, and transformer-based, covering most of the popular model structures. 
%See $\S$~\ref{sec:expsetup} for more details on datasets and model architectures.
% Table~\ref{tab:dataset} (column 5) lists the model architectures we used for each dataset. SimpleNet contains two convolution layers and two fully-connected layers. AlexNet~\cite{krizhevsky2012imagenet}, ResNet18~\cite{he2016deep}, VGG16~\cite{simonyan2014very}, and ResNet34~\cite{he2016deep} are different architectures of convolution networks. 
% DNN contains two fully-connected layers. Vision Transformer (ViT)~\cite{dosovitskiyimage} is a Transformer based model for image recognition. 

% We summarize the datasets and model structures in Table~\ref{tab:dataset1} (columns 2-6). 

\vspace{0.5mm}

\noindent \textbf{Learning parameters:} 
We set up the standard training process using the AdamW/SGD optimizer for ViT/other model structures for local training, and the global learning rate is $\eta=1/n$.
Table~\ref{tab:dataset1} (columns 8-13) shows parameters.
% the number of total clients ($N$), participant clients each round ($n$), the percentage of malicious clients, and the percentage of poisoned data.
% Here malicious $lr$/Epochs is different from benign $lr$/epochs because malicious clients can modify hyperparameters to attack more effectively. 
% Malicious $lr$/Epochs can be modified to attack more effectively.
Following the setup in~\cite{shejwalkar2021manipulating, bagdasaryan2020backdoor}, we launch untargeted attacks at the beginning of training and backdoor attacks in the $200$th round when the model has converged. 
See Appendix~\ref{sec:expsetup} for more details.

\begin{table}[]
\setlength\tabcolsep{2pt}
\caption{Specific attacks in attack assessment. } %Left-untargeted attack and right-targeted attack.
\vspace{-2mm}
\resizebox{\linewidth}{!}{
\begin{threeparttable}
\begin{tabular}{|ll||ll|}
\hline
\multicolumn{2}{|c||}{Untargeted attack type / Specific attack} & \multicolumn{2}{c|}{Targeted attack type / Specific attack} \\ \hline
\multicolumn{1}{|l|}{Add noise~\cite{wu2020federated}} & Add noise & \multicolumn{1}{l|}{BadNets~\cite{gu2017badnets}} & Patch-BN, Noise\cite{chen2017targeted}-BN \\ \hline
\multicolumn{1}{|l|}{Sign-flipping\cite{li2019rsa}} & Sign-flipping & \multicolumn{1}{l|}{\multirow{2}{*}{\begin{tabular}[c]{@{}l@{}} Model\cite{bagdasaryan2020backdoor} \\ replacement\end{tabular}}} & \multirow{2}{*}{Patch-MRA, Noise-MRA} \\ \cline{1-2}
\multicolumn{1}{|l|}{\multirow{2}{*}{ \begin{tabular}[c]{@{}l@{}}Dirty label \\~\cite{biggio2012poisoning} \end{tabular} }} & \multirow{2}{*}{\begin{tabular}[c]{@{}l@{}}Fix-Fix, Fix-Rnd, \\ Rnd-Fix, Rnd-Rnd\end{tabular}} & \multicolumn{1}{l|}{} &  \\ \cline{3-4} 
\multicolumn{1}{|l|}{} &  & \multicolumn{1}{l|}{\multirow{2}{*}{\begin{tabular}[c]{@{}l@{}}Distributed\\ backdoor\cite{xie2019dba}\end{tabular}}} & \multirow{2}{*}{Patch-DBA, DBA-MRA} \\ \cline{1-2}
\multicolumn{1}{|l|}{\multirow{3}{*}{ \begin{tabular}[c]{@{}l@{}}MB attack$^*$\\ ~\cite{shejwalkar2021manipulating}\end{tabular} }} & \multirow{3}{*}{\begin{tabular}[c]{@{}l@{}}MB MKrum, MB Bulyan,\\ MB Median, MB Trmean,\\ MB Max, MB Sum\end{tabular}} & \multicolumn{1}{l|}{} &  \\ \cline{3-4} 
\multicolumn{1}{|l|}{} &  & \multicolumn{1}{l|}{\multirow{2}{*}{\begin{tabular}[c]{@{}l@{}}Blind~\cite{bagdasaryan2020blind}\\ backdoor$^*$\end{tabular}}} & \multirow{2}{*}{Patch-Blind, Noise-Blind} \\
\multicolumn{1}{|l|}{} &  & \multicolumn{1}{l|}{} &  \\ \hline
\multicolumn{1}{|l|}{ \begin{tabular}[c]{@{}l@{}}Fang attack$^*$\\~\cite{2019Local} \end{tabular} } & Fang Mkrum, Fang Trmean & \multicolumn{1}{l|}{\begin{tabular}[c]{@{}l@{}} Feature \\ replacement$^*$\end{tabular}} & \begin{tabular}[c]{@{}l@{}}FRA combine with the \\ above eight attacks\end{tabular} \\ \hline
\end{tabular}
\begin{tablenotes} 
\item {``$*$'' denotes untargeted adaptive or adaptive backdoor attacks. } 
\end{tablenotes}
\end{threeparttable}
}
\label{tab:attack_specific}
\vspace{-4mm}
\end{table}

\vspace{0.5mm}

\noindent\textbf{Evaluation Metrics:} 
\label{sec:metrics}
First, we assess 14 untargeted attacks of 5 types and 16 backdoor attacks of 5 types, as listed in Table~\ref{tab:attack_specific}. Table~\ref{tab:performance_metrics} presents 9 metrics to access the performance of the normal task and backdoor task against attacks from three aspects: effectiveness, stability, and robustness. For the normal task, we measure the overall \emph{accuracy} and \emph{model stability} of the global model. The effectiveness of the attack is measured by \emph{clean accuracy drop}. In addition, a set of category metrics is introduced to further evaluate the effectiveness and stability of untargeted attacks. These metrics include \emph{category accuracy}, \emph{worst category accuracy}, \emph{best category accuracy}, and \emph{category accuracy stability}. %Lower accuracy, best category accuracy, and worst category accuracy indicate stronger attacks, and higher clean accuracy drop, model stability, and category accuracy stability also indicate stronger attacks. 
For the backdoor task, the \emph{attack success rate} is used to assess the effectiveness of backdoor attacks. Additionally, the \emph{backdoor injection rounds} and \emph{backdoor removal rounds} are utilized to evaluate the robustness of backdoor attacks. %injection and removal speed. %Higher accuracy, attack success rate, and backdoor removal rounds indicate stronger attacks, and lower backdoor injection rounds and model stability imply stronger attacks.
%
% We employ nine metrics (as shown in Table~\ref{tab:performance_metrics}) to evaluate the effectiveness and robustness of attacks. We use accuracy (ACC) and model stability (MS) to measure the impact of each attack on the global model.
% For untargeted attacks, clean accuracy drop (CAD) and Category accuracy (CA) measure the effectiveness of the attack, i.e., the impact on the global model accuracy. 
% We incorporate three category metrics: worst category accuracy (WCA), best category accuracy (BCA), and category accuracy stability (CAS). Lower accuracy, best category accuracy, and worst category accuracy indicate stronger attacks, while higher clean accuracy drop, model stability, and category accuracy stability indicate stronger attacks. 
% For backdoor attacks, we use the attack success rate (ASR) to evaluate the effectiveness of the backdoor injection. In addition, we observe that the speed of backdoor injection and the speed of backdoor disappearance after the attack vary significantly across attacks. We incorporate backdoor injection rounds (BIR) and backdoor removal rounds (BRR) to measure the robustness of the backdoor attacks. Higher accuracy, attack success rate, and backdoor removal rounds indicate stronger attacks, while lower backdoor injection rounds and model stability imply stronger attacks.
%  and to evaluate the velocity of backdoor implantation and the natural disappearance speed of backdoors during normal training with terminating attacks. 

Second, to evaluate the detection performance, we select representative attacks (11 untargeted and 5 backdoor attacks). We assess detection effectiveness based on three aspects: detection results, performance impact, and backdoor-proofing of the global model. We use TPR and FPR for detection results. %Higher TPRs and lower FPRs indicate more effective detection methods. 
We use \emph{clean accuracy drop} to measure the accuracy decrease of the global model after attack and detection when compared to the clean model. For the highly imbalanced BDD100K dataset (where the number of images in the largest class is $100\times$ that of the smallest class), we use \emph{area under the curve} (AUC) to show model accuracy. The \emph{attack success rate} is used to assess the effectiveness of backdoor detection. %for the security of the global model. %A lower clean accuracy drop, a higher area under the curve, and a lower attack success rate indicate better detection performance.
% For untargeted attacks, higher TPR and lower FPR of untargeted attacks indicate that malicious clients have less influence on the main tasks of global models.

% eight metrics introduced in $\S$~\ref{sec:metrics}, accuracy, \textit{WCA/BCA}, model stability and category accuracy stability for untargeted attacks, accuracy, model stability, attack success rate, backdoor injection rounds and \textit{BDR} for backdoor attacks. 

% \vspace{-4mm}

% \vspace{-4mm}
\begin{table}[]
\centering
\scriptsize
\setlength\tabcolsep{2pt}
\caption{Evaluation metrics}
\vspace{-2mm}
\resizebox{\linewidth}{!}{
\begin{threeparttable}  
\begin{tabular}{@{}|l|l|@{}}
\hline %toprule
\textbf{Metrics}      & \textbf{Definition}               \\ % & Demand
\hline %midrule
(E) Accuracy & \begin{tabular}[c]{@{}c@{}}\# correct\ predictions / \# total\ predictions \end{tabular}                \\ %& High
(S)  Model stability &
  \begin{tabular}[c]{@{}c@{}} Standard deviation of accuracy \end{tabular} \\ %, \\ Accuracy stability &Low  for the last ten rounds \(\sqrt{\frac{1}{10}\sum_{r\in[\rho]}(ACC^r-\overline{ACC})^2}\)
% \hline 
% \midrule
(E, D) Clean accuracy drop & \# clean vanilla accuracy (w/o attack and detection) - \\
& \quad \# accuracy after the attack (and detection) \\
(E)  $\textrm{Category accuracy}_c$ &
  \begin{tabular}[c]{@{}c@{}} \# correct predictions of class $c$ / \end{tabular}  \\ %,\\ Accuracy on category & High
  & \quad \# total predictions of class $c$ \\
(E) Worst category accuracy & Minimum $\textrm{category accuracy}_c$ in all categories  \\ %, \\ Worst category accuracy  &  High \(\min\limits_{c\in[m]}CA_c\), where $m$ is the total number of categories
(E) Best category accuracy & Maximum $\textrm{category accuracy}_c$ in all categories \\
  %\begin{tabular}[c]{@{}c@{}}\(\max\limits_{c\in[m]}CA_c\)\end{tabular}, where $m$ is the total number of categories  \\ %, \\ Best category accuracy & High
(S) Category accuracy stability &
  \begin{tabular}[c]{@{}c@{}} Standard deviation of all category accuracies \end{tabular} \\  %, \\ Category accuracy stability & Low    \(\sqrt{\frac{1}{10}\sum_{r\in[\rho],c\in[m]}(CA_c^r-\overline{CA})^2}\) for the last ten rounds
% \hline 
% \midrule
(E, D)  Attack success rate &  \# correct poisoned predictions / \# total predictions  \\ % & High
(R)  Backdoor injection rounds  & Number of rounds to reach $90\%$ attack success rate \\ % & 
(R)  Backdoor removal rounds  & \begin{tabular}[c]{@{}l@{}}Number of rounds to remove backdoors (attack  \end{tabular} \\%& High   \multirow{1}{*}{\begin{tabular}[c]{@{}l@{}}\\ (reach the benchmark attack success rate) \\ \end{tabular}} 
& \quad success rate$=\!10\%$) when training with clean data \\
(D) True positive rate (TPR) & \begin{tabular}[c]{@{}l@{}}\# correctly identified malicious clients / \end{tabular} \\
& \quad \# total malicious clients \\
(D) False positive rate (FPR) & \# benign clients incorrectly identified as malicious /  \\
& \quad \# total benign updates \\
(D) Area under the curve & the area under the Receiver Operating \\ & \quad Characteristic (ROC) curve\\
\hline %bottomrule
\end{tabular}
\begin{tablenotes}
    \item{(E)-Effectiveness; (S)-Stability; (R)-Robustness; (D)-Detection performance}
\end{tablenotes}
\end{threeparttable}
}
\vspace{-4mm}
\label{tab:performance_metrics}
\end{table}

%----------------------------------------------------

% \noindent \textbf{Attacks and detection methods:}
% As summarized in Table~\ref{tab:attacktax1} (column 3), we evaluate 14 types of untargeted attacks and 16 types of backdoor attacks. 
% We detect untargeted and backdoor attacks by three different detection and defense methods: DnC~\cite{shejwalkar2021manipulating}, Median~\cite{yin2018byzantine}, and Fine-pruning~\cite{liu2018fine}, which represent gradient anomaly detection method, Byzantine-robust aggregation, and model sanitization method, respectively. 
% We use three levels (i.e., high, medium, and low) to measure the effectiveness of the detection methods. \textit{High} indicates that the method is effective in detection, i.e., it achieves a high TPR at a low FPR. \textit{Low} indicates that the method performs poorly in detection, i.e., it achieves high FPR when TPR is high. The \textit{Medium} detection effectiveness is between \textit{high} and \textit{low}.

\vspace{0.5mm}
\noindent\textbf{Baselines:} We compare FLTracer with four SOTA methods, including all three categories in $\S$~\ref{sec:defense}: Byzantine-robust aggregation, update anomaly detection, and model sanitization. 1) \emph{MKrum}\,\cite{blanchard2017machine} detect untargeted and backdoor attacks. Its initial strategy selects $n\!-\!2m\!-\!2$ clients for aggregation, but we reduce its high FPR by selecting $n\!-\!m\!-\!2$ clients.
2) \emph{DnC}\,\cite{shejwalkar2021manipulating} detect untargeted and backdoor attacks. We adjust the tolerance rate $c$ to detect the maximum number of malicious clients. 
3) \emph{FLAME}\,\cite{nguyen2022flame} is designed to detect backdoor attacks, and we adopt its parameters for detection.
4) \emph{Fine-pruning}\,\cite{liu2018fine} is a defense method used to mitigate the effectiveness of backdoor attacks. We choose $10\%$ clean samples from the test dataset randomly, then prune $50\%\!-\!70\%$ neurons with a $5\%$ change interval. 
We examine the effectiveness of the Differential Privacy (DP)-based method to counter backdoor attacks in Appendix~\ref{sec:DP_defense}. The results show that the adversary can alter their attack strategy to successfully inject a backdoor.

Results for non-IID settings are reported here, while results for IID settings can be found in Appendix~\ref{sec:appendix_evaluation}. The runtime of FLTracer is analyzed in Appendix~\ref{sec:appendix_runtime}. Our method's TPR and FPR are instantaneous, allowing erroneously classified clients to participate in subsequent training rounds.
% , unless stated otherwise

%% ---------------------------------------------------------------------------------------

\vspace{-2mm}

\subsection{Assessment of Previous Attacks and Detection} 
\label{sec:performance_attack}

% \vspace{-1.5mm}

We assess the effectiveness, stability, and robustness of untargeted and backdoor attacks and detection methods with 9 metrics. Here, we present important remarks, and for detailed analysis and results, please refer to Appendix~\ref{sec:assess_appendix}. 

\textbf{Untargeted attacks}, like Add noise, Sign-flipping, and Fix-Fix attacks, significantly impact the model's stability. \textbf{Untargeted adaptive attacks}, like MB and Fang attacks, have a significant impact on the model's accuracy and category accuracy.  
Out of 16 \textbf{backdoor attacks} (including adaptive attacks), backdoors can quickly be injected in approximately 30 rounds and persist for over 200 rounds. Injecting backdoors becomes faster and more successful as the global model converges. 
\textbf{Current detection methods} only achieve high accuracy in detecting a small fraction of specific attacks, but at the cost of a high FPR in non-IID settings.

% Detection Performance -------------------------------------------------------------------------------------

\vspace{-2mm}

\subsection{Performance of FLTracer} \label{sec:performance}

% \vspace{-1.5mm}

Based on the assessment results, we select effective and representative attacks for the detection evaluation, covering five types of untargeted and five types of backdoor attacks in Table~\ref{tab:attack_specific}.
FLTracer achieves the design goals in $\S$~\ref{sec:design_goal}.

\begin{figure}[]
  \centering
  % \subfigcapskip=-1cm
  \subfigure{
  \includegraphics[width=0.85\linewidth]{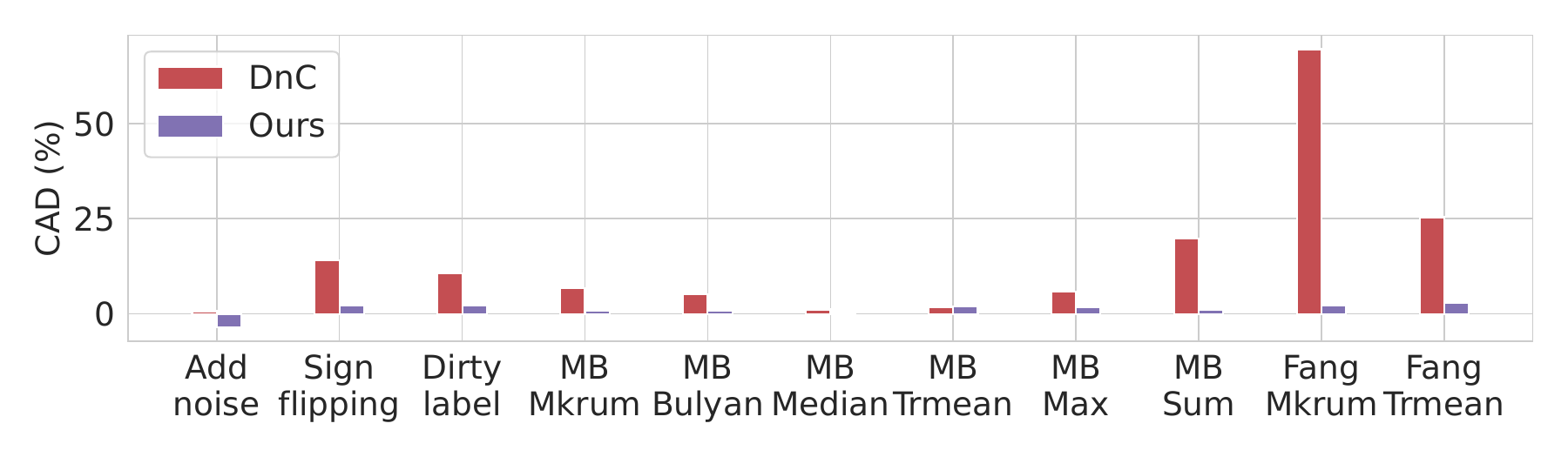}}
  \vskip -3.5mm
  \subfigure{
  \includegraphics[width=0.85\linewidth]{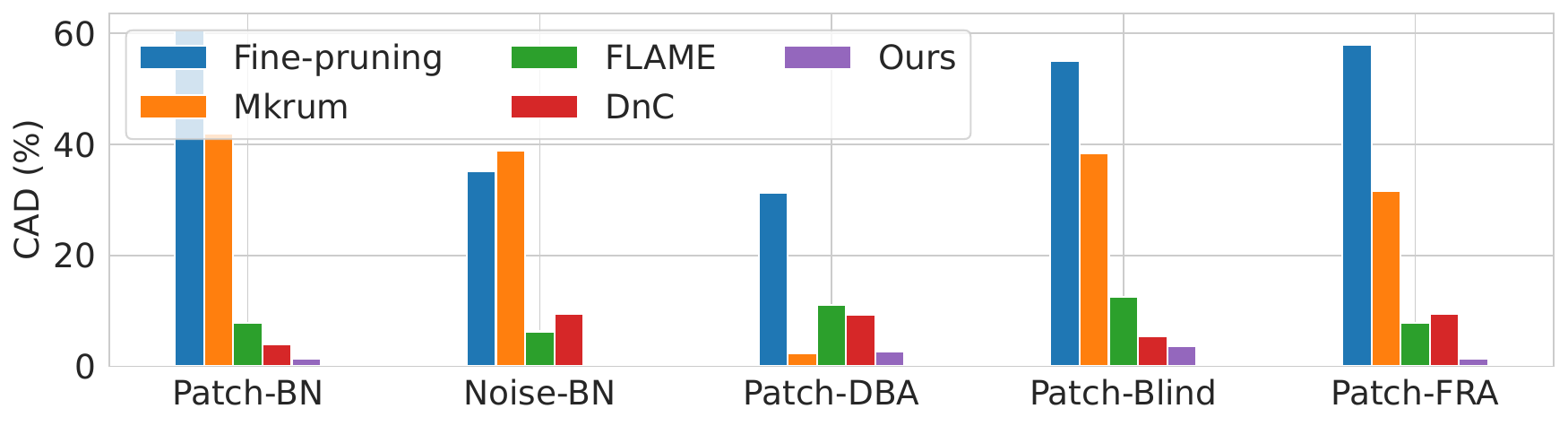}}
  \vspace{-3.5mm}
  \caption{Clean accuracy drop (CAD) of different methods on CIFAR10 (ResNet18). Top: against untargeted attacks. Bottom: against backdoor attacks. FLAME is unable to prevent backdoor injections against Patch-FRA even if only two clients remain for aggregation (attack success rate $\!=\!99.97\%$). }
  \vspace{-5mm}
  \label{fig:robust_evaul_cad}
\end{figure}

\vspace{0.5mm}
%Higher TPR and lower FPR for untargeted attacks indicate that malicious clients have less influence on the accuracy of global models. 
\noindent\textbf{Accuracy against untargeted attacks:} 
Table \ref{tab:tpr_untarget} displays the TPR and FPR of detection methods against untargeted attacks. The results show that, on average, our FLTracer achieves a significantly higher TPR ($63.42\%$ higher than MKrum and $21.05\%$ higher than DnC) with the lowest FPR ($29.54\%$ less than MKrum and $19.77\%$ less than DnC). In $10$ training rounds, FLTracer uses data from seven fewer malicious clients and 26 more benign clients than DnC, improving global model accuracy. Fig.~\ref{fig:robust_evaul_cad} (top) illustrates the global model's clean accuracy drop under untargeted attacks, indicating that FLTracer outperforms DnC with an average improvement of $13.5\%$, which is consistent with our analysis. 
The clean accuracy drop for FLTracer, even with a TPR as low as 75\% under the Dirty label attack on CIFAR10 (ResNet18), is only 2.2\%. This is due to the Dirty label attacl's relatively ineffective (see Fig.~\ref{fig:robustness_un}). 
In addition, we observe that FLTracer outperforms the clean vanilla model with a clean accuracy drop of $-3.59\%$. We will explain the reason behind this improvement in no-attack settings.
Moreover, FLTracer shows greater stability in comparison to MKrum and DnC, accurately detecting most settings across two datasets, three models, and $11$ attacks. However, DnC can only achieve precise detection on simple datasets/models, like EMNIST/SimpleNet. It exhibits a high FPR on CIFAR10. This is due to the heightened sensitivity of our \texttt{sortv}, \texttt{signv}, and \texttt{classv} in comparison to the updates
concatenation in Mkrum and DnC. Thus FLTracer can handle stealthy modifications even on complex models.

% \usepackage{multirow}
\begin{table}[]
\caption{Comparing TPR ($\%$) and FPR ($\%$) of  MKrum, DnC, and Ours against untargeted attacks. The \textbf{bold} and \textcolor{darkblue2}{blue} mark the best and second performances, respectively.}
\vspace{-2mm}
\centering
\scriptsize
\setlength\tabcolsep{2pt}
\resizebox{\linewidth}{!}{
\begin{tabular}{clcccccc}
\toprule
 &
  \multicolumn{1}{c}{} &
  \multicolumn{2}{c}{MKrum~\cite{blanchard2017machine}} &
  \multicolumn{2}{c}{DnC~\cite{shejwalkar2021manipulating}} &
  \multicolumn{2}{c}{FLTracer (Ours)} \\ \cmidrule(r){3-4} \cmidrule(r){5-6} \cmidrule(r){7-8} 
\multirow{-2}{*}{\begin{tabular}[c]{@{}c@{}}Dataset\\  (Model)\end{tabular}} &
  \multicolumn{1}{c}{\multirow{-2}{*}{Attack}} &
  TPR &
  FPR &
  TPR &
  FPR &
  TPR &
  FPR \\ \midrule
 &
  Add Noise &
  88.50 &
  27.88 &
  \textbf{100.0} &
  \textbf{0.00} &
  \textbf{100.0} &
  \textbf{0.00} \\
 &
  Sign-flipping &
  62.50 &
  34.38 &
  \textcolor{darkblue2}{87.00} &
  \textcolor{darkblue2}{28.63} &
  \textbf{90.50} &
  \textbf{1.88} \\
 &
  Dirty label (Fix-Fix) &
  68.00 &
  33.00 &
  \textcolor{darkblue2}{86.00} &
  \textcolor{darkblue2}{23.13} &
  \textbf{99.50} &
  \textbf{3.25} \\
 &
  MB Mkrum &
  25.00 &
  31.25 &
  \textcolor{darkblue2}{92.00} &
  \textcolor{darkblue2}{7.00} &
  \textbf{99.00} &
  \textbf{0.25} \\
 &
  MB Bulyan &
  25.00 &
  31.25 &
  \textbf{100.0} &
  \textbf{0.00} &
  \textbf{100.0} &
  \textbf{0.00} \\
 &
  MB Median &
  25.00 &
  31.25 &
  \textbf{100.0} &
  \textbf{0.00} &
  \textbf{100.0} &
  \textbf{0.00} \\
 &
  MB Trmean &
  25.00 &
  31.25 &
  \textbf{100.0} &
  \textbf{0.00} &
  \textcolor{darkblue2}{99.50} &
  \textcolor{darkblue2}{0.13} \\
 &
  MB Max &
  25.00 &
  31.25 &
  \textbf{100.0} &
  \textbf{0.00} &
  \textbf{100.0} &
  \textcolor{darkblue2}{0.13} \\
 &
  MB Sum &
  24.75 &
  31.31 &
  \textcolor{darkblue2}{93.00} &
  \textcolor{darkblue2}{1.75} &
  \textbf{98.00} &
  \textbf{0.13} \\
 &
  Fang Mkrum &
  23.50 &
  31.63 &
  \textcolor{darkblue2}{73.00} &
 \textcolor{darkblue2}{21.13} &
  \textbf{89.00} &
  \textbf{2.63} \\
\multirow{-11}{*}{\begin{tabular}[c]{@{}c@{}}EMNIST\\ (Simple-\\ Net)\end{tabular}} &
  Fang Trmean &
  25.00 &
  31.25 &
  \textbf{100.0} &
  \textbf{0.13} &
  \textcolor{darkblue2}{88.00} &
  \textcolor{darkblue2}{4.25} \\ \midrule
 &
  Add Noise &
  83.50 &
  29.13 &
  \textbf{100.0} &
  \textbf{0.00} &
  \textbf{100.0} &
  \textcolor{darkblue2}{0.13} \\
 &
  Sign-flipping &
  53.00 &
  36.75 &
  \textcolor{darkblue2}{72.00} &
  \textcolor{darkblue2}{33.25} &
  \textbf{100.0} &
  \textbf{0.00} \\
 &
  Dirty label (Fix-Fix) &
  42.50 &
  39.38 &
  \textbf{70.25} &
  \textcolor{darkblue2}{32.63} &
  \textcolor{darkblue2}{67.50} &
  \textbf{7.81} \\
 &
  MB Mkrum &
  11.50 &
  \textcolor{darkblue2}{34.63} &
  \textcolor{darkblue2}{57.25} &
  45.31 &
  \textbf{92.00} &
  \textbf{4.50} \\
 &
  MB Bulyan &
  12.50 &
  \textcolor{darkblue2}{34.38} &
  \textcolor{darkblue2}{70.25} &
  34.94 &
  \textbf{96.00} &
  \textbf{4.81} \\
 &
  MB Median &
  25.00 &
  31.25 &
  \textbf{100.0} &
  \textbf{0.00} &
  \textbf{100.0} &
  \textcolor{darkblue2}{0.38} \\
 &
  MB Trmean &
  25.00 &
  31.25 &
  \textbf{100.0} &
  \textbf{0.00} &
  \textbf{100.0} &
  \textcolor{darkblue2}{0.13} \\
 &
  MB Max &
  24.25 &
  31.44 &
  \textcolor{darkblue2}{82.25} &
  \textcolor{darkblue2}{30.38} &
  \textbf{95.00} &
  \textbf{7.00} \\
 &
  MB Sum &
  5.250 &
  \textcolor{darkblue2}{36.19} &
  \textcolor{darkblue2}{29.00} &
  45.31 &
  \textbf{94.00} &
  \textbf{5.13} \\
 &
  Fang Mkrum &
  1.000 &
  \textcolor{darkblue2}{37.25} &
  \textcolor{darkblue2}{2.50} &
  50.88 &
  \textbf{100.0} &
  \textbf{5.38} \\
\multirow{-11}{*}{\begin{tabular}[c]{@{}c@{}}CIFAR10\\  (AlexNet)\end{tabular}} &
  Fang Trmean &
  2.000 &
  37.00 &
  \textcolor{darkblue2}{84.41} &
  \textcolor{darkblue2}{30.96} &
  \textbf{98.00} &
  \textbf{3.13} \\ \midrule
 &
  Add Noise &
  71.50 &
  32.13 &
  \textcolor{darkblue2}{93.50} &
  \textcolor{darkblue2}{10.00} &
  \textbf{100.0} &
  \textbf{0.63} \\
 &
  Sign-flipping &
  61.00 &
  \textcolor{darkblue2}{34.75} &
  \textcolor{darkblue2}{65.50} &
  39.75 &
  \textbf{82.50} &
  \textbf{6.00} \\
 &
  Dirty label (Fix-Fix) &
  \textbf{82.00} &
  \textcolor{darkblue2}{14.25} &
  52.30 &
  45.28 &
  \textcolor{darkblue2}{75.00} &
  \textbf{6.56} \\
 &
  MB Mkrum &
  15.75 &
  \textcolor{darkblue2}{33.56} &
  \textcolor{darkblue2}{55.75} &
  43.94 &
  \textbf{100.0} &
  \textbf{5.81} \\
 &
  MB Bulyan &
  17.25 &
  \textcolor{darkblue2}{33.19} &
  \textcolor{darkblue2}{39.75} &
  40.69 &
  \textbf{94.00} &
  \textbf{7.13} \\
 &
  MB Median &
  24.75 &
  31.31 &
  \textbf{100.0} &
  \textcolor{darkblue2}{0.06} &
  \textbf{100.0} &
  \textbf{2.38} \\
 &
  MB Trmean &
  25.00 &
  31.25 &
  \textbf{100.0} &
  \textbf{0.00} &
  \textcolor{darkblue2}{100.0} &
  \textbf{4.19} \\
 &
  MB Max &
  14.00 &
  \textcolor{darkblue2}{34.00} &
  \textcolor{darkblue2}{31.50} &
  41.50 &
  \textbf{100.0} &
  \textbf{6.31} \\
 &
  MB Sum &
  4.250 &
  \textcolor{darkblue2}{36.44} &
  \textcolor{darkblue2}{31.75} &
  51.81 &
  \textbf{99.00} &
  \textbf{2.00} \\
 &
  Fang Mkrum &
  16.00 & 
   \textcolor{darkblue2}{33.50}&
  \textcolor{darkblue2}{48.44} &
  39.30 &
  \textbf{100.0} &
  \textbf{3.00} \\
\multirow{-11}{*}{\begin{tabular}[c]{@{}c@{}}CIFAR10\\  (ResNet18)\end{tabular}} &
  Fang Trmean &
  15.75 &
  \textcolor{darkblue2}{33.56} &
  \textcolor{darkblue2}{30.83} &
  52.08 &
  \textbf{86.50} &
  \textbf{2.44} \\\midrule
\textbf{Average} &
   &
  31.82 &
  32.49 &
  \textcolor{darkblue2}{74.19} &
  \textcolor{darkblue2}{22.72} &
  \textbf{95.24} &
  \textbf{2.95} \\\bottomrule
\end{tabular}
}
% \begin{tablenotes}
%     \item{$^*$ report the results of the Fix-Fix attack considering its best attack performance. }
% \end{tablenotes}
\label{tab:tpr_untarget}
\vspace{-3mm}
\end{table}

\vspace{0.5mm}
\noindent\textbf{Accuracy against backdoor attacks:} 
Table~\ref{tab:tpr_target} displays the TPR and FPR of detections against backdoor attacks. Our FLTracer outperforms three SOTA methods with the highest TPR and lowest FPR on CIFAR10, GTSRB, and HAR datasets, demonstrating its capability to accurately distinguish between malicious and benign clients even in complex datasets/models. Specifically, we achieve an average TPR of $99.13\%$ at an extremely low FPR of $2.29\%$, which is $23.08\%$ lower than MKrum, $27.05\%$ lower than FLAME, and $16.04\%$ lower than DnC. Per round, FLTracer detects the most malicious clients while using helpful data from two more benign clients out of $10$. We report the lowest clean accuracy drop for each method that can prevent the backdoor injection (see Fig.~\ref{fig:robust_evaul_cad} (bottom)). The results reveal that FLTracer's clean accuracy drop averages only $1.8\%$, which is $46.1\%, 28.8\%, 7.3\%$, and $5.7\%$ lower than Fine-pruning, MKrum, FLAME, and DnC, respectively. This indicates that our FLTracer can prevent backdoor injection while maintaining global model accuracy.

\vspace{0.5mm}
\noindent\textbf{Accuracy under no-attack} are analyzed in Appendix~\ref{sec:no_attack}.

\begin{table}[]
\caption{Comparing TPR ($\%$) and FPR ($\%$) of  MKrum, FLAME, DnC, and Ours against backdoor attacks. ``BN'', ``DBA'', and ``FRA'' denote the BadNets, distributed backdoor, and feature replacement attack, respectively. The \textbf{bold} and \textcolor{darkblue2}{blue} mark the best and second performances, respectively.}
\vspace{-2mm}
\centering
\scriptsize
\setlength\tabcolsep{2pt}
\resizebox{\linewidth}{!}{
\begin{tabular}{clcccccccc}
\toprule
 &
   &
  \multicolumn{2}{c}{MKrum~\cite{blanchard2017machine}} &
  \multicolumn{2}{c}{FLAME~\cite{nguyen2022flame}} &
  \multicolumn{2}{c}{DnC~\cite{shejwalkar2021manipulating}} &
  \multicolumn{2}{c}{FLTracer (Ours)} \\ \cmidrule(r){3-4} \cmidrule(r){5-6} \cmidrule(r){7-8} \cmidrule(r){9-10} 
\multirow{-2}{*}{\begin{tabular}[c]{@{}c@{}}Dataset \\ (Model)\end{tabular}} &
  \multirow{-2}{*}{Attack} &
  TPR &
  \multicolumn{1}{c}{FPR} &
  TPR &
  \multicolumn{1}{c}{FPR} &
  TPR &
  \multicolumn{1}{c}{FPR} &
  TPR &
  FPR \\ \midrule
 &
  \begin{tabular}[c]{@{}c@{}}Patch-BN\end{tabular} &
  91.00 &
  \multicolumn{1}{c}{23.22} &
  67.00 &
  \multicolumn{1}{c}{31.11} &
  \textbf{100.0} &
  \multicolumn{1}{c}{\textbf{0.22}} &
  \textbf{100.0} &
  \textcolor{darkblue2}{2.33} \\
 &
  \begin{tabular}[c]{@{}c@{}}Noise-BN\end{tabular} &
   \textcolor{darkblue2}{92.00} &
  \multicolumn{1}{c}{23.11} &
  54.00 &
  \multicolumn{1}{c}{30.56} &
  80.00 &
  \multicolumn{1}{c}{ \textcolor{darkblue2}{9.11}} &
  \textbf{100.0} &
  \textbf{3.89} \\
 &
  Patch-DBA &
  80.50 &
  \multicolumn{1}{c}{29.88} &
  54.50 &
  \multicolumn{1}{c}{24.22} &
   \textcolor{darkblue2}{ 94.00} &
  \multicolumn{1}{c}{ \textcolor{darkblue2}{ 11.13}} &
  \textbf{100.0} &
  \textbf{2.13} \\
 &
  Patch-Blind &
 
   \textcolor{darkblue2}{ 73.00} &
  \multicolumn{1}{c}{ \textcolor{darkblue2}{ 25.22}} &
  40.00 &
  \multicolumn{1}{c}{31.33} &
  50.00 &
  \multicolumn{1}{c}{16.67} &
  \textbf{99.00} &
  \textbf{6.78} \\
\multirow{-5}{*}{\begin{tabular}[c]{@{}c@{}}CIFAR10\\  (ResNet18)\end{tabular}} &
  Patch-FRA &
   \textcolor{darkblue2}{ 75.00} &
  \multicolumn{1}{c}{ \textcolor{darkblue2}{ 25.00}} &
  35.00 &
  \multicolumn{1}{c}{29.67} &
  55.00 &
  \multicolumn{1}{c}{28.22} &
  \textbf{100.0} &
  \textbf{4.78} \\ \midrule 
 &
  \begin{tabular}[c]{@{}c@{}}Patch-BN\end{tabular} &
  80.00 &
  \multicolumn{1}{c}{24.44} &
  58.00 &
  \multicolumn{1}{c}{31.78} &
   \textcolor{darkblue2}{ 98.00} &
  \multicolumn{1}{c}{ \textcolor{darkblue2}{ 11.33}} &
  \textbf{99.00} &
  \textbf{3.89} \\
 &
  \begin{tabular}[c]{@{}c@{}}Noise-BN\end{tabular} &
   \textcolor{darkblue2}{ 81.00} &
  \multicolumn{1}{c}{24.33} &
  60.00 &
  \multicolumn{1}{c}{32.89} &
  79.00 &
  \multicolumn{1}{c}{ \textcolor{darkblue2}{ 19.00}} &
  \textbf{98.00} &
  \textbf{2.56} \\
 &
  Patch-DBA &
  80.50 &
  \multicolumn{1}{c}{29.88} &
  50.00 &
  \multicolumn{1}{c}{28.11} &
   \textcolor{darkblue2}{ 95.50} &
  \multicolumn{1}{c}{ \textcolor{darkblue2}{ 8.38}} &
  \textbf{100.0} &
  \textbf{0.11} \\
 &
  Patch-Blind &
  68.00 &
  \multicolumn{1}{c}{25.78} &
  47.00 &
  \multicolumn{1}{c}{33.44} &
   \textcolor{darkblue2}{ 72.00} &
  \multicolumn{1}{c}{ \textcolor{darkblue2}{ 21.33}} &
  \textbf{93.00} &
  \textbf{5.56} \\
\multirow{-5}{*}{\begin{tabular}[c]{@{}c@{}}CIFAR10\\  (VGG16)\end{tabular}} &
  Patch-FRA &
   \textcolor{darkblue2}{ 75.00} &
  \multicolumn{1}{c}{25.00} &
  51.00 &
  \multicolumn{1}{c}{33.67} &
  52.00 &
  \multicolumn{1}{c}{ \textcolor{darkblue2}{ 22.00}} &
  \textbf{99.00} &
  \textbf{3.44} \\ \midrule 
 &
  \begin{tabular}[c]{@{}c@{}}Patch-BN\end{tabular} &
  80.00 &
  \multicolumn{1}{c}{ \textcolor{darkblue2}{ 24.44}} &
  44.00 &
  \multicolumn{1}{c}{29.33} &
   \textcolor{darkblue2}{ 93.00} &
  \multicolumn{1}{c}{24.56} &
  \textbf{100.0} &
  \textbf{0.00} \\
 &
  \begin{tabular}[c]{@{}c@{}}Noise-BN\end{tabular} &
  81.00 &
  \multicolumn{1}{c}{ \textcolor{darkblue2}{ 24.33}} &
  49.00 &
  \multicolumn{1}{c}{27.22} &
   \textcolor{darkblue2}{ 92.00} &
  \multicolumn{1}{c}{24.67} &
  \textbf{100.0} &
  \textbf{0.00} \\
 &
  Patch-DBA &
  69.00 &
  \multicolumn{1}{c}{32.75} &
  58.50 &
  \multicolumn{1}{c}{ \textcolor{darkblue2}{ 28.13}} &
   \textcolor{darkblue2}{ 91.00} &
  \multicolumn{1}{c}{33.75} &
  \textbf{99.50} &
  \textbf{0.13} \\
 &
  Patch-Blind &
  75.00 &
  \multicolumn{1}{c}{25.00} &
  42.00 &
  \multicolumn{1}{c}{29.22} &
   \textcolor{darkblue2}{ 99.00} &
  \multicolumn{1}{c}{ \textcolor{darkblue2}{ 21.56}} &
  \textbf{100.0} &
  \textbf{0.00} \\
\multirow{-5}{*}{\begin{tabular}[c]{@{}c@{}}GTSRB\\  (ResNet34)\end{tabular}} &
  Patch-FRA &
   \textcolor{darkblue2}{ 74.00} &
  \multicolumn{1}{c}{ \textcolor{darkblue2}{ 25.11}} &
  35.00 &
  \multicolumn{1}{c}{30.11} &
  71.00 &
  \multicolumn{1}{c}{31.56} &
  \textbf{100.0} &
  \textbf{0.22} \\ \midrule 
 &
  \begin{tabular}[c]{@{}c@{}}Patch-BN\end{tabular} &
   \textcolor{darkblue2}{ 99.00} &
  \multicolumn{1}{c}{ \textcolor{darkblue2}{ 25.25}} &
  28.50 &
  \multicolumn{1}{c}{31.75} &
  85.50 &
  \multicolumn{1}{c}{30.25} &
  \textbf{99.50} &
  \textbf{2.13} \\
 &
  Patch-DBA &
   \textcolor{darkblue2}{ 97.50} &
  \multicolumn{1}{c}{ \textcolor{darkblue2}{ 25.63}} &
  53.50 &
  \multicolumn{1}{c}{27.00} &
  83.50 &
  \multicolumn{1}{c}{34.75} &
  \textbf{99.00} &
  \textbf{3.38} \\
 &
  Patch-Blind &
   \textcolor{darkblue2}{ 94.50} &
  \multicolumn{1}{c}{ \textcolor{darkblue2}{ 26.38}} &
  56.00 &
  \multicolumn{1}{c}{26.38} &
  92.50 &
  \multicolumn{1}{c}{33.25} &
  \textbf{98.50} &
  \textbf{4.25} \\
\multirow{-4}{*}{\begin{tabular}[c]{@{}c@{}}HAR\\ (DNN)\end{tabular}} &
  Patch-FRA &
   \textcolor{darkblue2}{ 98.50} &
  \multicolumn{1}{c}{25.38} &
  91.00 &
  \multicolumn{1}{c}{ \textcolor{darkblue2}{ 20.38}} &
  81.50 &
  \multicolumn{1}{c}{32.38} &
  \textbf{100.0} &
  \textbf{4.63} \\ \midrule 

  &
  \begin{tabular}[c]{@{}c@{}}Patch-BN\end{tabular} &
  90.00 &
  23.33 &
  35.00 &
  30.56 &
  \textbf{100.0} &
  \textbf{0.00} &
  \textbf{100.0} &
  \textcolor{darkblue2}{0.22} \\
 &
  \begin{tabular}[c]{@{}c@{}}Noise-BN\end{tabular} &
   83.00 &
  24.11 &
  39.00 &
  30.67 &
  \textbf{100.0} &
  \textbf{0.00} &
  \textbf{100.0} &
  \textcolor{darkblue2}{0.11}  \\
 &
  Patch-DBA &
  88.00 &
  \textcolor{darkblue2}{23.56} &
  34.00 &
  26.25 &
  \textcolor{darkblue2}{96.50} &
  25.88 &
  \textbf{99.50} &
  \textbf{3.75} \\
 &
  Patch-Blind &
  87.00 &
  23.67 &
  31.00 &
  29.89 &
  \textbf{100.0} &
  \textbf{0.00} &
  \textcolor{darkblue2}{97.00} &
  \textcolor{darkblue2}{0.33}  \\
\multirow{-5}{*}{\begin{tabular}[c]{@{}c@{}}BDD100K\\  (ViT)\end{tabular}} &
  Patch-FRA & 84.00 & 24.00 & 31.00 & 30.56 & \textbf{100.0} & \textbf{0.00} & \textcolor{darkblue2}{98.00} & \textcolor{darkblue2}{0.33}  \\ \midrule 
\textbf{Average} &
  \multicolumn{1}{l}{} &
  83.19 &
  25.37 &
  47.67 &
  29.34 &
   \textcolor{darkblue2}{85.88} &
  \textcolor{darkblue2}{18.33} &
  \textbf{99.13} &
  \textbf{2.29} \\ \bottomrule  
\end{tabular}
}
\begin{tablenotes}
    \item{Results against MRA are not reported because the malicious updates scale up to $10\times$, so extreme anomalies are easily captured for almost all detections.}
\end{tablenotes}
\label{tab:tpr_target}
\vspace{-4mm}
\end{table}

\begin{figure}[] %!ht
  \centering
% \vspace{-2mm}
  \includegraphics[width=0.8\linewidth]{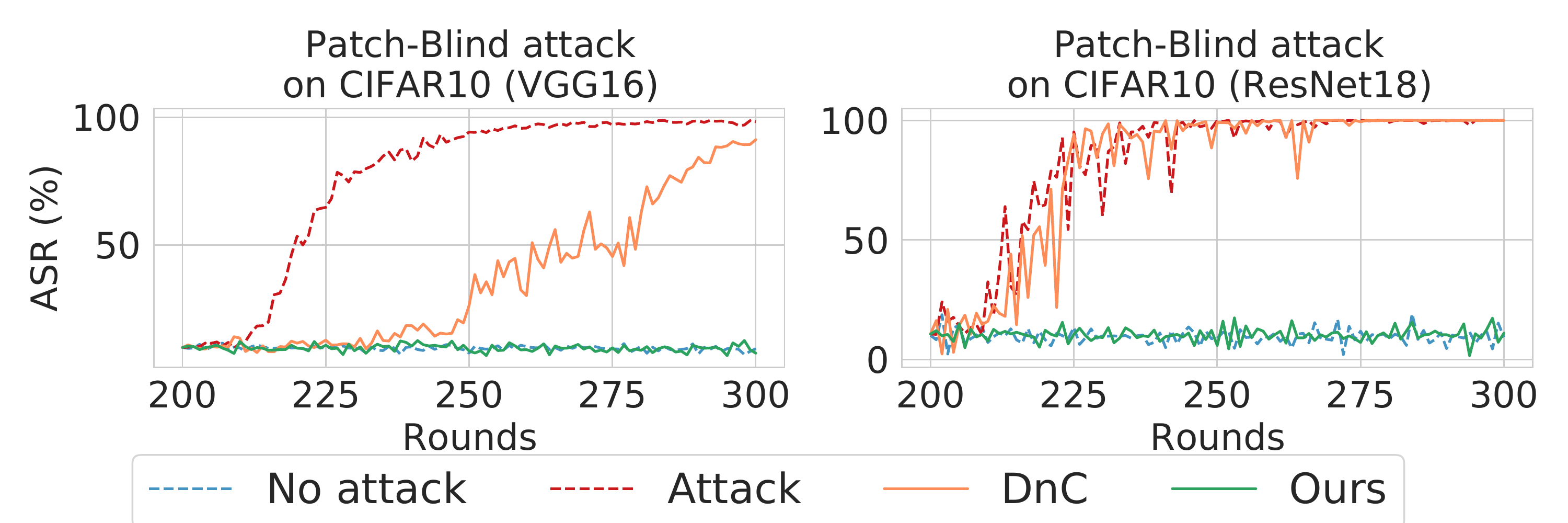}
  \vspace{-2mm}
  \caption{Attack success rate (ASR) of the different attacks. }
  \vspace{-5mm}
  \label{fig:acc_t_cifar_backdoor}
\end{figure}

\vspace{0.5mm}
\noindent\textbf{Backdoor-proof:} 
% SwS successfully prevents 40 backdoor injections out of 40 backdoor attacks without sacrificing the accuracy of the global model in IID and non-IID settings.
Recall that backdoor accuracy can achieve $90\%$ with an average of $36$ backdoor injection rounds (see Fig.~\ref{fig:Robustness_tar}). This means at least $70$ malicious clients must be detected (TPR$>\!70\%$) to reduce the attack success rate below $90\%$. Table~\ref{tab:tpr_target} shows that FLTracer can successfully address 25 backdoor injections in 25 settings while marginally impairing the global model accuracy. In contrast, other methods, such as FLAME, are unable to prevent Patch-FRA backdoor injections unless every update is excluded from the aggregation. The attack success rate of various attacks is shown in Fig.~\ref{fig:acc_t_cifar_backdoor}. FLTracer can successfully prevent backdoor injection even at the lowest TPR, particularly Patch-Blind attacks on CIFAR10 (VGG16), whereas DnC cannot.

% In our experiments,  with two failures against BL (A-BM) classical and adaptive attacks on CIFAR10 (ResNet18). This is mainly due to the weak attack effect of BL trigger combined with A-BM attack. We select three settings to show the security analysis of FLTracer and DnC on CIFAR10, as shown in Fig.~\ref{fig:acc_t_cifar_backdoor}. Specifically, in BN (A-M) classical attack, due to the 100\% TPR and 0\% FPR, both DnC and FLTracer are effective, their attack success rates are consistent with the baseline. In BN (A-BM) adaptive attack, the attack success rate of DnC is 100\% at only 32\% TPR, implying the backdoor can be successfully implanted. In contrast, the attack success rate of FLTracer is still close to the baseline. In BL (A-BM) adaptive attack, despite the worst detection result (TPR: 93\%) of FLTracer, FLTracer can prevent backdoor injection successfully. 

% We also compare FLTracer with Fine-pruning (see Fig.~\ref{fig:fine_pruning_acc}). FLTracer significantly outperforms Fine-pruning. In our experiments, when the attack success rate of the global model by Fine-pruning is the same as the attack success rate of the normal model, the performance of the model is in ruin, where accuracy drops from 30\% to 50\%. This means that there is a trade-off between backdoor tasks and main tasks. However, the final global model trained with FLTracer achieves a win-win, whose attack success rate and accuracy are all extremely close to the normal model.

\vspace{-2mm}

\subsection{Impact of Different Data Distribution} \label{sec:exp_diff_system}

% \vspace{-1.5mm}

\noindent\textbf{Different degrees of non-IID:} 
Table~\ref{tab:diff_d} displays the TPR and FPR of detections under varying degrees of non-IID. It indicates that the detection performance is worse as $d$ decreases due to a more heterogeneous data distribution and greater obscurity of the adversary. FLTracer surpasses the other three methods in attaining the highest TPR and FPR across all degrees of $d$. In particular, under the one-class setting, FLTracer achieves a TPR that is $5\%$ higher and an FPR that is $20\%$ lower compared to the second best method.

% \vspace{-2mm}

\begin{table}[] %!ht
\caption{TPR ($\%$) and FPR ($\%$) of detection against Patch-BN attack on CIFAR10 (ResNet18) under different degrees of non-IID. ``one class'' represents each client has only one class. The \textbf{bold} and \textcolor{darkblue}{blue} mark the best and second performances.}
\vspace{-2mm}
\centering
\scriptsize
\setlength\tabcolsep{3pt}
\resizebox{\linewidth}{!}{
\begin{tabular}{lcccccccc}
\toprule
\multicolumn{1}{c}{\multirow{2}{*}{\begin{tabular}[c]{@{}c@{}}Data\\ distribution\end{tabular}}} &
  \multicolumn{2}{c}{MKrum~\cite{blanchard2017machine}} &
  \multicolumn{2}{c}{FLAME~\cite{nguyen2022flame}} &
  \multicolumn{2}{c}{DnC~\cite{shejwalkar2021manipulating}} &
  \multicolumn{2}{c}{FLTracer (Ours)} \\ 
  \cmidrule(r){2-3} \cmidrule(r){4-5} \cmidrule(r){6-7} \cmidrule(r){8-9} 
\multicolumn{1}{c}{} & TPR   & FPR            & TPR   & FPR   & TPR            & FPR            & TPR            & FPR            \\ \midrule
one class              & 77.00 & \textcolor{darkblue}{24.78} & 20.00 & 32.44 & \textcolor{darkblue}{89.00} & 27.78          & \textbf{94.00} & \textbf{7.67} \\
$d=0.5$                & 91.00 & 23.22          & 67.00 & 31.11 & \textbf{100.0} & \textcolor{darkblue}{2.33} & \textbf{100.0} & \textbf{0.22} \\
$d=5$                  & 90.00 & 23.33          & 68.00 & 26.22 & \textbf{100.0} & \textbf{0.00} & \textbf{100.0} & \textbf{0.00} \\
$d\!\to\! +\infty$ (IID)                & 88.00 & 23.56          & 96.00 & 26.56 & \textbf{100.0} & \textbf{0.00} & \textbf{100.0} & \textbf{0.00} \\ \bottomrule
\end{tabular}
}
\vspace{-4mm}
\label{tab:diff_d}
\end{table}

\begin{figure}[!ht]
\vspace{-2mm}
 \includegraphics[width=0.98\linewidth]{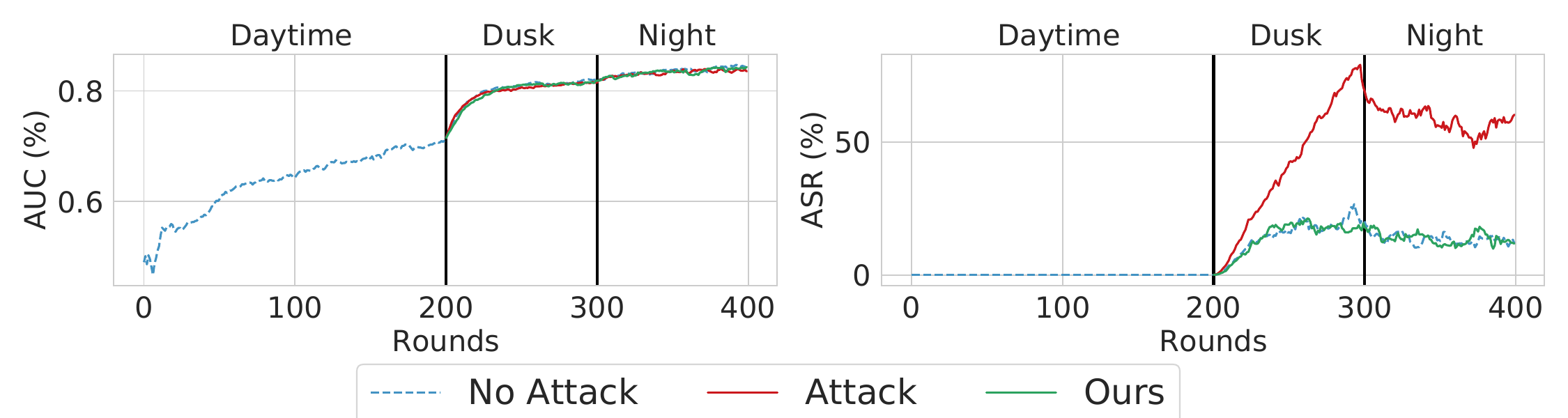}
 \vspace{-2mm}
  \caption{Area under the curve (AUC) of the main task and attack success rate (ASR) on BDD100K. Data distribution of clients change over time (daytime, dusk, and night). We launch Patch-BN attack after round 200.}
  \label{fig:acc_data_change}
\vspace{-4mm}
\end{figure}

\vspace{0.5mm}
\noindent\textbf{Training data over Time:}
While clients' data changes gradually, we detect and update the dataset parameter $\mathrm{\Omega}$ of clients identified as benign per round, see $\S$~\ref{sec:kf_model}. During rounds 0-200, every client normally trains using only daytime data. In rounds 200-300, each client adds the dusk data, while adversaries launch a backdoor attack. In rounds 300-400, each client adds nighttime data, while adversaries continue their backdoor attack. Fig.~\ref{fig:acc_data_change} shows the successful injection of the backdoor in a real-world scenario where the dataset changes gradually. Our detection effectively thwarts the injection of the backdoor and achieves an AUC value comparable to no-attack.

\vspace{-2mm}

\subsection{Impact of Selecting Different $\tau$} \label{sec:tau}
\begin{figure}[!ht]
  \centering
  \includegraphics[width=0.9\linewidth]{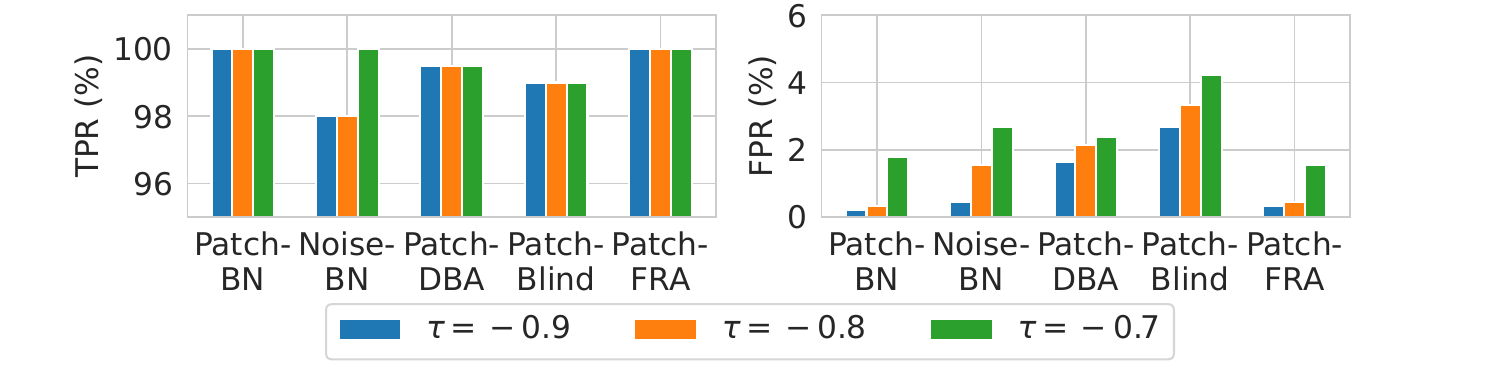}
  \vspace{-2mm}
  \caption{The impact of different $\tau$ on CIFAR10 (ResNet18).}
  \vspace{-4mm}
  \label{fig:certain_attacker_threshold}
\end{figure}

% \vspace{-1.5mm}
In FLTracer, we utilize a threshold $\tau\sim -1$ in \textsc{TaskDetect} function (line 15 of Alg.~\ref{alg:detect_alg}) to identify malicious clients. Fig.~\ref{fig:certain_attacker_threshold} presents the TPR and FPR of FLTracer at different $\tau$. The results indicate that smaller $\tau$ corresponds to lower TPR and lower FPR, resulting in fewer malicious clients. However, the impact of $\tau$ on TPR and FPR is not significant, with a maximum change of $\leq 2\%$ for both TPR and FPR. Hence, the choice of $\tau$ can be adjusted based on the desired trade-off between accuracy and backdoor-proof (larger $\tau$ for higher accuracy and a smaller $\tau$ for enhanced backdoor-proof).

\vspace{-3mm}

\vspace{-2mm}
\subsection{Impact of Percentage of Malicious Client}

\begin{figure}[!ht]
  \centering
  % \subfigcapskip=-1cm
  \subfigure{
  \includegraphics[width=0.9\linewidth]{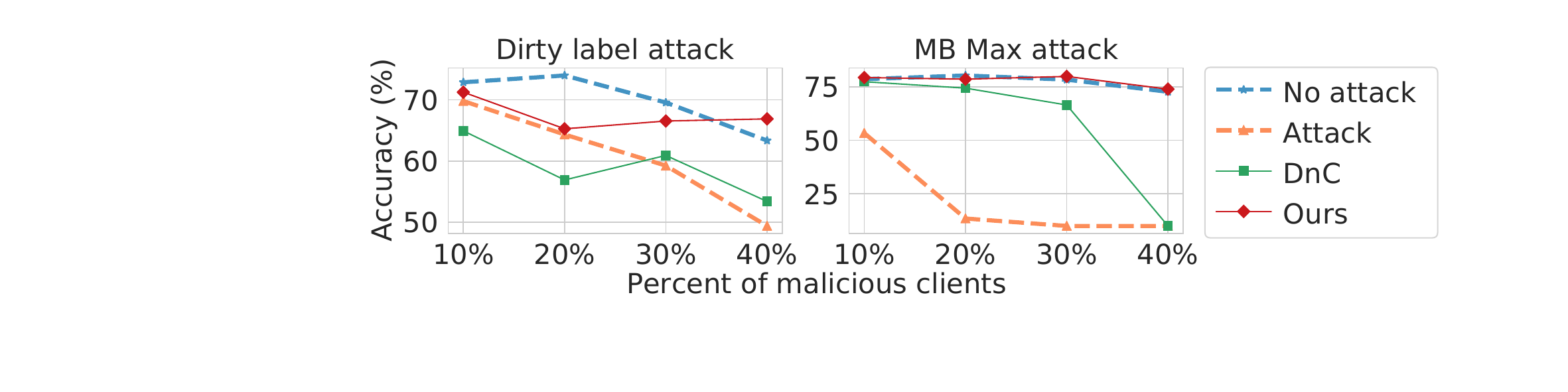}}
  \vskip -3mm
  \subfigure{
  \includegraphics[width=0.9\linewidth]{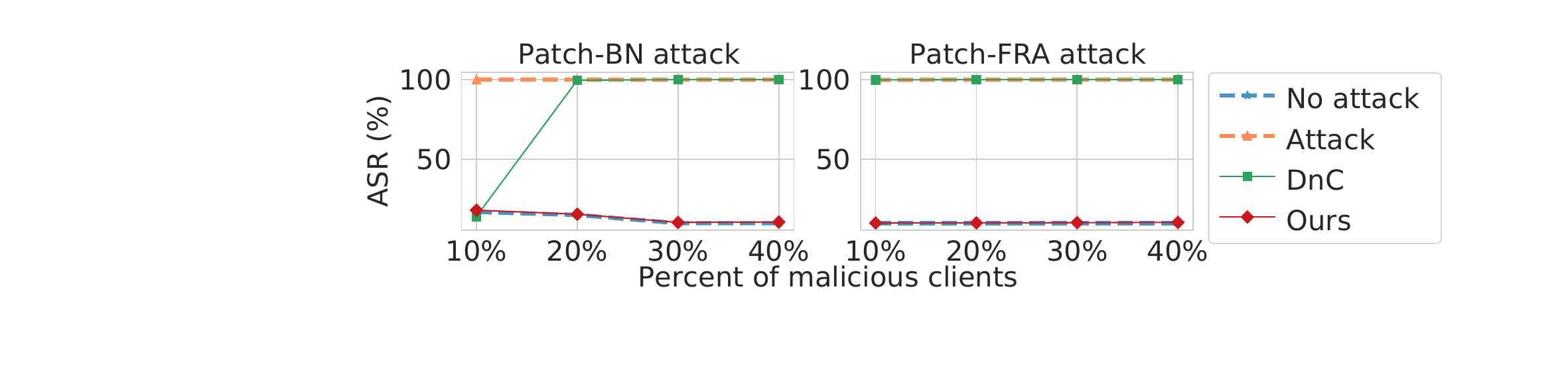}}
  \vspace{-4mm}
  \caption{Impact of the percentage of malicious clients on accuracy and attack success rate (ASR) in CIFAR10 (ResNet18). Top-untargeted attacks. Bottom-targeted attacks.}
  \vspace{-4mm}
  \label{fig:attacker_ratio}
\end{figure}

%\noindent\textbf{Impact of the percentage of malicious clients:} 
Fig.~\ref{fig:attacker_ratio} compares FLTracer and DnC's accuracy and attack success rate for varying percentages of malicious clients. We select two untargeted and two backdoor attacks. For untargeted attacks, DnC experiences a significant decrease in accuracy. However, this has minimal effect on FLTracer. Specifically, even with the worst detection result in the Dirty label attack, FLTracer experiences a smaller decline in accuracy than DnC ($2.45\%$ vs. $10.9\%$). In the MB Max attack, FLTracer can effectively prevent the accuracy drop, which sharply decreases under DnC, particularly with $40\%$ of malicious clients (accuracy: $10\%$). For backdoor attacks, backdoors are injected more rapidly as the number of malicious clients increases. In the Patch-BN attack, FLTracer is robust with varying numbers of malicious clients, whereas DnC is ineffective with more than $10\%$ of malicious clients. FLTracer also has a more stable detection effect against the Patch-FRA attack.

% \vspace{-1mm}

% \begin{figure}[h]
%  %\vspace{-4mm}
%   \centering
%   \subfigcapskip=-4pt
%   \subfigure[Untargeted attacks under different datasets and model structures\vspace{-3mm}]{
%   \includegraphics[width=\linewidth]{./figures/evaul_untar_attacker_ratio}}\vskip -4pt
%   \subfigure[Targeted attacks under different datasets and model structures]{
%   \includegraphics[width=\linewidth]{./figures/evaul_tar_attacker_ratio}
%   }
%   \vspace{-4mm}
%   \caption{Impact of the percentage of malicious clients on accuracy and attack success rate in different attacks.}
%   \vspace{-5mm}
%   \label{fig:attacker_ratio}
% \end{figure}

\vspace{-3mm}

\subsection{Ablation Study} \label{sec:abalation}
Table~\ref{tab:ablation} shows the impact of removing different components of FLTracer against untargeted attacks (Sign-flipping and Fang Mkrum) and backdoor attacks (Patch-BN and Patch-Blind). The results indicate that all three detections are crucial in enhancing effectiveness, specifically in achieving high TPR and low FPR. We observe that Local Anomaly Detection contributes the most to the high TPR in untargeted attack detection. Task Detection and Domain Detection effectively increase TPR and decrease FPR in backdoor attack detection. 

\begin{table}[!h]
\centering
\scriptsize
\vspace{-1mm}
\caption{TPR/ FPR of FLTracer for removing certain detections on CIFAR10 (ResNet18). ``J'' denotes the joint decision-making with all detections.}
\vspace{-2mm}
\setlength\tabcolsep{5pt}
\begin{tabular}{cccccc}
\toprule
 & Attacks & TPR(\%) & FPR(\%) & TPR-J(\%) & FPR-J(\%) \\
\midrule
\multirow{4}{*}{\begin{tabular}[c]{@{}c@{}}Remove Local \\ Anomaly \\ Detection\end{tabular}} 
 & Sign-flipping & 14.00 & 3.25 & 82.50 & 6.00 \\
 & Fang Mkrum & 34.50 & 2.81 & 100.0& 3.00 \\
 & Patch-BN & 100.0& 2.33 & 100.0& 2.33 \\
 & Patch-Blind & 99.00 & 6.78 & 99.00 & 6.78 \\
\midrule
\multirow{4}{*}{\begin{tabular}[c]{@{}c@{}}Remove Task \\ Detection\end{tabular}}
 & Sign-flipping & 82.50 & 6.00 & 82.50 & 6.00 \\ 
 & Fang Mkrum & 87.00 & 1.19 & 100.0& 3.00 \\
 & Patch-BN & 97.00 & 11.00 & 100.0& 2.33 \\
 & Patch-Blind & 93.00 & 13.90 & 99.00 & 6.78 \\
\midrule
\multirow{4}{*}{\begin{tabular}[c]{@{}c@{}}Remove \\ Domain \\ Detection\end{tabular}}
 & Sign-flipping & 82.50 & 6.00 & 82.50 & 6.00 \\
 & Fang Mkrum & 87.00 & 1.19 & 100.0& 3.00 \\
 & Patch-BN & 100.0& 0.78 & 100.0& 2.33 \\
 & Patch-Blind & 100.0& 14.70 & 99.00 & 6.78 \\
\bottomrule
\end{tabular}
\vspace{-4mm}
\label{tab:ablation}
\end{table}

% *Adaptive strategies against LAD and TCD are implemented by adding additional loss functions during training. The adversary adapts to all the metrics to evade detection.
 \vspace{-3mm}
\subsection{Advanced Adaptive Attacks} 
 % \vspace{-2mm}
We assume that the adversary knows our FLTracer and designs advanced attacks to evade individual detection strategies. We assess their efficacy on CIFAR10 (ResNet18). Notably, combining two evasion strategies weakens the attack strength, as evading multiple detections entails adjusting fewer parameters (otherwise, it can be easily detected).

\begin{figure}[!ht]
\vspace{-9mm}
\centering
\subfigcapskip=-10pt
\subfigbottomskip=1pt
\subfigure[Evade local anomaly detection]{
\begin{minipage}[t]{1\linewidth}
\centering
\includegraphics[width=0.4\linewidth]{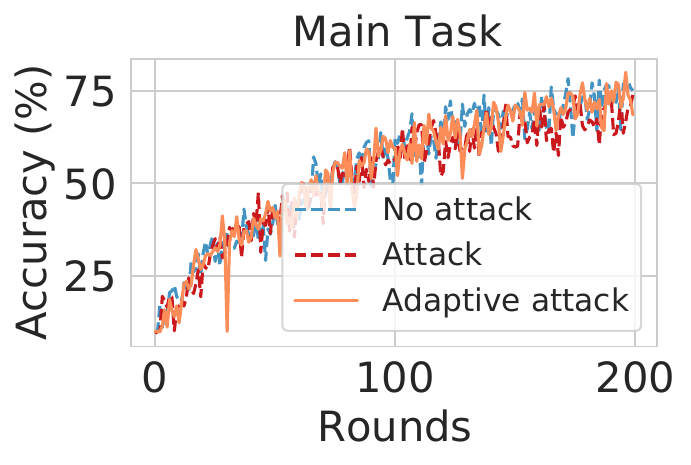}
\end{minipage}
}
\vskip -0.5mm
\subfigure[Evade task and domain detection]{
\includegraphics[width=0.8\linewidth]{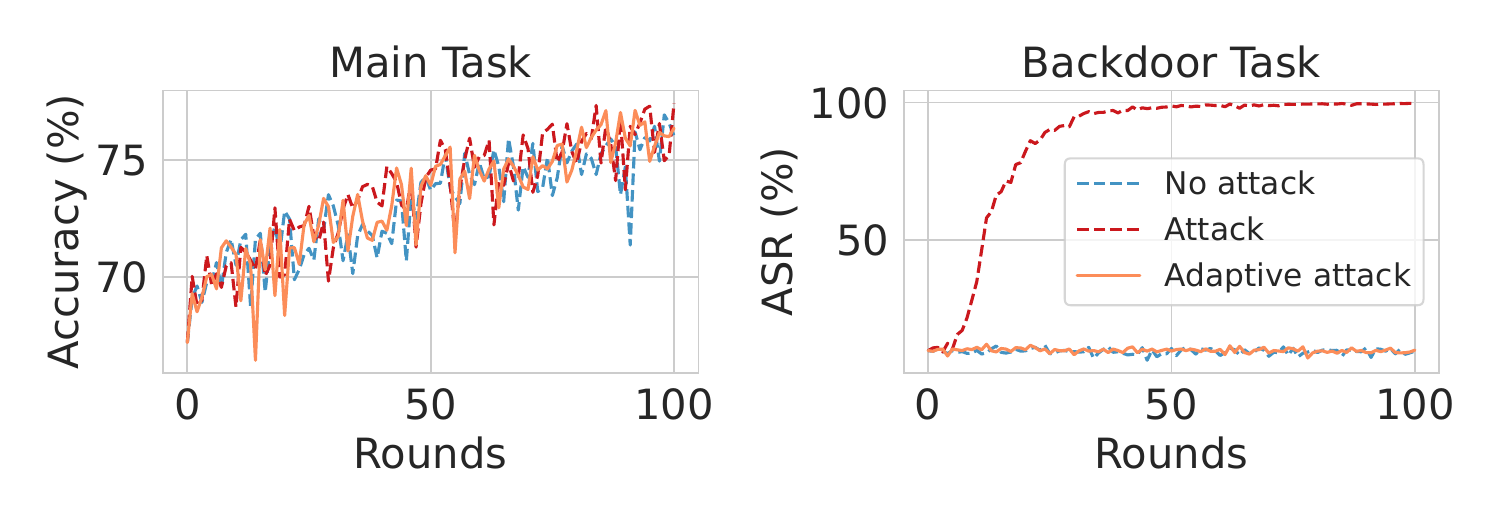}}
\vskip -0.5mm
\subfigure[Evade task detection]{
\includegraphics[width=0.8\linewidth]{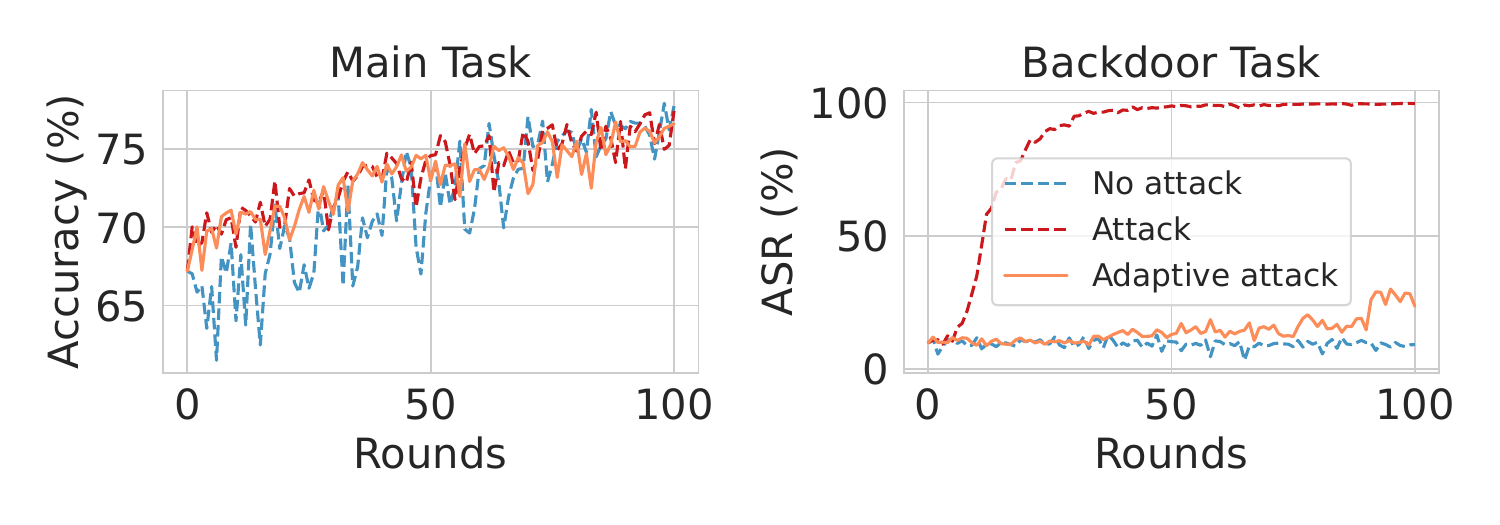}}
\vskip -0.5mm
\subfigure[Evade domain detection]{
\includegraphics[width=0.8\linewidth]{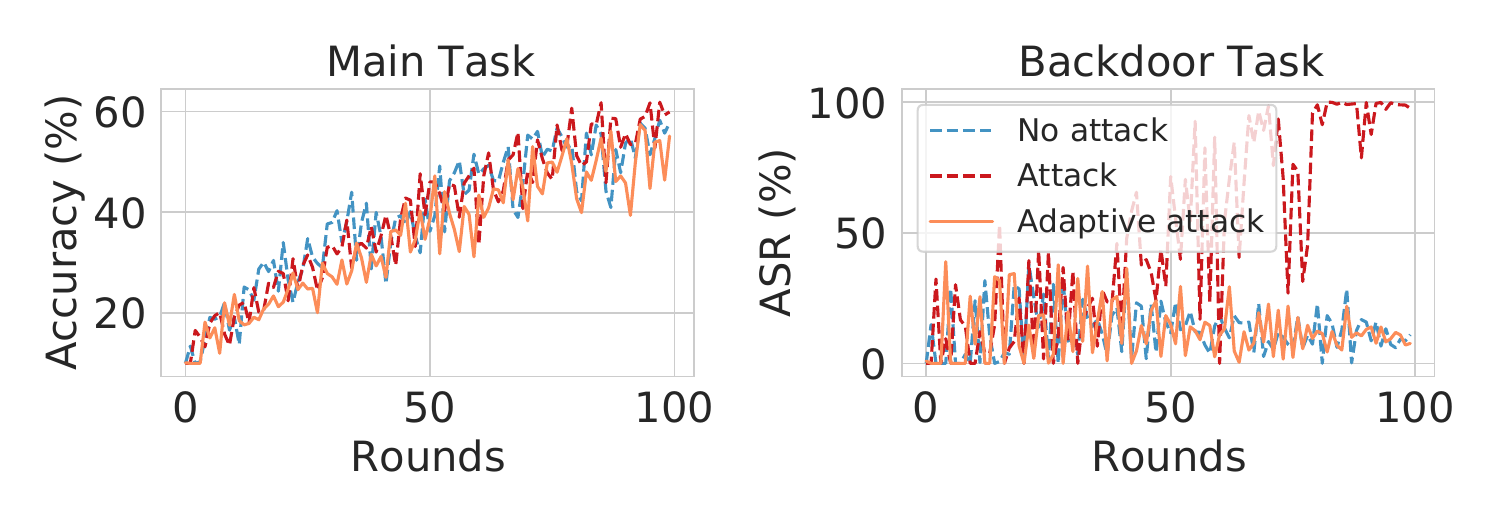}}
% \vskip -4mm
\vspace{-2mm}
\caption{Accuracy and attack success rate (ASR) of FLTracer against advanced adaptive attacks.
% The attack success rate of BadNets attacks on CIFAR10 (VGG16) and GTSRB (ResNet34). Backdoor attack is the baseline that modifies all layers. Fixed conv attack means freezing the weight of convolution layers. Conv replacement attack only modifies the weight of convolution layers.
}
\label{fig:fixed_conv}
\end{figure}

\vspace{0.5mm}
\noindent\textbf{Evade local anomaly detection:}
We assume that the adversary knows all benign client updates to evade anomaly detection by \texttt{signv}, \texttt{sortv}, and \texttt{classv}. The adversary first trains normally to obtain benign updates ($u^0$) and computes the original $\texttt{signv}^0$, $\texttt{sortv}^0$, and $\texttt{classv}^0$ based on all other benign updates. Then, the adversary modifies the update to the maximum extent possible while ensuring these three features remain unchanged to counter detection. The attack against our FLTracer only cause 
a clean accuracy drop of $0.09\%$ as the limited modification of updates (see Fig. \ref{fig:fixed_conv} (a)).

\vspace{0.5mm}
\noindent\textbf{Evade task\,and\,domain detection:}
A direct advanced backdoor attack is to refine the benign model with frozen feature extractor layers. For example, the adversary replaces only the bias and classifier with the backdoor model. This attempted backdoor injection almost fails, despite many rounds of training for the bias and classifier. The attack success rate is only $9.48\%$, and the clean accuracy drops by $1.31\%$ (see Fig.\ref{fig:fixed_conv}(b)).

\vspace{0.5mm}
\noindent\textbf{Evade task detection:} We assume the adversary knows all benign clients' updates. The adversary can selectively retain the updates of certain layers in a backdoor model to reduce alterations of updates. The adversary first trains a backdoor model and calculates TSim ($\alpha^\kappa$) for each layer. If $\alpha^\kappa\!\leq\!0$, the attack on layer $\kappa$ is revoked. Additionally, if the adjusted TSim ($\alpha_{new}^\kappa$) shows a downward trend, the attack on all convolutional layers at the current round is revoked. Under FLTracer, after 100 rounds of training, the attack's success rate is $27.09\%$ and clean accuracy drop is $0.52\%$ (see Fig.\ref{fig:fixed_conv}(c)).

\vspace{0.5mm}
\noindent\textbf{Evade domain detection:} The adversary can launch the backdoor attack at the first round of training to counter the domain detection. 
However, injecting the backdoor into the global model while simultaneously converging the global model is difficult when its accuracy is low. Moreover, at the beginning of training, benign clients provide more general patterns, resulting in adversary updates being more anomalous. Thus, our Task Detection is sufficient to identify the adversary accurately, and the malicious candidates by Task Detection can be directly considered adversaries, i.e., $M_d=M_c\cup M_d$. This attack achieves an attack success rate of only $13.99\%$ with a clean accuracy drop of $2.41\%$, as shown in Fig.~\ref{fig:fixed_conv} (d).

\vspace{0.5mm}
\noindent\textbf{Evading a combination} of strategies one, three, and four leads to a clean accuracy drop of 0.301\% for new untargeted attack; an attack success rate of 10.31\%, and a clean accuracy drop of 2.172\% for new backdoor attack, fails to inject backdoors.

\section{Related Work}
% \vspace{-2mm}
\label{sec:defense}
% Next, we describe the existing defenses to poisoning attacks and adaptive attacks counter the defenses.

% \subsubsection{Detection methods on FL} aim to directly remove the malicious updates or mitigate the effect of malicious updates generated by different poisoning attacks based on three lines of techniques. 
% \vspace{-1.5mm}
\noindent\textbf{Byzantine-robust Aggregation Algorithms:}
give provable convergence guarantees for each round of FL aggregation against poisoning attacks caused by Byzantine client failures~\cite{chen2017distributed, hashemi2021byzantine, guerraoui2018hidden}. Blanchard et al.~\cite{blanchard2017machine} proposed Krum and Multi-Krum (MKrum), which assume IID training data across clients and have knowledge of the number of malicious clients ($m$). Krum selects one update with the minimum Euclidean distance to its other $n-m-2$ closet updates as benign ones, while MKrum iteratively selects $n-2m-2$ updates with the minimum Euclidean distance from its closet update as benign. 
Yin et al.~\cite{yin2018byzantine} proposed Median, which aggregates all updates by computing the median of the updates in all the dimensions. 
% This method has been empirically proven more robust than Krum~\cite{2019Local}. Blanchard et al.
% However, these methods assume the clients' training data are IID~\cite{bagdasaryan2020backdoor}.

\vspace{0.5mm}
\noindent\textbf{Updates Anomaly Detection:} aims to detect and remove malicious updates caused by poisoning attacks in each round of FL~\cite{awan2021contra, shen2016auror, rieger2022deepsight}. Fung et al.~\cite{fung2020limitations} proposed FoolsGold, which assumes IID training data for malicious clients and non-IID data for benign clients. %FoolsGold selects updates that have smaller cosine similarities to other updates as benign ones. %munoz2019byzantine
Cao et al.\cite{cao2020fltrust} proposed FLTrust, which assumes an extra clean training dataset on the server and assigns lower trust scores to updates that have anomaly similarity with the clean updates trained by the server. Shejwalkar et al.~\cite{shejwalkar2021manipulating} proposed DnC, which assumes that the number of malicious clients is known, computes the anomaly of each update by the singular value decomposition, and then selects the top updates with the highest anomaly as malicious ones. %where $c\geq 1$ is the tolerance rate. 
Nguyen et al.~\cite{nguyen2022flame} proposed FLAME, which selects benign updates by calculating the cosine similarity and clustering their similarities, then aggregates the clipped updates to mitigate backdoors. Zhang et al.~\cite{zhang2022fldetector} and Cao et al.~\cite{cao2023fedrecover} proposed FLDetector and FedRecover, both of which assume that all clients participate in FL training in each round, and use the stored historical information of each client to estimate and recover their model update in each round. These methods assume that the data of benign clients is constant; otherwise, the updates of benign clients are also considered adversaries and are recovered to the updates before the data was changed.

% Previous detection methods (DnC~\cite{shejwalkar2021manipulating} and FLAME~\cite{nguyen2022flame}) and our work apply to the more general and practical scenarios where only a small fraction of clients participate in training each round. Due to the different settings, we do not consider FedRecover and FLDetector as comparison methods.
% improved their method to adapt it to our scenario and made a simple comparison in Sec~\ref{}. %and requires the server to predict model updates for all clients each round. based on their historical model updates. 
% Moreover, we conduct more experiments using a broader range of datasets, attack types, and comparison detection methods. 

\vspace{0.5mm}
\noindent\textbf{Model Sanitization:} is developed on centralized training to detect if models have backdoors and sanitize them by retraining~\cite{li2021neural,liu2019abs,wang2019neural,liu2018trojaning,shen2021backdoor,truong2020systematic, andreina2021baffle,zhao2020shielding}. Liu el at.~\cite{liu2018fine} proposed fine-pruning, which mitigates the backdoor by pruning neurons with low activation for clean samples and then fine-tunes the pruned model with clean samples to recover the model accuracy. Xu et al.~\cite{xu2021detecting} proposed MNTD, which trains a binary meta-classifier to predict whether a given model has backdoors. The meta-classifier is trained using a set of clean and backdoor shadow models trained on the same tasks as the given model. Another type of model sanitization is based on DP. Naseri et al.~\cite{naseri2020local} showed that the DP mechanism could mitigate the backdoors by bounding updates and adding perturbations. However, our study in Appendix~\ref{sec:DP_defense} shows that the adversary can successfully inject backdoors in a DP-based FL system by adjusting the attack strategy.

\section{Conclusion} \label{sec:conclusion}

% \vspace{-3mm}

%We first conduct a comprehensive study of prior FL attacks and detection methods and summarize two major security threats to FL. To address these issues, we develop a reliable and explainable detection framework, FLTracer, which detects attacks accurately and explicate the details of attack backtracks. Employing new features to analyze the crucial impacts of different attacks, FLTracer can detect various stealthy attacks accurately. It further leverages task and domain detection approaches to avoid the false positives caused by data heterogeneity. Our extensive evaluations show that FLTracer can achieve a high TPR at a low FPR for a variety of attacks and successfully prevents backdoor injections without sacrificing the accuracy of the global model in both IID and non-IID settings. 
In this paper, we comprehensively study the prior FL attacks and detection methods and then propose the first attack provenance framework, FLTracer. It can accurately detect various attacks and provide attack tracing by exploiting new features of various attacks. We develop task and Kalman filter-based domain detection to reduce false positives caused by data heterogeneity. Our extensive evaluations show that FLTracer has a high TPR and low FPR against various attacks and successfully addresses backdoors while only minimally affecting the global model in both IID and non-IID settings.

% Our extensive evaluations show that, in both IID and non-IID settings, FLTracer can achieve a high TPR at a low FPR for various attacks, and successfully prevents backdoor injections without sacrificing the accuracy of the global model, which significantly outperforms existing detection methods. 

% \input{acknowledgments}

\bibliographystyle{IEEEtran}
\bibliography{refer_simp}

% \section{Appendix}

% \section{Code Reference}
% % For the defense methods we compare with, we leverage the officially released code for them if available, including
% % For the attacks used for evaluation, we refer to several officially released code bases and the original papers for the implementation, including
% Blind backdoor~\cite{ebagdasa56:online}
% DBA~\cite{AIsecure57:online}
% MTB~\cite{vrt1shjw70:online}
% DnC~\cite{vrt1shjw70:online}

% \appendix

\newpage

\appendices

%-------------------------------------------------------------------------
\section{\texttt{CAM} Extraction Details} \label{sec:details_CAA}
\begin{figure}[!h]
\vspace{-2mm}
  \centering
  \includegraphics[width=0.8\linewidth]{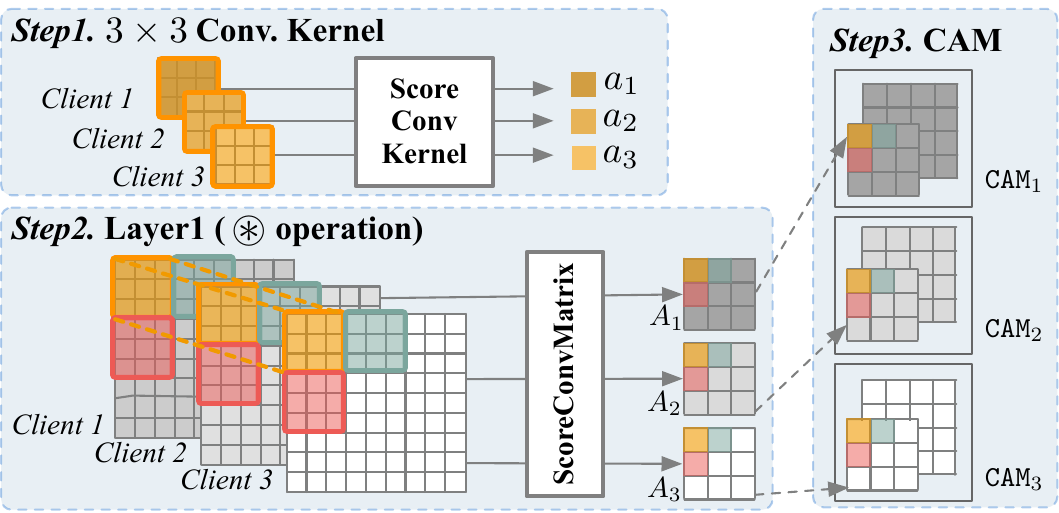}
  \vspace{-2mm}
  \caption{The \texttt{CAM} extraction for each client in a round.} %, including the convolution kernels, layers, and anomaly matrices in order
  \label{fig:pca_cc}
 \vspace{-3mm}
\end{figure}

Fig.~\ref{fig:pca_cc} depicts the process of \texttt{CAM} extraction. Here three clients are present, each having a model consisting of two convolutional layers. The first and second layers convolutional layers have sizes of $3\!\times\! 3 \!\times\! 3 \!\times\! 3$ and $4 \!\times\! 4\!\times\! 3\!\times\! 3$, respectively. The convolution kernel's size is $L_1\!\times\! L_2=3\!\times\! 3$. 

% \vspace{-1mm}
\noindent \textbf{Step1.} 
We collect the first convolution kernel of each client and compute the anomaly score (i.e., $a_1$, $a_2$, and $a_3$) using the function \textsc{ScoreConvKernel} of Alg.~\ref{alg:cam}.

% \vspace{-1mm}
\noindent \textbf{Step2.} ($\circledast$ operation)
We compute the anomaly scores for the convolution kernels of the same channels in layer one using the function \textsc{ScoreConvKernel}. After obtaining all anomaly scores for client 1, we combine them into an anomaly vector $A_1$. The anomaly vectors $A_2$ and $A_3$ for client 2 and client 3 can be obtained in a similar method.

% \vspace{-1mm}
\noindent \textbf{Step3.}
After computing the anomaly matrices for all layers of client 1, we merge them into a convolution matrix $\texttt{CAM}_1$ using the function \textsc{ScoreConvMatrix} of Alg.~\ref{alg:cam}. Similarly, we can derive $\texttt{CAM}_2$ and $\texttt{CAM}_3$ for client~2 and client~3.
\vspace{-2mm}

% \textbf{Update decoupling for backdoor attacks: Convolution Anomaly Analysis (CAA) to construct \texttt{CAM}.} 
Alg.~\ref{alg:cam} described functions used in \texttt{CAM} extraction. The function \textsc{ScoreConvKernel} computes anomaly scores for a set of convolution kernels ($w_{\{i\in [n]\}}$) in the same channel, where $u_i$ is a convolution kernel for client $i$. PCA and Mahalanobis Distance\cite{de2000mahalanobis} are employed to obtain the outliers. 
% The normalized result is the anomaly score $a_i$. 
The \textsc{Main} function computes \texttt{CAM} for $n$ clients. The loop in lines 7-9 computes the anomaly scores for $n$ clients in layer $\kappa$. The loop in lines 10-11 generates $A_i^\kappa$ for each client. %i.e., the operation $\circledast$.
%which can be expressed as:
% \begin{equation}
% \small
%     CAM_i=\{A^1_i, A^2_i, \cdots, A^k_i\} %CAA(\theta_i)=  , A_i^1=\{a_i^\iota| \iota \mathrm{\ from\ all\ channels}\}
%     \label{eq:cam}
% \vspace{-3mm}
% \end{equation}

\vspace{-2mm}
\begin{algorithm}[htb]  
\small
  \caption{\textsc{ScoreConvMatrix} Algorithm}  
  \begin{algorithmic}[1]
    \Require
    $\theta_{\{i\in [n]\}}$: updates of $n$ clients; $k$: total number of conv. layers
    \Ensure 
    $\texttt{CAM}_{\{i\in [n]\}}$: convolution anomaly matrices of $n$ clients
    \Function{ScoreConvKernel}{$w_{\{i\in [n]\}}$}{:}
        % \State $u \gets \mathrm{Standardize}(u)$
        \State $w_{pca}\gets \textsc{PCA}(u, \mathrm{componenets}=2)$
        \State $w_{md}\gets \mathrm{MahalanobisDistance} (u_{pca})$
        % \State $U_{a}\gets \textsc{MadScore}(U_{md})$
        \State $a_{\{i\in [n]\}} \gets \mathrm{Normalize}(u_{md})$ \quad \Return $a_{\{i\in [n]\}}$  %\Comment{{\small Anomaly scores}}
    \EndFunction
    \Function{Main}{}{:} %\Comment{{\small Main function}} %$\theta_{\{i\in [n]\}}, k$
    \For{conv. layer $\kappa \in [k]$} \Comment{the operation $\circledast$}
        \For{each channel $\iota$}
            \State $w^\iota_{\{i\in[n]\}} \gets$ Conv. kernel set of channel $\iota$ for $n$ clients
            \State $a^\iota_{\{i\in[n]\}} \gets \textsc{ScoreConvKernel}(w^\iota_{\{i\in[n]\}})$
        \EndFor
        \For{$i\in [n]$}
        \State $A^\kappa_i \gets$ Combine all $a^\iota_i$ of client $i$ in layer $\kappa$%
        \EndFor
    \EndFor
    \State $\texttt{CAM}_i\gets$ Generate using $A^\kappa_i$ and Eq.(\ref{eq:cam}), $i\in[n]$
    \State \Return $\texttt{CAM}_{\{i\in [n]\}}$
    \EndFunction
  \end{algorithmic}  
  \label{alg:cam}
\end{algorithm}
\vspace{-1mm}

%---------------------------------------------------------------------------------------

\vspace{-2mm}
\section{Experimental Setup Details} \label{sec:expsetup}

\noindent\textbf{Details on datasets and model architectures:}
MNIST~\cite{lecun1998gradient} is a 10-class class-balanced digit image classification dataset. 
EMNIST~\cite{cohen2017emnist} is a 47-class class-imbalanced digit image classification dataset. 
CIFAR10~\cite{krizhevsky2009learning} is a 10-class class-balanced color image classification dataset. 
German Traffic Sign Recognition Benchmark (GTSRB)~\cite{Houben-IJCNN-2013} is a 43-class class-imbalanced traffic sign dataset with varying light conditions and rich backgrounds.
For more details on datasets, please refer to Table~\ref{tab:dataset1} (columns 3-5).
% number of training inputs, categories, and input sizes

Table~\ref{tab:dataset1} (column 7) lists the model architectures utilized for each dataset. SimpleNet comprises two convolution layers and two fully-connected layers. AlexNet~\cite{krizhevsky2012imagenet}, ResNet18~\cite{he2016deep}, VGG16~\cite{simonyan2014very}, and ResNet34~\cite{he2016deep} are different architectures of convolution networks. 
DNN contains two fully connected layers. Vision Transformer (ViT)~\cite{dosovitskiyimage} is a Transformer based model for image recognition.

\vspace{0.5mm}
\noindent\textbf{Details on learning parameters:}
Following the standard setup in~\cite{mcmahan2017learning, bagdasaryan2020backdoor}, we train local models with local learning rate $lr$, local epochs 2, and batch size 64. At each round, the $n$ selected clients train a local model to aggregate.
For untargeted attacks, we start with a sustained attack from scratch of the training. We train MNIST and EMNIST with SimpleNet using $lr=0.01$ for rounds 0-100. We train CIFAR10 with AlexNet and ResNet18 using $lr=0.1$ for rounds 0-200. 
For backdoor attacks, we begin the attack when the global model is convergent, which is round 200 for CIFAR10, 100 for GTSRB (IID), and 200 for GTSRB (non-IID). We train CIFAR10 with ResNet18 and VGG16 using $lr=0.1$ and $lr=0.05$ for rounds 0-400. We train GTSRB with ResNet34 using $lr=0.1$ for rounds 0-200 in the IID settings and $lr=0.1$ for rounds 0-300 in the non-IID settings. 
% We use the AdamW optimizer for the ViT model on the BDD100K dataset.
% In our experimental experience, to facilitate backdoor injection, the $lr$ of malicious clients can be reduced to half of the benign learning rate. The epoch can be increased to twice the benign epoch.

% We report mainly results on CIFAR10 using ResNet18 in non-IID settings due to space constraints, and we postpone the results of other datasets and models to Appendix~\ref{}.

\begin{table*}[]
\centering
\scriptsize
\caption{Datasets, model structures, and parameters}
\vspace{-2mm}
\begin{threeparttable}
\begin{tabular}{|c|cccccc||cccccc|}
\hline
 & Dataset & \begin{tabular}[c]{@{}c@{}}Training\\      input\end{tabular} & Class & \begin{tabular}[c]{@{}c@{}}Input \\      size\end{tabular} & \begin{tabular}[c]{@{}c@{}}Data \\      distribution\end{tabular} & \begin{tabular}[c]{@{}c@{}}Model \\      structure\end{tabular} & $N$ & $n^*$ & $P_m^*$ & $P_p^*$ & \begin{tabular}[c]{@{}c@{}}Benign\\      $lr$/epochs\end{tabular} & \begin{tabular}[c]{@{}c@{}}Malicious\\      $lr$/epochs\end{tabular} \\
 \hline
\multirow{4}{*}{\begin{tabular}[c]{@{}c@{}}Un\\      targeted\end{tabular}} & MNIST & 60000 & 10 & $28\times28$ & IID & SimpleNet & 100 & 10 & 20\% & 0\% & 0.01/2 & 0.01/2 \\
 & EMNIST & 73168 & 47 & $28\times28$ & non-IID & SimpleNet & 100 & 10 & 20\% & 0\% & 0.01/2 & 0.01/2 \\
 & CIFAR10 & 50000 & 10 & $32\times32$ & IID\& non-IID & AlexNet  & 100 & 10 & 20 & 0\% & 0.1/2 & 0.1/2 \\
 & CIFAR10 & 50000 & 10 & $32\times32$ & IID\& non-IID & ResNet18  & 100 & 10 & 10\%-40\% & 0\% & 0.1/2 & 0.1/2 \\
 \hline
\multirow{5}{*}{Targeted} & CIFAR10 & 50000 & 10 & $32\times32$ & IID\& non-IID & ResNet18  & 100 & 10 & 10\%-40\% & 3\% & 0.01/2 & 0.005/4 \\
 & CIFAR10 & 50000 & 10 & $32\times32$ & IID \& non-IID & VGG100 & 100 & 10 & 10\% & 3\% & 0.01/2 & 0.005/4 \\
 & GTSRB & 50000 & 43 & $32\times32$ & IID \& non-IID & ResNet34 & 50 & 10 & 10\% & 3\% & 0.01/2 & 0.005/4 \\
 & HAR & 7353 & 6 & $1\times561$ & non-IID & DNN & 21 & 10 & 10\% & 3\% & 0.01/2 & 0.005/4 \\
 & BDD100K & 474706 & 10 & $64\times64$ & non-IID & ViT (Transformer) & 100 & 10 & 10\% & 3\% & 0.0001/4 & 0.00005/6 \\
 \hline
\end{tabular}
\begin{tablenotes}
%     %\item [1] Attack rounds of backdoor attacks in GTSRB in IID and non-IID setting is 101-200 and 201-300, respectively. 
\item{$^* n=20$ for MB and Fang attacks, $P_m=20\%$ for DBA, $P_p=5\%$ for dirty label attack.} 
\item{In MRA, the adversary attacks for one round and scales the updates by 10$\times$. }%performs in the first round of attack, with one malicious client.
\end{tablenotes}
\end{threeparttable}
\vspace{-4mm}
\label{tab:dataset1}
\end{table*}

\vspace{0.5mm}
\noindent\textbf{Details on attacks:} 
For Add noise attack, we use Gaussian noise $\delta\sim \CN(\mu,\sigma^2)$, where $\mu\!=\!0$ and $\sigma\!=\!0.3$. For Dirty label attacks, we increase the number of sources and target labels to enhance the attack's impact. Based on the different strategies applied to ``source-target label'' pairs, we evaluate four types of dirty label attacks, including the Fix-Fix attack, the Fix-Rnd attack, the Rnd-Fix attack, and the Rnd-Rnd attack. We set the proportion of poisoned samples for each malicious client to 50\%. 
In the Fix-Fix attack, we mislabel the samples in categories 1 to 5 as category 3. In the Fix-Rnd attack, we mislabel the samples in categories 1 to 5 as random categories. In the Rnd-Fix attack, we mislabel half of the samples to category 3. In the Rnd-Rnd attack, we mislabel half of the samples into random categories. 
For MB and Fang attacks, we set $n\!=\!20$ because these attacks require more than four malicious clients per round to generate malicious updates. 
For the backdoor attacks, we set $P_m\!=\!10\%$ for most backdoor attacks and increase the $P_m\!=\!20\%$ for DBA because of the distributed trigger and injection strategy. 
Following the setup in~\cite{bagdasaryan2020backdoor, xie2019dba}, we attack at the beginning of the training for MRA and at round 200 for other backdoor attacks. 
In all attacks, adversaries modify their local $lr/$epochs to attack efficiently. 
% Moreover, we launch attacks at round 100 for GTSRB because there are only 50 clients under this dataset. Each client has more training data under GTSRB, so the global model converges faster. 
% We launch an attack at round 200 of the training because the attacks are more effective after the model convergences~\cite{bagdasaryan2020backdoor}. 
% For multiple-attack backdoor attack (i.e., A-M and A-BM), the adversary attack 200 rounds and scales the updates by a factor of 1. 
% For a single-attack backdoor attack (i.e., A-S), the adversary attack only one round and scales the updates by a factor of 10. 
% Specifically, for A-S attacks combined with DB trigger, we set the global learning rate as $\eta=1/n$. 

\vspace{-2mm}
\section{Assessment of Previous Attacks and Detection} \label{sec:assess_appendix}

\begin{figure}[]
  \centering
    \vspace{-2mm}
  % \subfigcapskip=-1cm
  \subfigure{
  \includegraphics[width=0.98\linewidth]{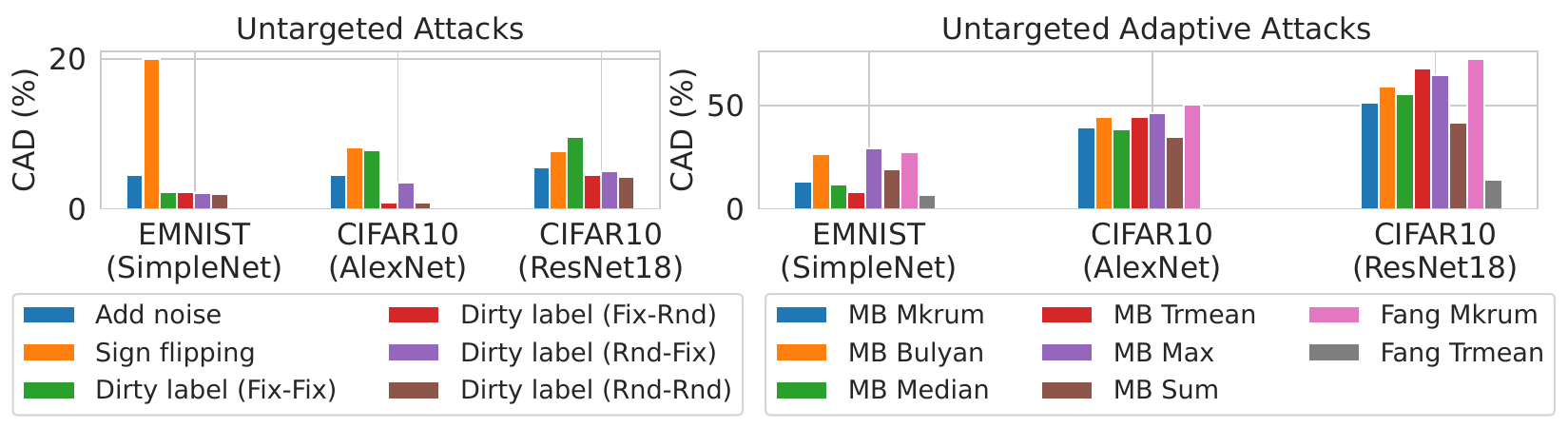}}
  \vspace{-5mm}
  \caption{Clean accuracy drop (CAD) of untargeted attacks in non-IID settings. The CAD of the Sign-flipping attack on EMNIST is $81.28\%$.}
  \vspace{-2mm}
  \label{fig:CAD_un}
\end{figure}

\begin{figure}[]
  \centering
    \vspace{-2mm}
  \subfigcapskip=-5pt
  \subfigure{
  \includegraphics[width=0.98\linewidth]{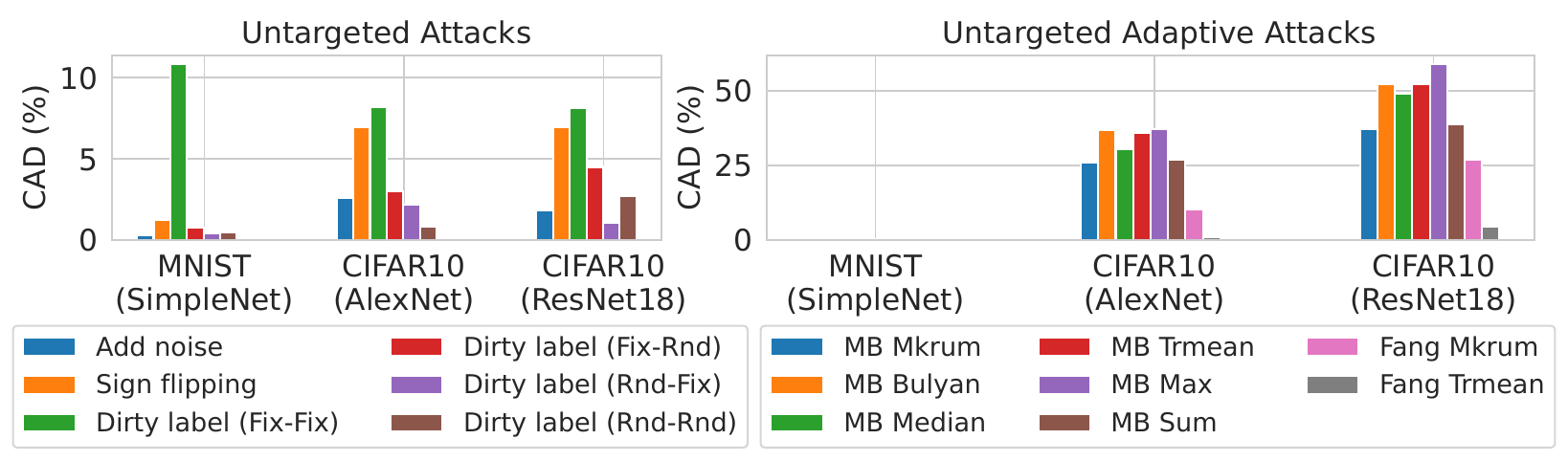}}
  \vskip -3mm
  \caption{Clean accuracy drop (CAD) of untargeted attacks in IID settings. The CAD of untargeted adaptive attacks on MNIST is less than $0.2\%$.}
  \vspace{-5mm}
  \label{fig:CAD_un_iid}
\end{figure}

\noindent\textbf{Untargeted attacks:} Fig.~\ref{fig:CAD_un} shows the clean accuracy drop of untargeted and adaptive attacks in non-IID settings. Most untargeted attacks achieve a high clean accuracy drop, except for three dirty label attacks (Fix-Rnd to Rnd-Rnd). Specifically, the clean accuracy drop of MB Mkrum and Fix-Rnd on CIFAR10 (AlexNet) is $39.29\%$ and $0.85\%$. 
% On the other hand, most untargeted adaptive attacks on MNIST in IID settings can only slow down the training process but not reduce the final model accuracy. 
We also observe that the performance of untargeted adaptive attacks is generally stronger than that of untargeted attacks due to the ability of adaptive attacks to generate optimized perturbations. The average clean accuracy drop for adaptive and untargeted attacks is $25.8\%$ and $3.77\%$, respectively. (see Fig.\ref{fig:CAD_un_iid} for IID).

\begin{figure}[]
  \centering
  % \subfigcapskip=-1cm
  \subfigure{
  \includegraphics[width=0.85\linewidth]{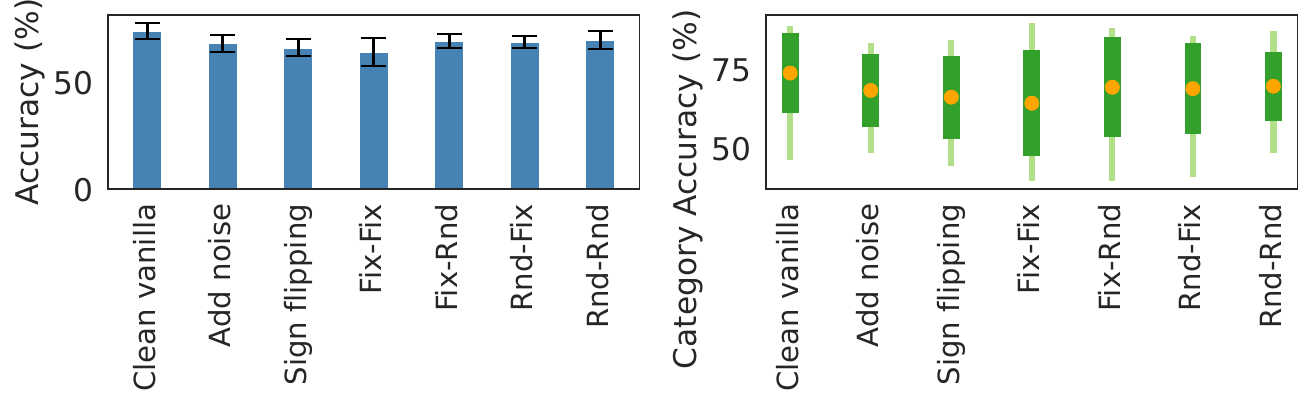}}
  \vskip -2mm
  \subfigure{
  \includegraphics[width=0.85\linewidth]{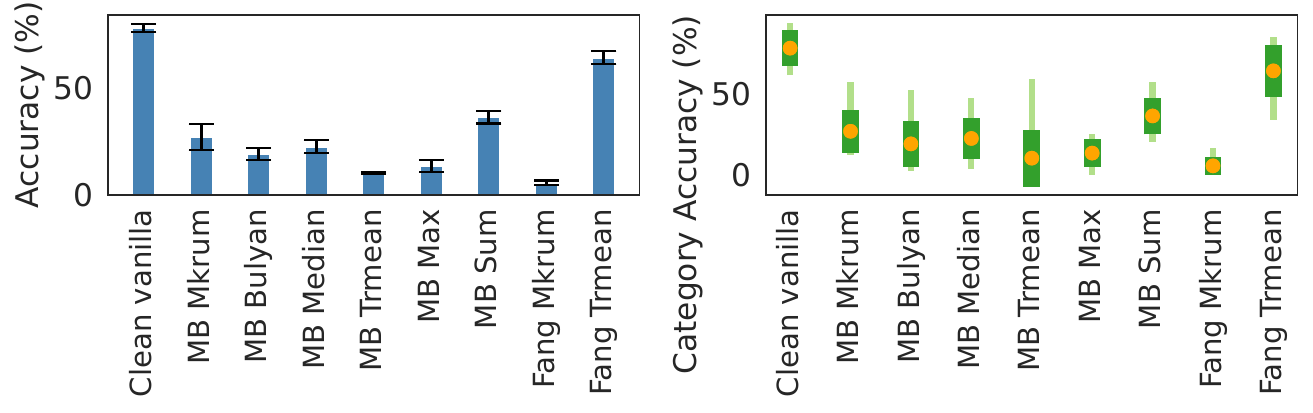}}
  \vspace{-3mm}
  \caption{Accuracy and category accuracy of untargeted attacks compared to the clean vanilla on CIFAR10 (ResNet18). Top-untargeted attacks mainly affect stability. Bottom-untargeted adaptive attacks mainly affect accuracy.}
  \vspace{-6mm}
  \label{fig:robustness_un}
\end{figure}

\begin{figure}[!ht]
  \centering
  \vspace{-2mm}
  \includegraphics[width=0.8\linewidth]{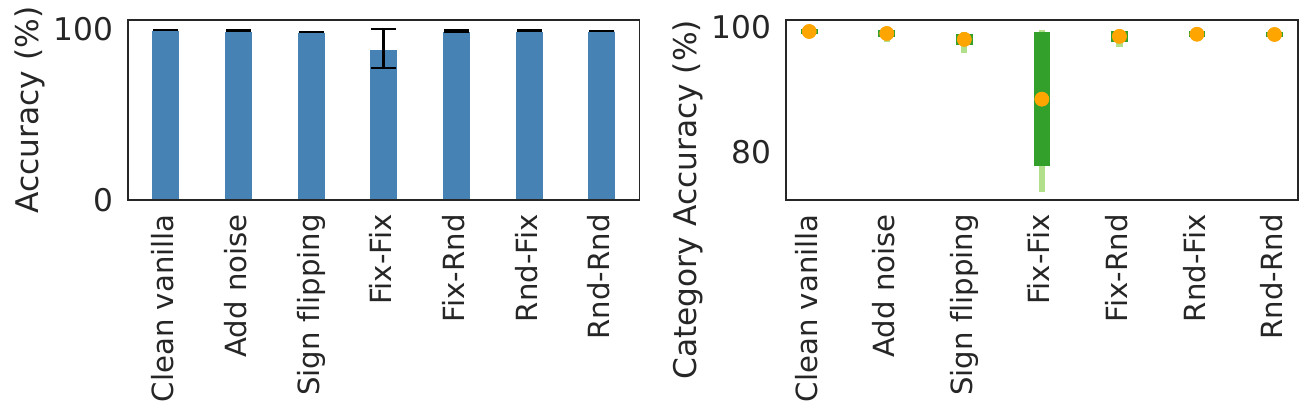}
  \vspace{-2mm}
  \subfigure[MNIST (SimpleNet)]{
  \includegraphics[width=0.8\linewidth]{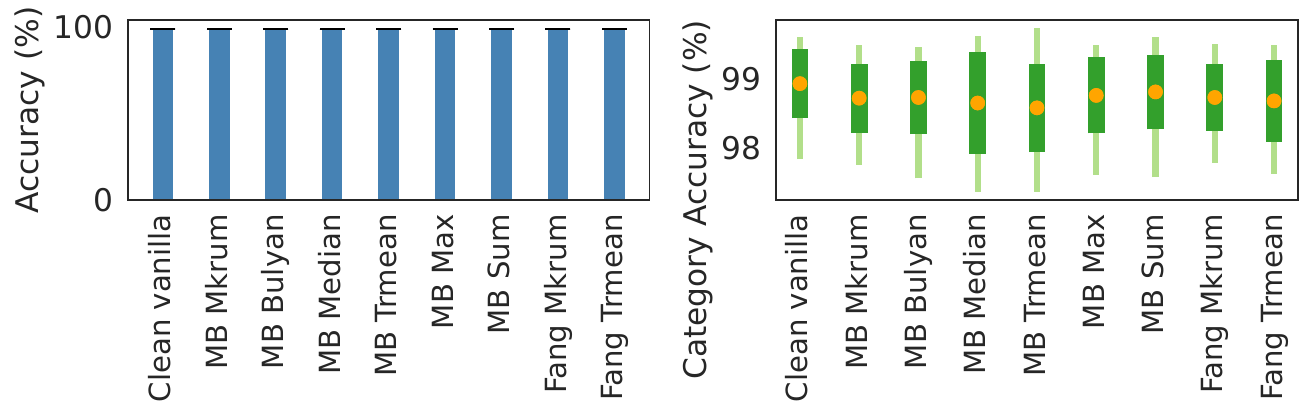}
  }
  \includegraphics[width=0.8\linewidth]{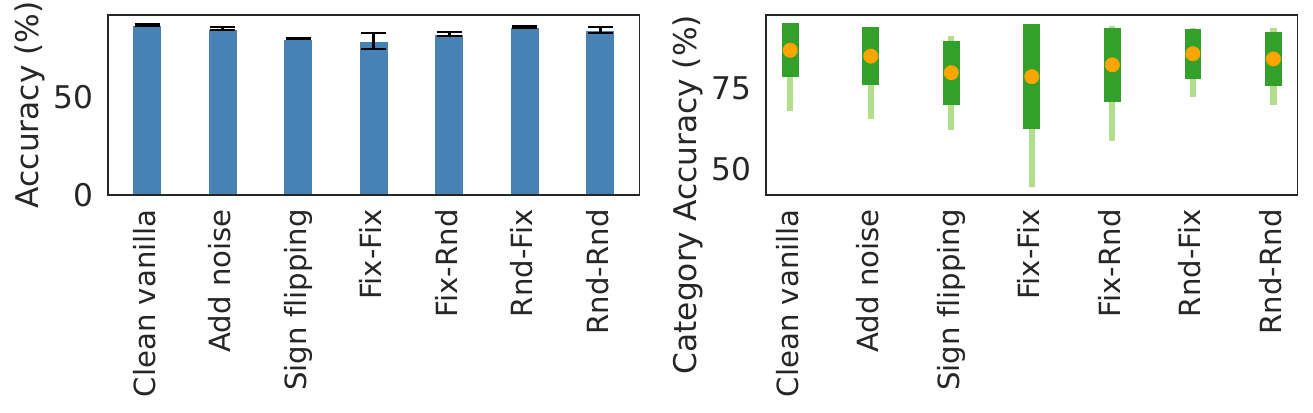}
  \vspace{-2mm}
  \subfigure[CIFAR10 (AlexNet))]{
  \includegraphics[width=0.8\linewidth]{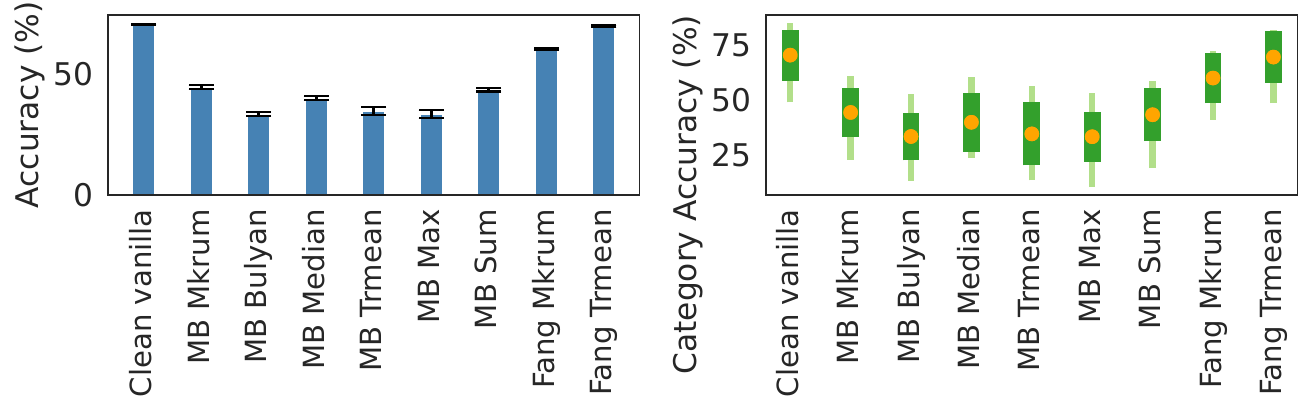}
  }
  \includegraphics[width=0.8\linewidth]{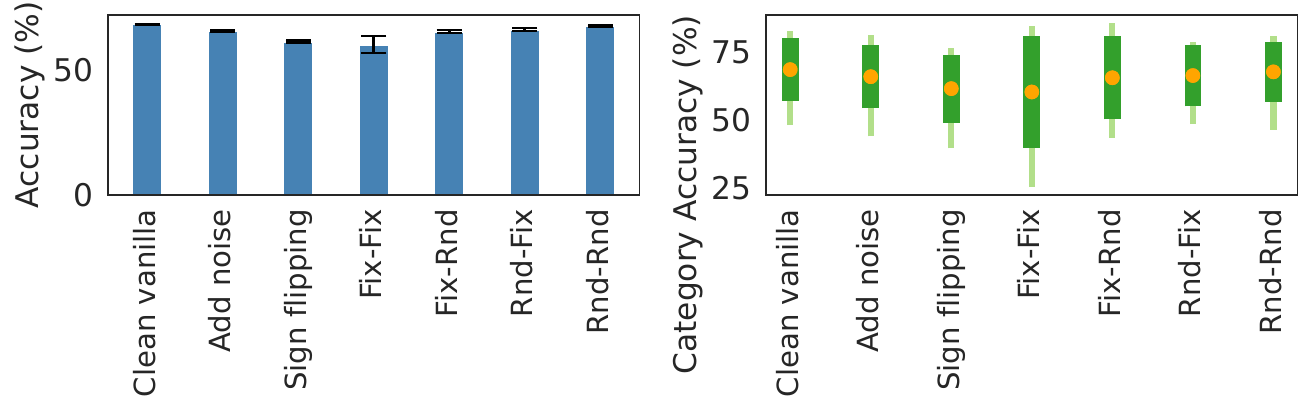}
  \vspace{-2mm}
  \subfigure[CIFAR10 (ResNet18)]{
  \includegraphics[width=0.8\linewidth]{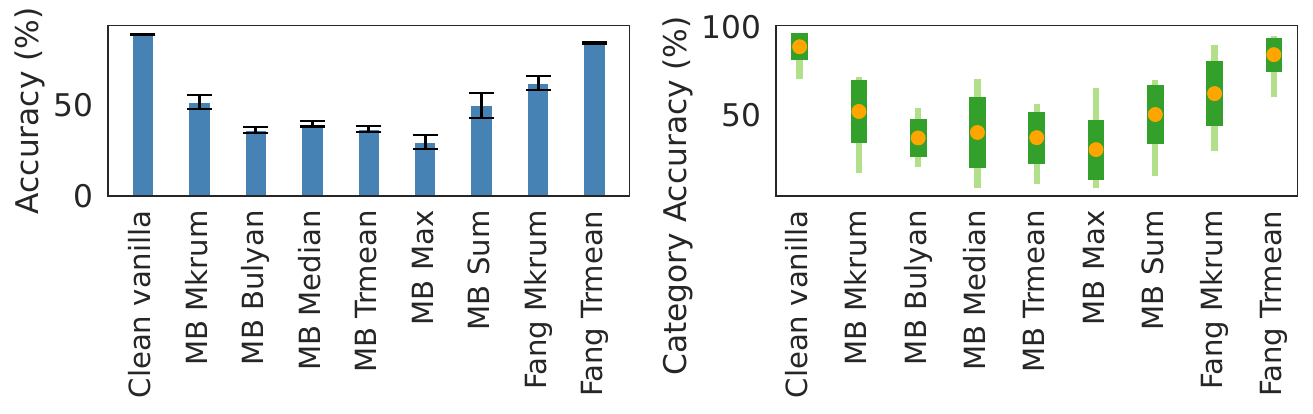}
  }
  \vspace{-2mm}
  \caption{The accuracy and category accuracy of untargeted attacks compared with clean vanilla in IID settings.} %under different datasets and models
  \label{fig:robustness_un_iid}
\end{figure}

\begin{figure}[!ht]
  \centering
  \vspace{-2mm}
  \includegraphics[width=0.8\linewidth]{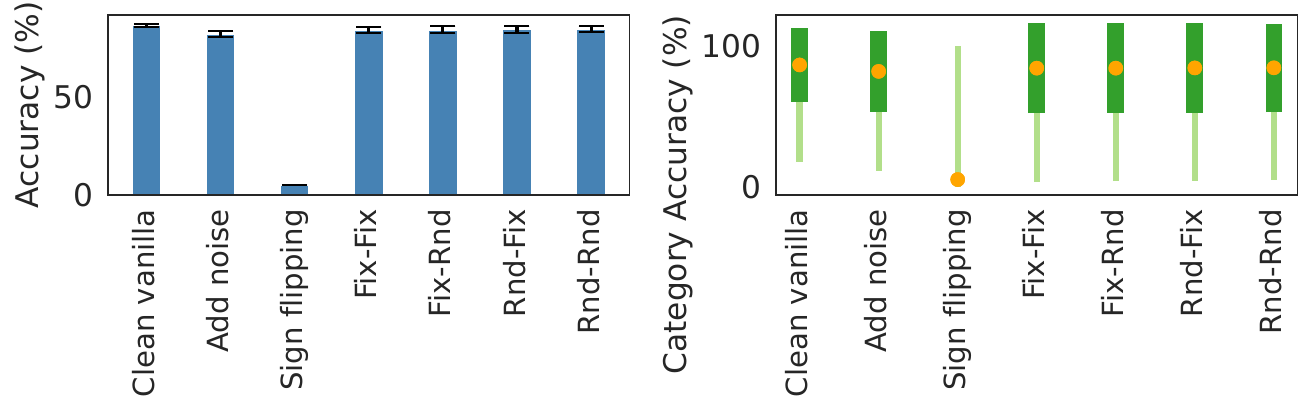}
  \vspace{-2mm}
  \subfigure[EMNIST (SimpleNet)]{
  \includegraphics[width=0.8\linewidth]{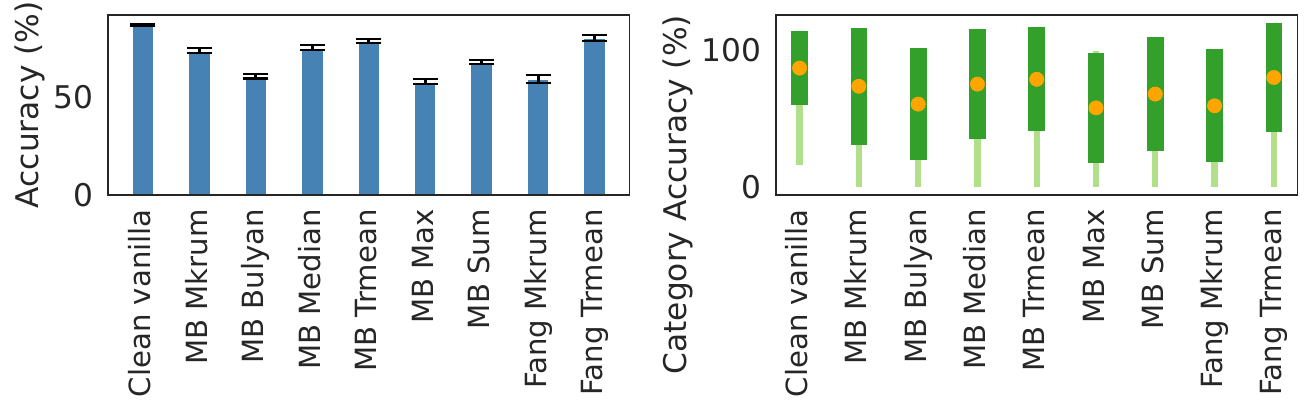}
  }
  \includegraphics[width=0.8\linewidth]{./figures/untarget_feasi_cl_noniid_robust_resnet.pdf}
  \vspace{-2mm}
  \subfigure[CIFAR10 (AlexNet)]{
  \includegraphics[width=0.8\linewidth]{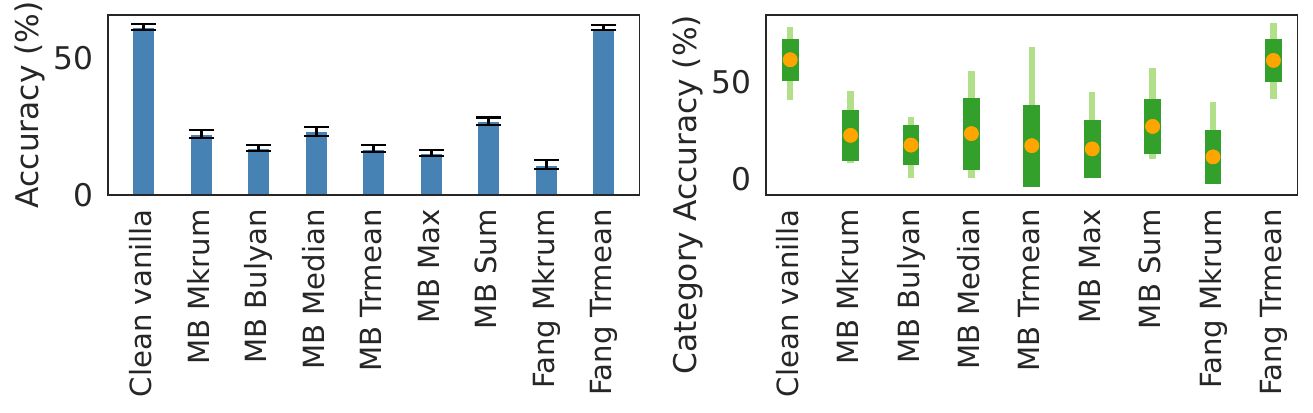}
  }
  \includegraphics[width=0.8\linewidth]{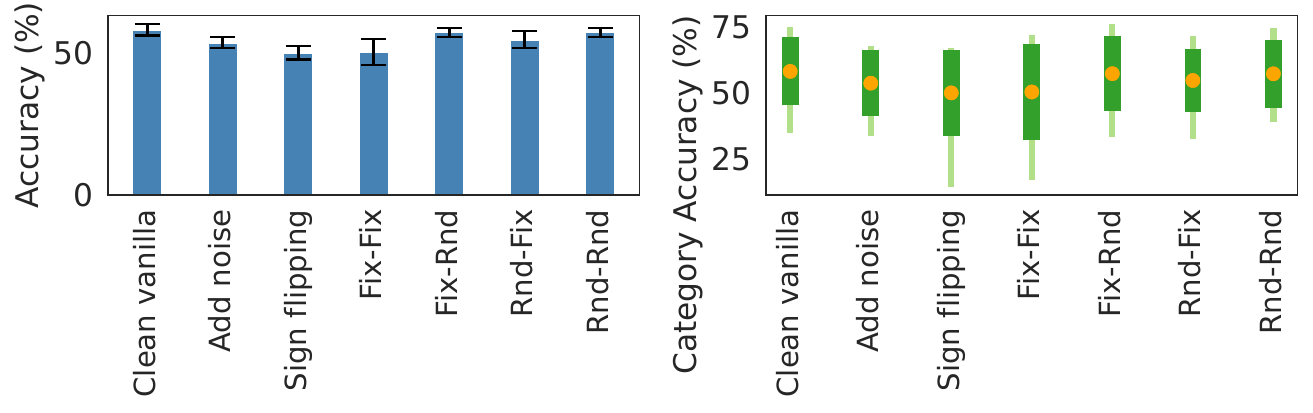}
  \vspace{-2mm}
  \subfigure[CIFAR10 (ResNet18)]{
  \includegraphics[width=0.8\linewidth]{./figures/untarget_feasi_ea_noniid_robust_resnet.pdf}
  }
  \vspace{-2mm}
  \caption{The accuracy and category accuracy of untargeted attacks compared with clean vanilla in non-IID settings.}
  \label{fig:robustness_un_noniid}
\end{figure}

Fig.~\ref{fig:robustness_un} shows the accuracy, model stability, worst/best category accuracy, and category accuracy stability of untargeted and adaptive attacks. We observe that untargeted attacks mainly affect the stability of the model. Specifically, the model stability for the Fix-Fix attack is $4.15$, much higher than the clean vanilla model ($0.7$). In addition, untargeted adaptive attacks significantly affect the accuracy and category accuracy. The accuracy and category accuracy of the MB attack and the Fang attack are smaller than the baseline by $41.82\%$ and $41.15\%$, respectively. (see Fig.\ref{fig:robustness_un_iid} and \ref{fig:robustness_un_noniid} for other datasets). % refer to \href{https://anonymous.4open.science/r/FLTracer-F7BB/FLTracer_Additional_Experimental_Results.pdf}{GitHub} % see Fig.\ref{fig:robustness_un_iid} and \ref{fig:robustness_un_noniid}
% the results on ResNet18 trained on CIFAR10 in non-IID settings

% \vspace{-2mm}
\begin{center}
\fcolorbox{darkblue}{gray!10}{\parbox{.96\linewidth}{\textbf{\textcolor{darkblue}{Remark 1}}. The 14 untargeted attacks (except for Fix-Rnd to Rnd-Rnd) lead to a significant decrease in the accuracy or stability of the global model.}}
\end{center}
% \vspace{-2mm}

\begin{figure}[!ht]
\vspace{-0.5cm}
  \centering
  % \subfigcapskip=-5pt
  \subfigure{
  \includegraphics[width=0.95\linewidth]{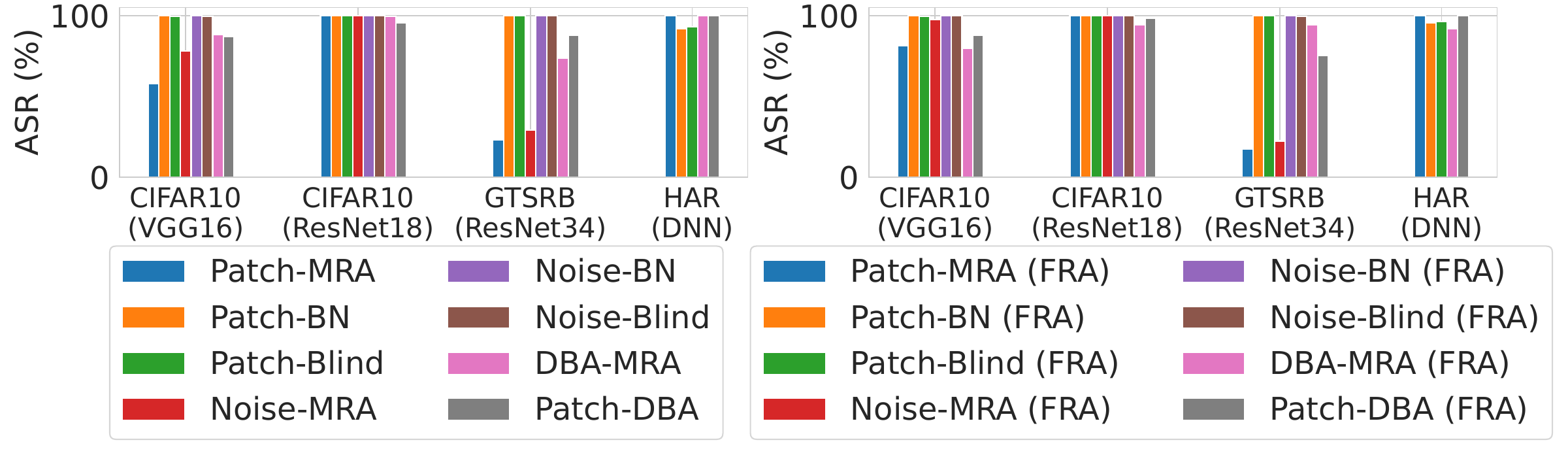}}
  \vspace{-4mm}
  \caption{Attack success rate (ASR) of backdoor attacks in non-IID settings. }
  \vspace{-3mm}
  \label{fig:ASR_tar}
\end{figure}

\begin{figure}[!ht]
\vspace{-0.4cm}
  \centering
  % \subfigcapskip=-1cm
  \subfigure{
  \includegraphics[width=0.95\linewidth]{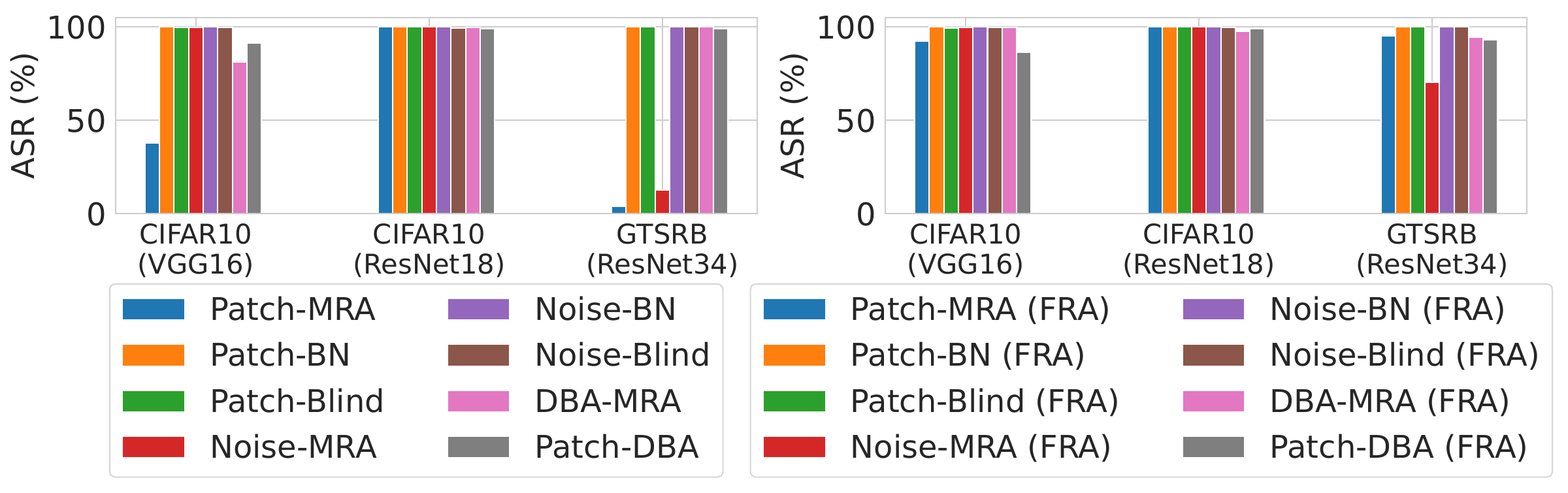}}
 \vspace{-4mm}
  \caption{Attack success rate (ASR) of backdoor attacks in IID settings. }
 \vspace{-3mm}
  \label{fig:ASR_tar_iid}
\end{figure}

\begin{figure}[!ht]
  \centering
\vspace{-0.5cm}
  \includegraphics[width=0.95\linewidth]{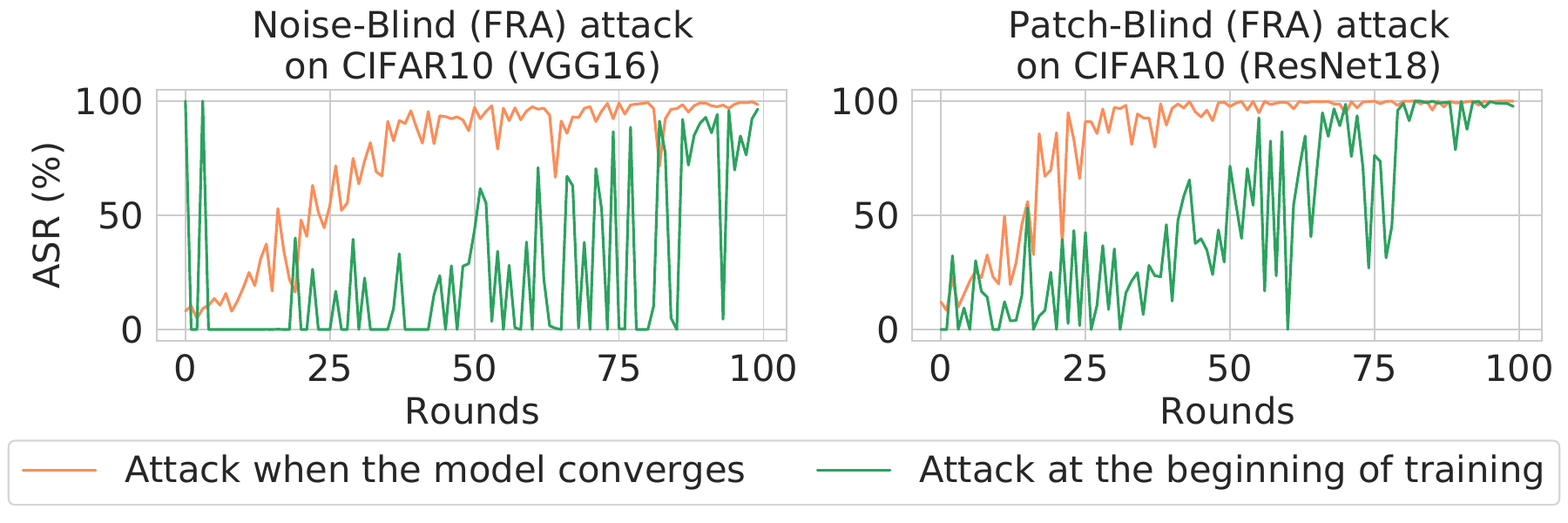}
  \vspace{-2mm}
  \caption{Comparison of attack success rate between the attack when the global model converges and the attack at the beginning of training.}
  \label{fig:asr_0_200}
  \vspace{-3mm}
\end{figure}

\vspace{0.5mm}
\noindent\textbf{Targeted attacks:} 
Fig.~\ref{fig:ASR_tar} shows the attack success rate of 16 backdoor attacks in non-IID settings. Most attacks can inject backdoors with almost $100\%$ attack success rate, except for Patch-MRA and Noise-MRA on GTSRB. For different attack strategies, MRA is more difficult to inject backdoors because MRA only injects a single round. For different datasets, those with fewer categories and more training inputs are easier to inject backdoors. Specifically, the average attack success rate of $96.21\%$ for CIFAR10 is higher than that of $76.11\%$ for GTSRB. 
In addition, the effect of our new adaptive strategy, FRA, achieves similar results as without it. Specifically, the average attack success rate w/ and w/o FRA is $88.3\%$ and $89.51\%$, respectively. We also observe that the backdoor injection is unstable and slow at the beginning of the training, as shown in Fig.~\ref{fig:asr_0_200}. Due to the low accuracy of the global model, it is difficult to converge the global model and inject backdoors into it simultaneously. (see Fig.~\ref{fig:ASR_tar_iid} for IID).
%, backdoors can be injected more quickly and successfully when the model is converged. 
% 
% A general strategy for backdoor attacks is to train as a benign client at the beginning of the training and launch the attack after model convergence.

% \vspace{-2mm}
\begin{center}
\fcolorbox{darkblue}{gray!10}{\parbox{.96\linewidth}{\textbf{\textcolor{darkblue}{Remark 2}}. When the global model converges, backdoors can be injected faster and more successfully, similar to the conclusion in \cite{bagdasaryan2020backdoor}.}} %, and the backdoor injection rounds are proportional to the backdoor removal rounds.
\end{center}
% \vspace{-2mm}

\begin{figure}[!ht]
  \centering
\vspace{-0.5cm}
  \includegraphics[width=0.95\linewidth]{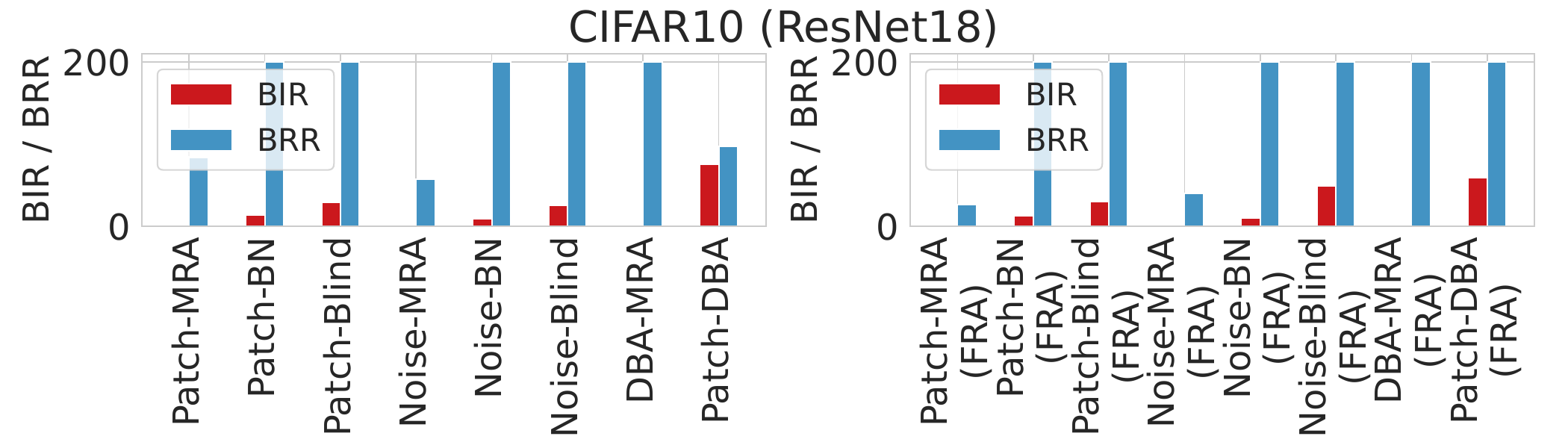}
  \vspace{-2mm}
  \caption{Backdoor injection rounds (BIR) and backdoor removal rounds (BRR)  of backdoor attack in non-IID settings. A BIR of 200 indicates that more than 200 training rounds are required to remove the backdoor. } %BIR and BRR assess the efficiency and robustness of backdoor attacks.
  \label{fig:Robustness_tar}
  \vspace{-2mm}
\end{figure}

\begin{figure}[!ht]
  \centering
  \includegraphics[width=0.85\linewidth]{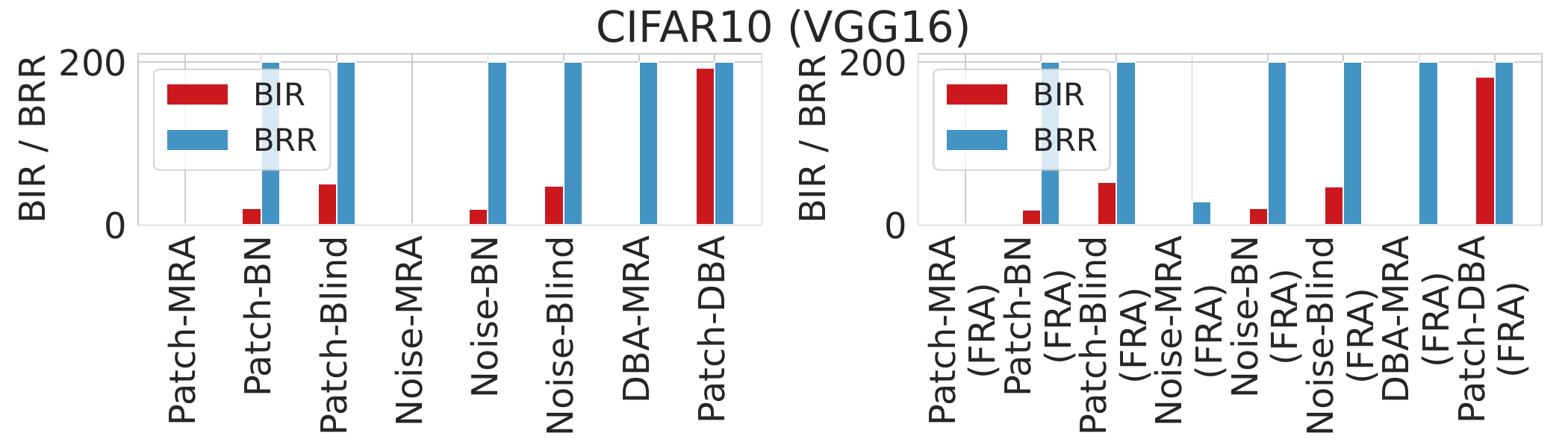}
  \includegraphics[width=0.85\linewidth]{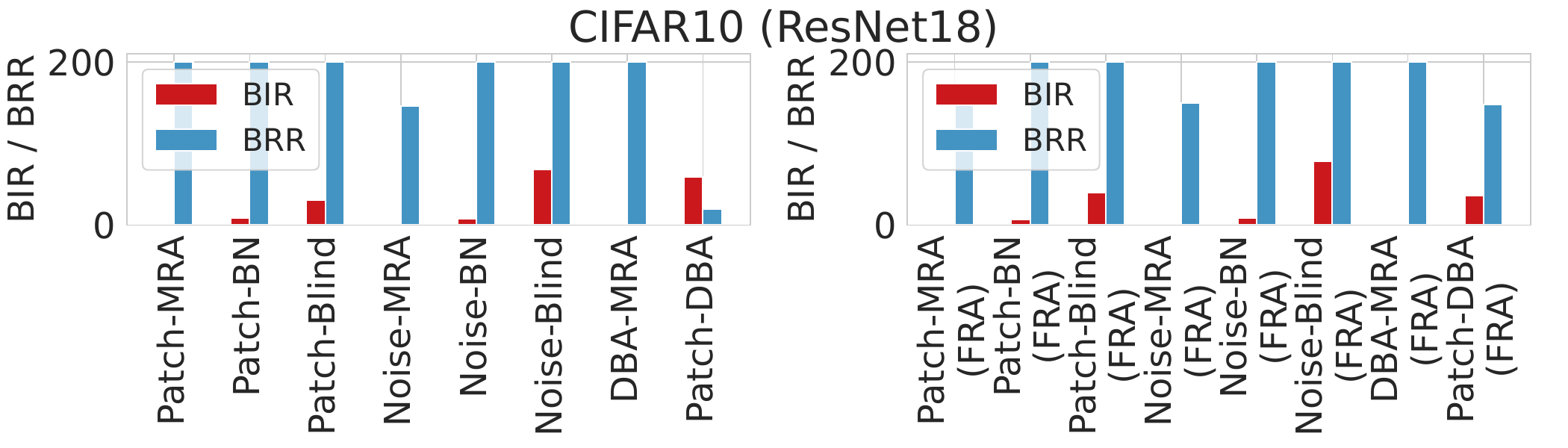}
  \includegraphics[width=0.85\linewidth]{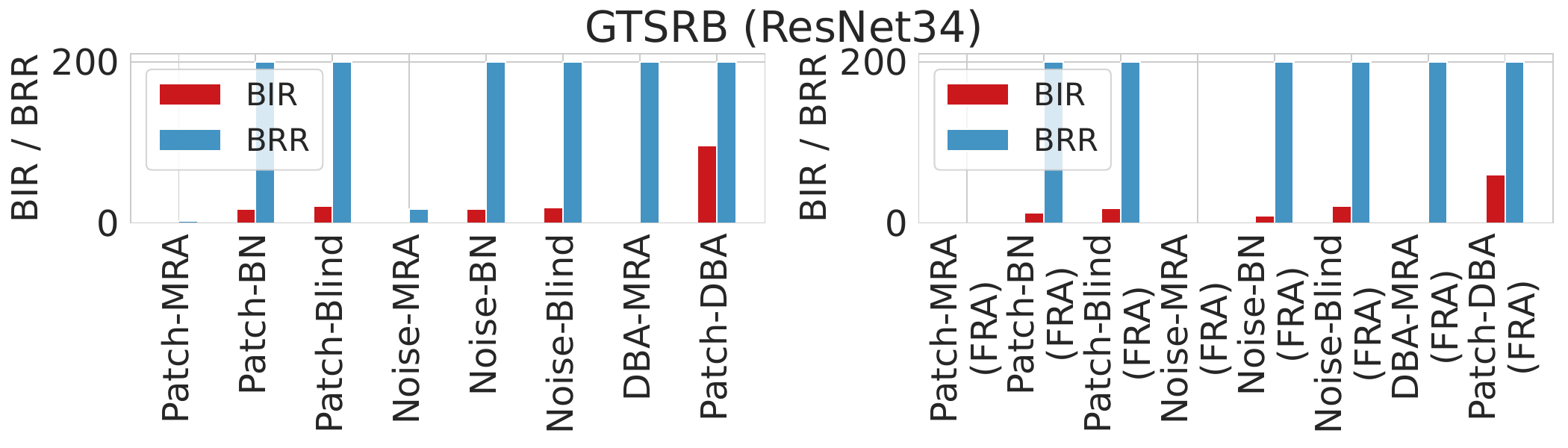}
%   
  % \subfigure[Targeted adaptive attacks]{
  % \includegraphics[width=0.32\linewidth]{./figures/target_feasi_robust_vgg_noniid.pdf}
  % \includegraphics[width=0.32\linewidth]{./figures/target_feasi_robust_vgg_iid.pdf}
  % \includegraphics[width=0.32\linewidth]{./figures/target_feasi_robust_vgg_iid.pdf}
  % }
    \vspace{-2mm}
    \caption{The backdoor injection rounds (BIR) and backdoor removal rounds (BRR) of backdoor attacks in IID settings.} % under different datasets and model structures
  \label{fig:Robustness_tar_iid}
\end{figure}

\begin{figure}[!ht]
  \centering
% \vspace{-0.5cm}
  \includegraphics[width=0.85\linewidth]{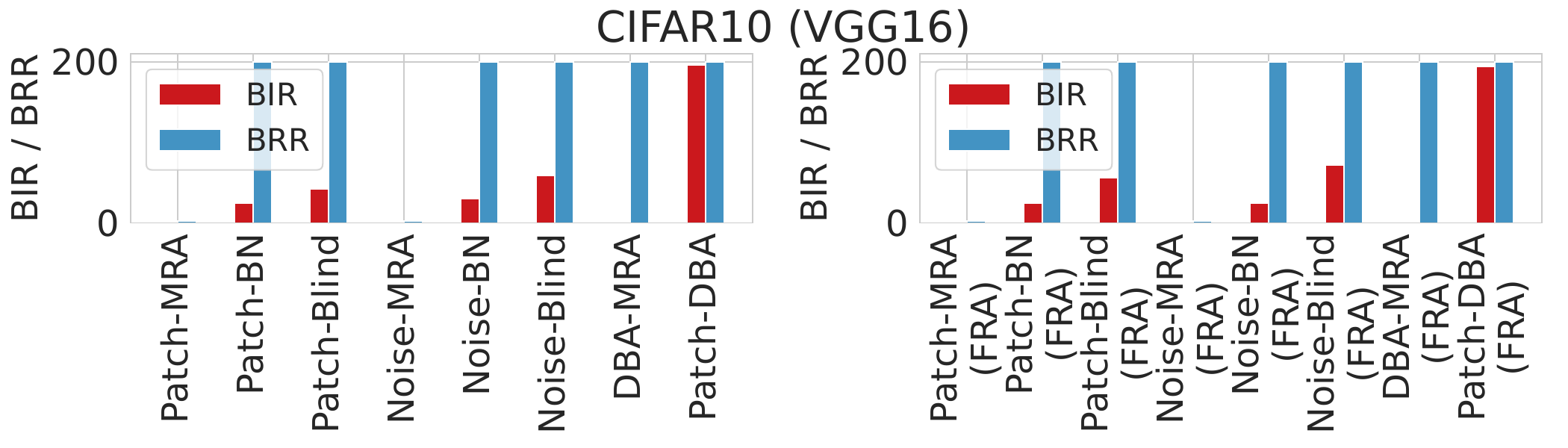}
  \includegraphics[width=0.85\linewidth]{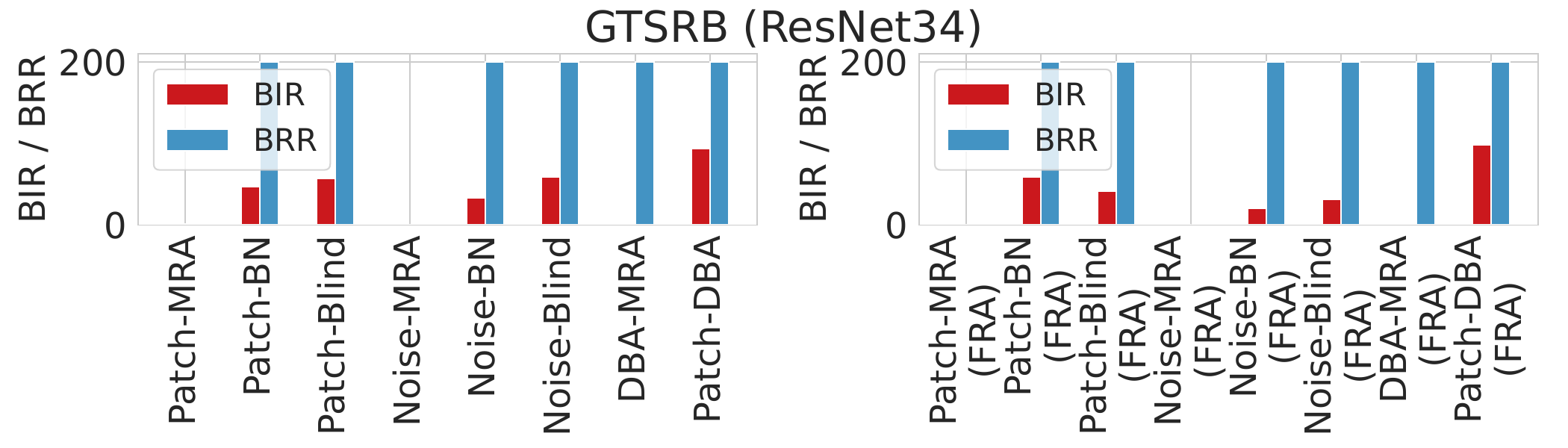}
  \includegraphics[width=0.75\linewidth]{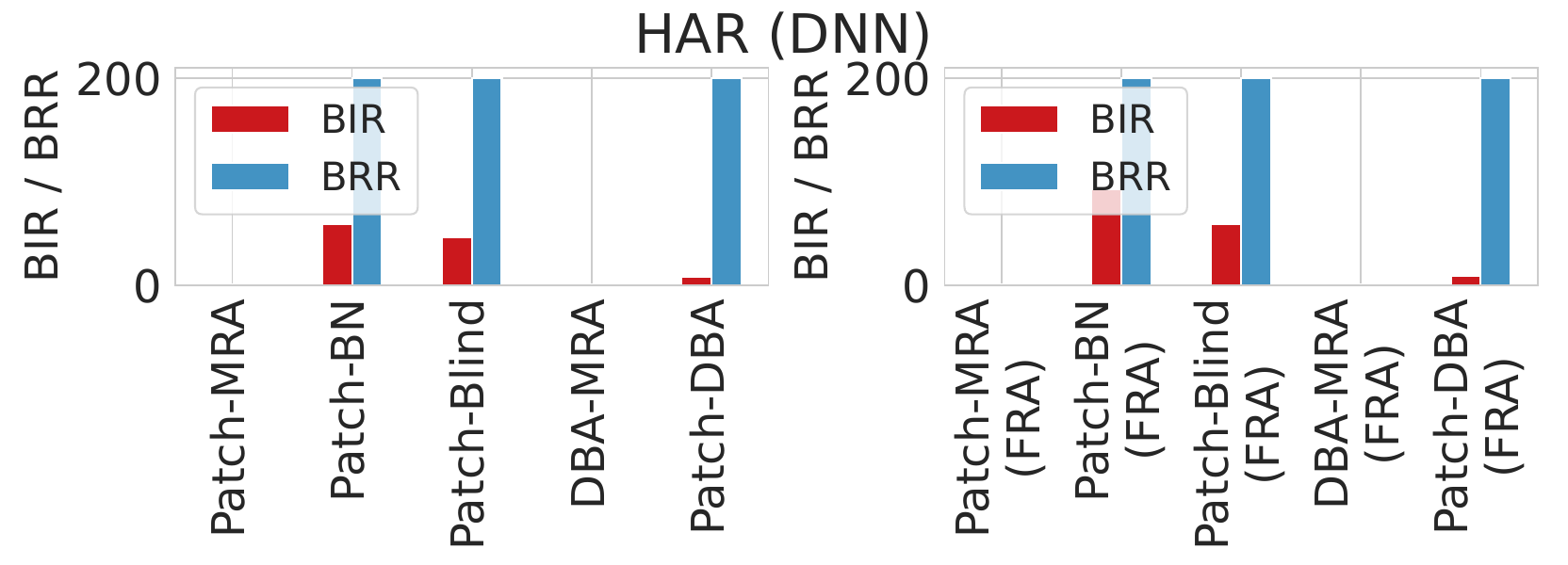}
  \vspace{-3mm}
  \caption{The backdoor injection rounds (BIR) and backdoor removal rounds (BRR) of backdoor attacks in non-IID settings.} % under different datasets and model structures
  \label{fig:Robustness_tar_noniid}
\end{figure}

% \textbf{Efficiency and robustness of backdoor attacks:}
Fig.~\ref{fig:Robustness_tar} shows the backdoor injection rounds and backdoor removal rounds of backdoor attacks in non-IID settings. All attacks with and without Patch-FRA (except MRA) achieve an attack success rate of over $90\%$, with an average of only $36$ and $39$ backdoors injection rounds required. In almost all attacks (except MRA), the backdoor removal rounds exceed more than $200$ rounds. In Patch-MRA, the adversary injects a backdoor for one round, and the global model takes more than $84$ rounds to remove the backdoor. 
%The low number of backdoor injection rounds and the high number of backdoor removal rounds imply that backdoor attacks are robust under FL, posing a significant security risk to FL. We also report the results for other datasets.
Such small backdoor injection rounds and large backdoor removal rounds imply that backdoor attacks are robust, posing a significant security risk to FL. (see Fig.~\ref{fig:Robustness_tar_iid} and Fig.~\ref{fig:Robustness_tar_noniid} for other datasets).  % please refer to \href{https://anonymous.4open.science/r/FLTracer-F7BB/FLTracer_Additional_Experimental_Results.pdf}{GitHub} % see Fig.~\ref{fig:Robustness_tar_iid} and Fig.~\ref{fig:Robustness_tar_noniid} for other datasets %

% \vspace{-2mm}
\begin{center}
\fcolorbox{darkblue}{gray!10}{\parbox{.96\linewidth}{\textbf{\textcolor{darkblue}{Remark 3}}. The 16 backdoor attacks can rapidly inject backdoors and persist for much more training rounds.}} %, and the backdoor injection rounds are proportional to the backdoor removal rounds.
\end{center}
% \vspace{-2mm}
% \textbf{Takeaway 1.} 
% All attacks considered (14 untargeted attacks and 8 classical backdoor attacks, except for C-D to C-F) pose a substantial threat to FL. 

% \begin{figure}[]
%   \centering
% %   \vspace{-3mm}
%   \subfigure[Targeted classical attacks]{
%   \includegraphics[width=0.9\linewidth]{./figures/target_cl_feasi_ASR}
  
%   }\vspace{-1mm}
%   \subfigure[Targeted adaptive attacks]{
%   \includegraphics[width=0.9\linewidth]{./figures/target_ea_feasi_ASR}
%   }
%   \vspace{-2mm}
%   \caption{Attack success rate of backdoor attacks. }
%   \label{fig:ASR_tar}
%   \vspace{-5mm}
% \end{figure}

% \vspace{-0.5mm}
\noindent\textbf{Detection against attacks:} 
In non-IID settings, the effectiveness of existing methods against untargeted and backdoor attacks is unstable, as shown in Table~\ref{tab:tpr_untarget} (column MKrum and DnC) and Table~\ref{tab:tpr_target} (column Mkrum, FLAME, and DnC). DnC is only effective against a few specific attacks, such as Add noise, MB Median, MB Trmean, and backdoor attacks under ViT, but is less effective against other attacks. The effectiveness of Mkrum is generally low under non-IID because Mkrum can only guarantee convergence of FL, which is not sufficient to detect malicious clients~\cite{bagdasaryan2020backdoor}. We also observe that existing methods are less effective on complex models and datasets (such as CIFAR10 and ResNet/VGG) because anomalies are more difficult to detect in complex settings. 
% The low performance of Fine-pruning is because the server under FL does not have a large dataset to fine-tune the model, resulting in a significant decrease in the classification accuracy of the global model. 
% DnC has high effectiveness only in MRA and Patch-BN attacks but has low effectiveness in Blind attacks due to the stealthiness of blind backdoor attacks. 

% \begin{figure}[]
%   \centering
%   % \vspace{-7mm}
%   \includegraphics[width=0.96\linewidth]{./figures/feasibility_defense_effect}
%   \vspace{-2mm}
%   \caption{Effectiveness of existing detection methods against attacks in non-IID settings (H:high; M:medium; L:low).}
%   %We apply DnC and Median against untargeted attacks. We apply DnC and Fine-pruning against backdoor attacks.} %CIFAR10, ResNet18
%   \vspace{-6mm}
%   \label{fig:resist}
% \end{figure}

% \vspace{-3mm}
\begin{center}
\fcolorbox{darkblue}{gray!10}{\parbox{.96\linewidth}{\textbf{\textcolor{darkblue}{Remark 4}}. Existing detection can only ensure high accuracy in detecting a small fraction of specific attacks, sacrificing high false positive rates (FPR) in non-IID settings.}}
\end{center}
% \vspace{-2mm}
% \textbf{Takeaway 2.} 
% Existing detection methods can only ensure high precision in detecting a small fraction of attacks and have high false positive rates in non-IID settings. %According to the assessment,

\vspace{-2mm}
\section{DP-based Defense against backdoor attacks} \label{sec:DP_defense}
\begin{figure}[!ht]
\centering
  \subfigure{
  \includegraphics[width=0.8\linewidth]{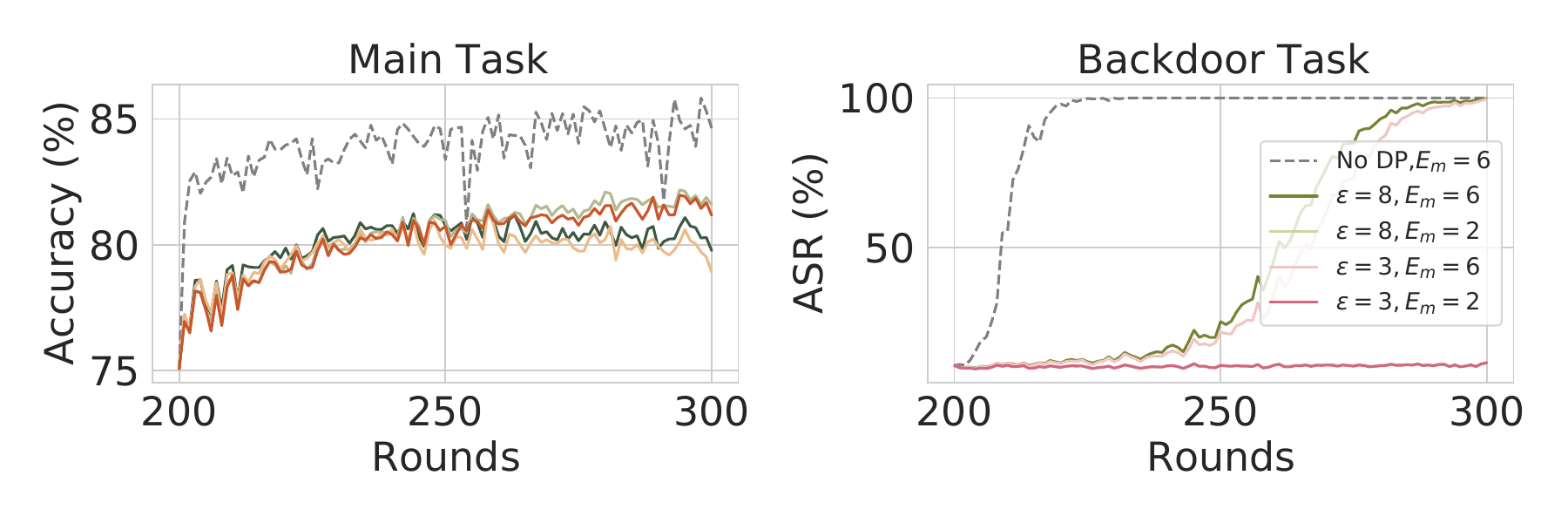}
  }
\vskip -4mm
  \centering
  \subfigure{
  \includegraphics[width=0.8\linewidth]{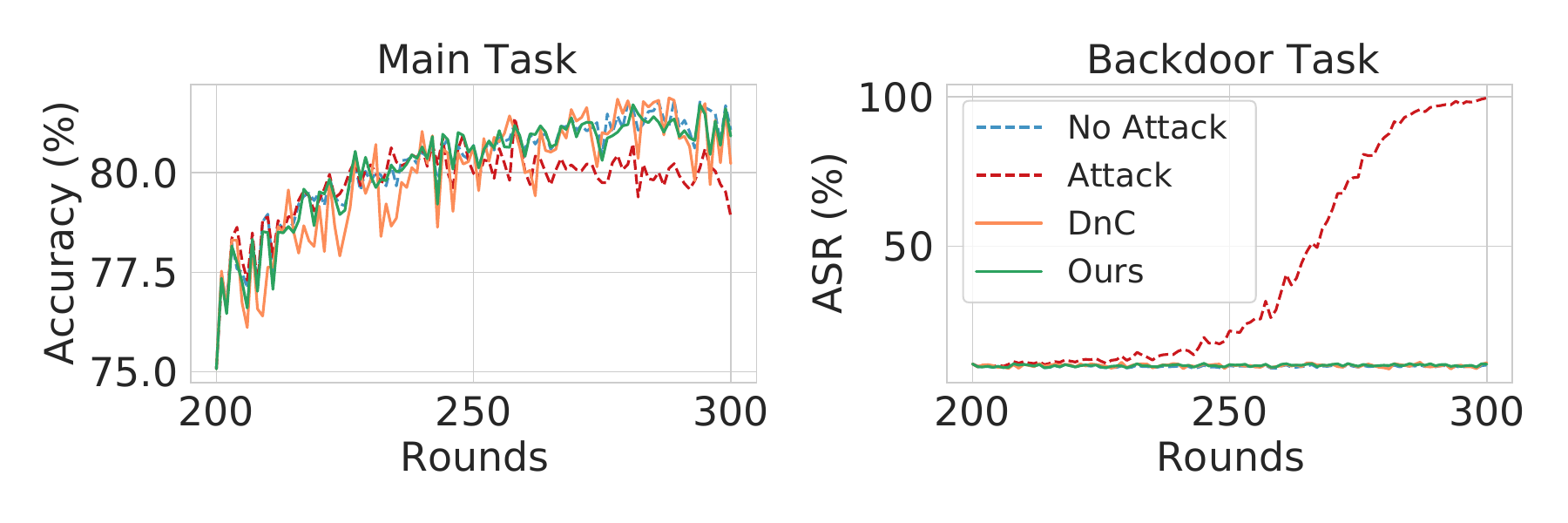}
  }
\vskip -4mm
  \subfigure{
  \includegraphics[width=0.8\linewidth]{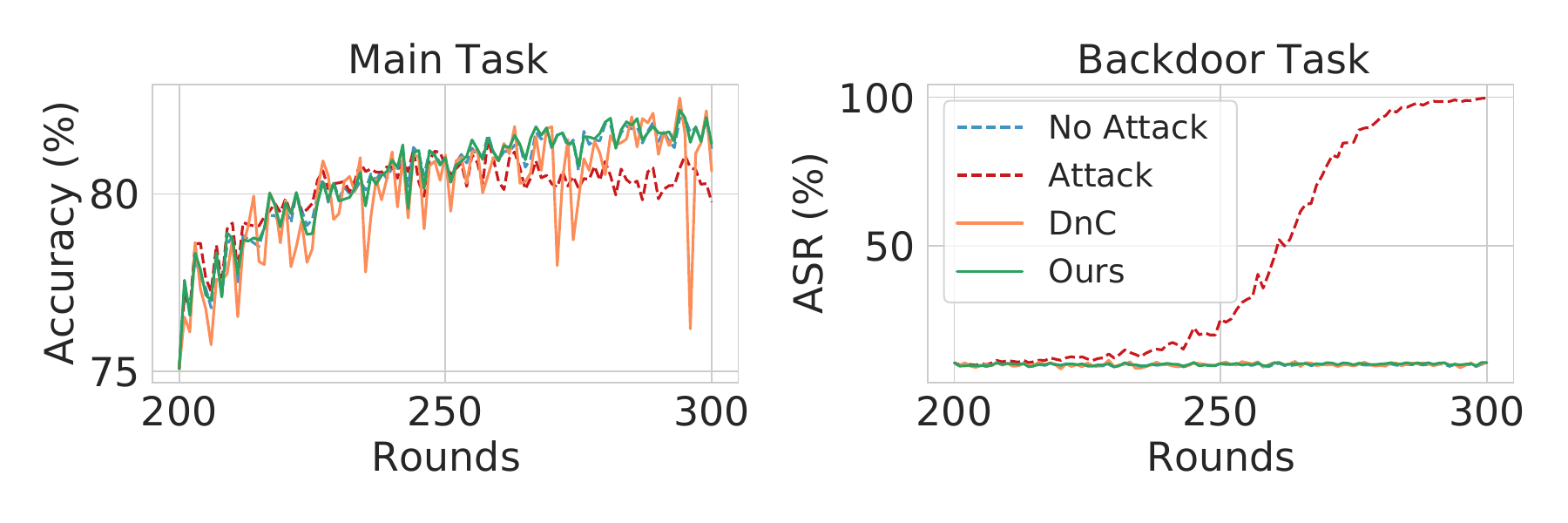}
  }
\vskip -5mm
  \caption{Accuracy and attack success rate (ASR) on CIFAR10 (ResNet18) under DP-based method. Top-under DP ($E_m$: malicious epochs). Middle-w/ detection under DP with sensitivity $\epsilon=3$. Bottom-w/ detection under DP with $\epsilon=8$.}
  \label{fig:acc_t_dp_detect}
 \vspace{-5mm}
\end{figure}

We evaluate the effectiveness of backdoor attacks launched under different DP strategies (see Fig.~\ref{fig:acc_t_dp_detect} (top)). We notice that adversaries can inject backdoors successfully through optimized attack strategies, such as increasing the adversary's local training epochs. Despite DP's inability to defend such optimized attacks, our method is still effective under DP.
Fig.\,\ref{fig:acc_t_dp_detect}\,(middle and bottom) show the effectiveness of our method against this attack. The results show that FLTracer can prevent backdoor injections (attack success rate$=\!10\%$) and the global model's accuracy can approximate the no-attack setting.

% \begin{figure}[ht]
%   \centering
%   \includegraphics[width=0.80\linewidth]{./figures/acc_attack_dp_lr=0.01}  \caption{The accuracy and attack success rate of backdoor attacks on CIFAR10 (ResNet-18) under LDP ($E_m$: poison epoch).}
%   \label{fig:acc_attack_dp}
% \end{figure}
\vspace{-5mm}

% \noindent\textbf{Results for data over time.} Our evaluations are based on typical FL settings. However, our system may be affected in a special Non-IID setting, that is, for a client, its local dataset changes over time, but the data distribution is consistent. For example, in some IoT applications, client datasets need to be updated regularly. Therefore, we evaluate the effectiveness of our system against backdoor attacks in a time-varying Non-IID setting. Note that, we do not consider Non-IID settings where the data distribution changes over time. Considering the amount of local dataset, we set a total of 50 clients, and the number of clients participating in each round is still 10, of which attackers account for 10\%. We follow the Non-IID setting of d = 0.5. A client divides the local dataset into two parts equally, first randomly selecting 50\%
% data for normal training lasting 200 rounds, and then using the remaining data to train a local model continually. An adversary launches a backdoor attack in round 200, lasting for 100 rounds. Fig.~\ref{fig:acc_data_change} shows the results for our system against A-M attack using BN trigger. TPR is 99\% and FPR is 0.56\%, and both accuracy and attack success rate are similar to normal models. That implies that our system is effective in the time-varying Non-IID setting.

\vspace{-2mm}
\section{Runtime Analysis} \label{sec:appendix_runtime}
We offer a baseline runtime for attack provenance, which includes the detection time and attack tracing time of FLTracer. In non-IID settings, the average runtime of FLTracer is $8.68$ seconds per round  (s/ r) for backdoor attacks on CIFAR10 (ResNet18). FLTracer has a relatively short attack tracing time overhead when compared to detection-only methods like MKrum ($5.51$ s/ r), FLAME ($1.2$ s/ r), and DnC ($4.95$ s/ r). 

Referring to the results in Table~\ref{tab:tpr_target} and Fig.~\ref{fig:robust_evaul_cad}, FLAME, despite its quick detection time, fails to prevent backdoor injection, such as Patch-FRA. 
On the other hand, while both DnC and MKrum successfully prevent backdoor injection, they significantly slow down the training of the global model due to their high FPR. 
For instance, to prevent Patch-BadNets attacks, FLTracer only requires 37 training rounds to achieve the same level of accuracy as a global model that is protected by DnC with 100 training rounds. The global model accuracy of Mkrum is lower than that of DnC. In summary, FLTracer achieves the predefined global model accuracy in the shortest time while ensuring backdoor-proof protection.

\vspace{-2mm}
\section{Performance of FLTracer under no attack} \label{sec:no_attack}
Table~\ref{tab:no_attack_noniid} lists the results of FLTracer under no attack in non-IID settings. It shows that FLTracer has little impact on the global model and increases its stability. In particular, FLTracer achieves a low $8\%$ FPR (under joint), which is close to the sum of the FPRs of the four features. This means that each feature detects malicious clients from different attack locations. The clean accuracy drop of FLTracer decreases by a total of $0.28\%$, and the model stability decreases by $0.4$. This is because without attacks, the updates of the clients with extremely heterogeneous data are significantly different from others, and aggregating these small fractions of updates decreases the global model performance. This observation matches the conclusion in~\cite{wang2020attack}. (see Table~\ref{tab:no_attack_iid} for IID).

\begin{table}[!ht]
\centering
\vspace{-2mm}
\footnotesize
\caption{Detection results under no attack in non-IID settings.}
\vspace{-2mm}
\setlength\tabcolsep{3pt}
\resizebox{\linewidth}{!}{
\begin{tabular}{|l|c||ccccc|}
\hline
      & Baseline & \texttt{signv}     & \texttt{sortv}     & \texttt{classv}     & \texttt{featv}     & Joint  \\ \hline
False positive rate & -         & 1.80\% & 3.30\% & 3.00\% & 1.70\% & 8.00\% \\ 
\begin{tabular}[c]{@{}l@{}}Accuracy\\Clean accuracy drop\end{tabular} &
  \begin{tabular}[c]{@{}c@{}}84.38\\0\end{tabular} &
  \begin{tabular}[c]{@{}c@{}}84.97\\-0.59\end{tabular} &
  \begin{tabular}[c]{@{}c@{}}84.99\\-0.61\end{tabular} &
  \begin{tabular}[c]{@{}c@{}}84.49\\-0.11\end{tabular} &
  \begin{tabular}[c]{@{}c@{}}84.37\\+0.01\end{tabular} &
  \begin{tabular}[c]{@{}c@{}}84.65\\-0.28\end{tabular} \\ 
Model stability  & 1.06      & 0.95   & 0.56   & 0.68   & 0.92   & 0.66   \\ \hline
\end{tabular}
}
\label{tab:no_attack_noniid}
\vspace{-3mm}
\end{table}

\begin{table}[!ht]
\centering
\vspace{-2mm}
\footnotesize
\caption{Detection results under no attack in IID settings.}
\vspace{-2mm}
\setlength\tabcolsep{3pt}
\resizebox{\linewidth}{!}{
\begin{tabular}{|l|c||ccccc|}
\hline
      & Baseline & \texttt{signv}     & \texttt{sortv}     & \texttt{classv}     & \texttt{featv}     & Joint  \\ \hline
False positive rate & -         & 0.20\% & 3.80\% & 0.60\% & 1.00\% & 4.50\% \\ 
\begin{tabular}[c]{@{}l@{}}Accuracy\\Clean accuracy drop\end{tabular} &
  \begin{tabular}[c]{@{}c@{}}89.76\\0\end{tabular} &
  \begin{tabular}[c]{@{}c@{}}89.86\\-0.10\end{tabular} &
  \begin{tabular}[c]{@{}c@{}}89.84\\-0.08\end{tabular} &
  \begin{tabular}[c]{@{}c@{}}90.04\\-0.27\end{tabular} &
  \begin{tabular}[c]{@{}c@{}}89.86\\+0.10\end{tabular} &
  \begin{tabular}[c]{@{}c@{}}89.85\\-0.09\end{tabular} \\ 
Model stability   & 0.11      & 0.08   & 0.07   & 0.08   & 0.08   & 0.14   \\ \hline
\end{tabular}
}
\label{tab:no_attack_iid}
\end{table}

\vspace{-4mm}
\section{Additional Experimental Results} \label{sec:appendix_evaluation}

\begin{table}[!h]
\caption{Comparing TPR($\%$) and FPR($\%$) of  DnC and FLTracer (Ours) against backdoor attacks in IID settings. } %``DBA'' and ``FRA'' denotes distributed backdoor and feature replacement attack.
\vspace{-2mm}
\centering
\scriptsize
\begin{tabular}{cccccc}
\toprule
\multirow{2}{*}{\begin{tabular}[c]{@{}c@{}}Dataset \\ (Model)\end{tabular}}    & \multirow{2}{*}{Attack} & \multicolumn{2}{c}{DnC~\cite{shejwalkar2021manipulating}} & \multicolumn{2}{c}{FLTracer (Ours)} \\ \cmidrule(r){3-4} \cmidrule(r){5-6}
                                                                             &                      & TPR   & FPR   & TPR   & FPR   \\ \midrule
\multirow{5}{*}{\begin{tabular}[c]{@{}c@{}}CIFAR10\\  (ResNet18)\end{tabular}} & Patch-BadNets           & 100.0      & 0.00      & 100.0       & 0.00      \\
                                                                             & Noise-BadNets        & 100.0 & 0.00 & 100.0 & 0.00 \\
                                                                             & Patch-DBA                  & 98.50 & 4.13 & 100.0 & 0.00 \\
                                                                             & Patch-Blind                & 92.00 & 0.89 & 100.0 & 0.00 \\
                                                                             & Patch-FRA                  & 100.0 & 0.00 & 100.0 & 0.11 \\ \midrule
\multirow{5}{*}{\begin{tabular}[c]{@{}c@{}}CIFAR10\\  (VGG16)\end{tabular}}  & Patch-BadNets        & 100.0 & 0.00 & 100.0 & 0.00 \\
                                                                             & Noise-BadNets        & 100.0 & 0.00 & 100.0 & 0.11 \\
                                                                             & Patch-DBA                  & 100.0 & 0.00 & 100.0 & 0.00 \\
                                                                             & Patch-Blind                & 100.0 & 0.00 & 100.0 & 0.00 \\
                                                                             & Patch-FRA                  & 100.0 & 0.00 & 100.0 & 0.00 \\ \midrule
\multirow{5}{*}{\begin{tabular}[c]{@{}c@{}}GTSRB\\  (ResNet34)\end{tabular}} & Patch-BadNets        & 100.0 & 0.00 & 100.0 & 0.00 \\
                                                                             & Noise-BadNets        & 100.0 & 0.00 & 100.0 & 0.11 \\
                                                                             & Patch-DBA                  & 93.00 & 10.25 & 99.00 & 0.00 \\
                                                                             & Patch-Blind                & 100.0 & 0.11 & 100.0 & 0.00 \\
                                                                             & Patch-FRA                  & 95.00 & 23.67 & 100.0 & 0.00 \\ \midrule
\textbf{Average}                                                             & \multicolumn{1}{l}{} & 98.57 & 2.60 & \textbf{99.93} & \textbf{0.02} \\ \bottomrule
\end{tabular}
\end{table}

% \vspace{-4mm}
\begin{table}[!ht]
\caption{Comparing TPR($\%$) and FPR($\%$) of  DnC, and FLTracer (Ours) against untargeted attacks in IID settings.}
\vspace{-2mm}
\centering
\scriptsize
\begin{tabular}{clcccc}
\toprule
\multirow{2}{*}{\begin{tabular}[c]{@{}c@{}}Dataset\\  (Model)\end{tabular}} &
  \multicolumn{1}{c}{\multirow{2}{*}{Attack}} &
  \multicolumn{2}{c}{DnC~\cite{shejwalkar2021manipulating}} &
  \multicolumn{2}{c}{FLTracer (Ours)} \\ \cmidrule(r){3-4} \cmidrule(r){5-6} 
                                                                                 & \multicolumn{1}{c}{} & TPR   & FPR   & TPR   & FPR   \\ \midrule
\multirow{11}{*}{\begin{tabular}[c]{@{}c@{}}MNIST\\  (SimpleNet)\end{tabular}} & Add Noise            & 100.0 & 0.00 & 100.0 & 0.00 \\
                                                                                 & Sign-flipping        & 74.00 & 10.13 & 100.0 & 0.00 \\
                                                                                 & Dirty label (Fix-Fix)          & 98.50 & 0.38 & 98.00 & 3.63 \\
                                                                                 & MB Mkrum             & 100.0 & 0.00 & 100.0 & 0.00 \\
                                                                                 & MB Bulyan            & 100.0 & 0.00 & 100.0 & 0.00 \\
                                                                                 & MB Median            & 100.0 & 0.00 & 100.0 & 0.00 \\
                                                                                 & MB Trmean            & 100.0 & 0.00 & 100.0 & 0.00 \\
                                                                                 & MB Max           & 100.0 & 0.00 & 100.0 & 0.00 \\
                                                                                 & MB Sum           & 100.0 & 0.00 & 100.0 & 0.00 \\
                                                                                 & Fang Mkrum           & 100.0 & 0.00 & 100.0 & 0.00 \\
                                                                                 & Fang Trmean          & 100.0 & 0.00 & 100.0 & 0.00 \\ 
\midrule
\multirow{11}{*}{\begin{tabular}[c]{@{}c@{}}CIFAR10\\  (AlexNet)\end{tabular}}   & Add Noise            & 100.0 & 0.00 & 100.0 & 0.00 \\
                                                                                 & Sign-flipping        & 99.50 & 22.50 & 100.0 & 0.00 \\
                                                                                 & Dirty label (Fix-Fix)          & 100.0 & 1.13 & 100.0 & 2.94 \\
                                                                                 & MB Mkrum             & 99.50 & 1.13 & 92.00 & 1.63 \\
                                                                                 & MB Bulyan            & 98.50 & 1.88 & 92.25 & 2.69 \\
                                                                                 & MB Median            & 100.0 & 0.00 & 100.0 & 0.00 \\
                                                                                 & MB Trmean            & 100.0 & 0.00 & 96.00 & 0.50 \\
                                                                                 & MB Max           & 99.75 & 1.50 & 92.00 & 2.25 \\
                                                                                 & MB Sum           & 97.50 & 4.25 & 94.50 & 4.50 \\
                                                                                 & Fang Mkrum           & 34.25 & 27.69 & 100.0 & 0.00 \\
                                                                                 & Fang Trmean          & 98.25 & 2.25 & 100.0 & 0.00 \\ \midrule
\multirow{11}{*}{\begin{tabular}[c]{@{}c@{}}CIFAR10\\  (ResNet18)\end{tabular}}  & Add Noise            & 99.00 & 0.25 & 99.50 & 0.38 \\
                                                                                 & Sign-flipping        & 68.00 & 23.75 & 100.0 & 0.13 \\
                                                                                 & Dirty label (Fix-Fix)          & 99.49 & 9.38 & 100.0 & 1.56 \\
                                                                                 & MB Mkrum             & 62.50 & 41.31 & 100.0 & 0.19 \\
                                                                                 & MB Bulyan            & 51.75 & 45.94 & 99.25 & 0.19 \\
                                                                                 & MB Median            & 100.0 & 0.00 & 100.0 & 0.00 \\
                                                                                 & MB Trmean            & 100.0 & 0.00 & 100.0 & 0.00 \\
                                                                                 & MB Max           & 50.25 & 42.25 & 100.0 & 0.06 \\
                                                                                 & MB Sum           & 33.75 & 52.69 & 100.0 & 0.00 \\
                                                                                 & Fang Mkrum           & 49.17 & 48.40 & 98.61 & 0.00 \\
                                                                                 & Fang Trmean          & 19.25 & 38.63 & 100.0 & 0.00 \\ 
\midrule
\textbf{Average} &
   &
  \multicolumn{1}{r}{85.85} &
  \multicolumn{1}{r}{11.38} &
  \multicolumn{1}{r}{\textbf{98.85}} &
  \multicolumn{1}{r}{\textbf{0.63}} \\ \bottomrule
\end{tabular}
\end{table}

% \vspace{-7mm}

%%%%%%%%%%%%%%%%%%%%%%%%%%%%%%%%%%%%%%%%%%%%%%%%%%%%%%%%% appendix in guthub

\end{document}